\documentclass[
		twoside,openright,titlepage,numbers=noenddot,headinclude,
                footinclude=true,cleardoublepage=empty,
                BCOR=5mm,paper=a4,fontsize=11pt, 
                american,polish 
                ]{scrreprt} 
                
\usepackage{graphicx}
\usepackage{epsf}
\usepackage{bm}
\usepackage{amsmath}
\usepackage{amsfonts}
\usepackage{amssymb}
\usepackage{epstopdf}
\usepackage{color}
                
\PassOptionsToPackage{eulerchapternumbers,listings, pdfspacing, subfig,beramono,eulermath,parts,dottedtoc}{classicthesis}

\newcommand{\myTitle}{Extended Theories of Gravity in cosmological and astrophysical applications\xspace}
\newcommand{\mySubtitle}{PhD Dissertation\xspace}

\newcommand{\myName}{Aneta Wojnar\xspace}
\newcommand{\myProf}{Prof. Dr hab. Andrzej Borowiec\xspace}
\newcommand{\myOtherProf}{Prof. Salvatore Capozziello\xspace}

\newcommand{\myInsitute}{Insitute for Theoretical Physics\xspace}
\newcommand{\myFaculty}{Department of Physics and Astronomy\xspace}

\newcommand{\myUni}{University of Wroc{\l}aw\xspace}

\newcommand{\myOtherFaculty}{Department of Physics}

\newcommand{\myOtherUni}{Univeristy of Napoli ``Federico II''\xspace}

\newcommand{\myVersion}{version 4.0\xspace}

\newcounter{dummy} 
\providecommand{\mLyX}{L\kern-.1667em\lower.25em\hbox{Y}\kern-.125emX\@}

\usepackage{lipsum} 

\PassOptionsToPackage{utf8}{inputenc} 
\usepackage{inputenc}
 
\usepackage{babel}

\PassOptionsToPackage{square,numbers}{natbib}
 \usepackage{natbib}
 
\PassOptionsToPackage{fleqn}{amsmath} 
 \usepackage{amsmath}
 
\PassOptionsToPackage{T1}{fontenc} 
\usepackage{fontenc}

\usepackage{xspace} 

\usepackage{mparhack} 

\usepackage{fixltx2e} 

\PassOptionsToPackage{smaller}{acronym} 
\usepackage{acronym} 

\PassOptionsToPackage{pdftex}{graphicx}
\usepackage{graphicx} 

\usepackage{tabularx} 
\usepackage{caption}
\usepackage{subfig}  

\usepackage{listings} 
\PassOptionsToPackage{pdftex,hyperfootnotes=false,pdfpagelabels}{hyperref}
\usepackage{hyperref}  
\hypersetup{
colorlinks=true, linktocpage=true, pdfstartpage=3, pdfstartview=FitV,
breaklinks=true, pdfpagemode=UseNone, pageanchor=true, pdfpagemode=UseOutlines,
plainpages=false, bookmarksnumbered, bookmarksopen=true, bookmarksopenlevel=1,
hypertexnames=true, pdfhighlight=/O, urlcolor=webbrown, linkcolor=RoyalBlue, citecolor=webgreen,
pdftitle={\myTitle},
pdfauthor={\textcopyright\ \myName, \myUni, \myFaculty},
pdfsubject={},
pdfkeywords={},
pdfcreator={pdfLaTeX},
pdfproducer={LaTeX with hyperref and classicthesis}
}   

\usepackage{ifthen} 
\newboolean{enable-backrefs} 
\setboolean{enable-backrefs}{false} 

\newcommand{\backrefnotcitedstring}{\relax} 
\newcommand{\backrefcitedsinglestring}[1]{(Cited on page~#1.)}
\newcommand{\backrefcitedmultistring}[1]{(Cited on pages~#1.)}
\ifthenelse{\boolean{enable-backrefs}} 
{
\PassOptionsToPackage{hyperpageref}{backref}
\usepackage{backref} 
\renewcommand*{\backref}[1]{}  
\renewcommand*{\backrefalt}[4]{
\ifcase #1 
\backrefnotcitedstring
\or
\backrefcitedsinglestring{#2}
\else
\backrefcitedmultistring{#2}
\fi}
}{\relax} 

\makeatletter
\@ifpackageloaded{babel}
{
\addto\extrasamerican{

}
\addto\extrasngerman{

}
}{\relax}
\makeatother

\usepackage{classicthesis} 

\usepackage[hmargin=4cm]{geometry}

\usepackage{multicol}

\begin{document}

\frenchspacing 

\raggedbottom 

\selectlanguage{american} 


\pagenumbering{roman} 

\pagestyle{plain} 

\newtheorem{theorem}{Theorem}
\newtheorem{defin}{Definition}



\begin{titlepage}

\begin{addmargin}{-3cm}
\begin{center}
\large

\hfill
\vfill

\begingroup
\color{Maroon}\spacedallcaps{\myTitle} \\ \bigskip
\endgroup

\spacedlowsmallcaps{Aneta Wojnar} 

\vfill
\vfill

\mySubtitle \\ \medskip 
\vfill

\begin{multicols}{2}
 \myProf \\[2ex]
 \myInsitute \\
 \myFaculty \\
 \myUni \\
\columnbreak
 \myOtherProf \\[2ex]
 \myOtherFaculty \\
 \myOtherUni \\
\end{multicols}
\bigskip



\end{center}
\end{addmargin}

\end{titlepage} 

\cleardoublepage

\thispagestyle{empty}
\refstepcounter{dummy}

\pdfbookmark[1]{Dedication}{Dedication} 

\vspace*{3cm}

\begin{center}
Mamie i Tacie 
 \\ \medskip
                -- Ziazia   
\end{center}

\medskip
\begin{center}
 Marcinie,\\ \smallskip
za Twoją nieskończoną (oby!) cierpliwość i miłość
\end{center}
\medskip
\begin{center}
 Babciom Bogusi, Marysi oraz Dziadzi Józiowi,\\ \smallskip
za bu{\l}ki na parze, jab{\l}ecznik i niekończące się poczucie humoru
\end{center}
\medskip
\begin{center}
Pamięci Stanis{\l}awa Graczyka. \\ \smallskip
1933\,--\,2013
\end{center} 


\cleardoublepage

\pdfbookmark[1]{Abstract}{Abstract} 

\begingroup
\let\clearpage\relax
\let\cleardoublepage\relax
\let\cleardoublepage\relax

\chapter*{Abstract} 

The main subjects of the PhD dissertation concern cosmological models considered in Palatini $f(\mathcal{R})$ gravity and scalar - tensor theories. We 
introduce a simple generalization of the $\Lambda$CDM model which 
is based on Palatini modified gravity with quadratic Starobinsky term. A matter source is provided by generalized Chaplygin gas. The statistical analysis of our 
model is investigated as well as we use dynamical system approach to study the evolution of the Universe. The model reaches a very good agreement with the newest 
experimental data and yields an inflationary epoch which is caused by a singularity of the type $III$. The present-day accelerated expansion is also provided
by the model.

We also show that the Lie and Noether symmetry approaches are very useful tools in cosmological considerations. We examine two other models of Extended Theories 
of Gravity (ETGs), that is, the novel hybrid metric-Palatini gravity and a minimally coupled to gravity scalar field as the simplest example of scalar-tensor theories.
The first one is applied to homogeneous and isotropic model while in the scalar - tensor theory we study anisotropic universes. We use Lie and Noether symmetries 
in order to find unknown forms of potential and to solve classical field equations in both models. The symmetries also are very helpful in searching exact and 
invariant solutions of Wheeler-DeWitt equations which are a quantized version of modified Einstein's equations.

In the last part we are interested in equilibrium configurations and stability conditions of relativistic stars in the framework of scalar - tensor theories. 
Firstly, we show that TOV-like form of the equilibrium equations can be obtained for a big class of ETGs if generalized energy density and pressure are defined. 
According to our studies, a neutron star is a stable system for the minimally coupled scalar field model.

There is a supplement including notes on symmetries as well as dynamical systems approach. The illustrative examples of applications are also provided.

\endgroup			

\vfill 
\cleardoublepage

\pdfbookmark[1]{Streszczenie}{Streszczenie} 

\begingroup
\let\clearpage\relax
\let\cleardoublepage\relax
\let\cleardoublepage\relax

\chapter*{Streszczenie} 
Główne problemy rozważane w przedstawionej rozprawie doktorskiej dotyczą kosmologicznych modeli w teoriach Palatiniego oraz skalarno - tensorowych. Pierwszym 
badanym modelem jest proste rozszerzenie modelu $\Lambda$CDM. Badany model oparty jest na dodaniu do lagranżjanu kwadratowego członu Starobinskiego oraz założeniu,
że metryka i koneksja są niezależnymi wielkościami. Źródłem materii jest uogólniony gaz Chaplygina. Statystyczna analiza pokazuje, że model zgadza z najnowszymi 
danymi obserwacyjnymi. Badanie punktów osobliwych pozwoliło na określenie rodzaju osobliwości kosmologicznych. Badanie Wszechświata jako dwuwymiarowego
dynamicznego układu pokazuje, że osobliwość typu $III$ przyczynia się do nagłej przyspieszonej ekspansji, tj. inflacji, we wczesnej fazie ewolucji. Co więcej, w 
rozważanym modelu mamy również epokę obecnej przyspieszonej ekspansji Wszechświata.

Dwa pozostałe kosmologiczne modele są badane za pomocą narzędzi, które dostarczają symetrie Noether oraz Liego. Model hybrydowej grawitacji, łączący formalizm 
metryczny z formalizmem Palatiniego, jest rozważany dla jednorodnych i izotropowych wszechświatów natomiast model z minimalnie sprzężonym do grawitacji polem 
skalarnym dotyczy anizotropowych czasoprzestrzeni. Symetrie Liego oraz Noether są używane do rozwiązywania klasycznych oraz kwantowych 
równań (tj. równań Wheelera-DeWitta).
Rozwiązania są dokładnie i niezmiennicze. Symetrie używane są również do znalezienia postaci potencjału pola skalarnego, którego określenie jest niezbędne do 
rozwiązania równań.

Ostatnia część rozprawy dotyczy relatywistycznych gwiazd, np. gwiazd neutronowych, w rozszerzonych teoriach grawitacji. Pokazano, że można otrzymać równania 
analogiczne do równań TOV (Tolmana - Oppenheimera - Volkoffa) z uogólnionym ciśnieniem i gęstością energii. Dodatko zbadano także stabilność takiego układu dla 
modelu z minimalnie sprzężonym polem skalarnym.

Końcowa część zawiera suplement, w którym zawarto główne metody matematyczne używane w rozprawie: symetrie oraz układy dynamiczne. Znajdują się tutaj również 
przykłady użycia tych metod.

\endgroup			

\vfill 

\cleardoublepage

\pdfbookmark[1]{Publications}{Publications} 

\chapter*{Publications} 
This PhD dissertation consists of research done at the Institute for Theoretical Physics, University of Wroc{\l}aw and Department of Physics, University of 
Napoli "Federico II" in collaboration with Prof. Dr hab. Andrzej Borowiec and Prof. Salvatore Capozziello.

Some ideas and figures have appeared previously in the following publications:
\begin{itemize}
 \item  A. Borowiec, S. Capozziello, M. De Laurentis, F.S.N Lobo, A. Paliathanasis, M. Paolella, A. Wojnar 
 "Invariant Solutions and Noether Symmetries in Hybrid Gravity"
 
PRD $91(2015):023517$ (arXiv: $1407.4313$)
\item   A. Borowiec, A. Stachowski, M. Szydłowski, A. Wojnar "Inflationary cosmology with Chaplygin gas in Palatini formalism" 

JCAP $01 (2016) 040$ (arXiv: $1512.01199$)
\item  M. Szyd{\l}owski, A. Stachowski, A. Borowiec, A. Wojnar "Do sewn singularities falsify the Palatini cosmology?" 

pre-print (arXiv:$1512.04580$)
\item  H. Velten, A. M. Oliveira, A. Wojnar "A free parametrized TOV: Modified Gravity from Newtonian to Relativistic Stars" 

PoS(MPCS$2015$)$025$  (arXiv:$1601.03000$)
\item A. Paliathanasis, L. Karpathopoulos, A. Wojnar, 
S. Capozziello \newline "Wheeler-DeWitt equation and Lie symmetries in Bianchi scalar - field cosmology" 

EPJ C $76(4) 2016$, (arXiv:$1601.06528$)
\item A. Wojnar and H. Velten "Equilibrium and stability of relativistic stars in extended theories of gravity"  

pre-print (arXiv:$1604.04257$) 
\end{itemize}

\bigskip


\cleardoublepage

\pdfbookmark[1]{Acknowledgements}{Acknowledgements} 


\bigskip


\begingroup

\let\clearpage\relax
\let\cleardoublepage\relax
\let\cleardoublepage\relax

\chapter*{Acknowledgements} 

\noindent Foremost, I would like to thank my advisors Prof. Dr hab. Andrzej Borowiec and Prof. Salvatore Capozziello. Undertaking this PhD would not have 
been possible to do without your guidance, patience, many stimulating discussions, and consistent encouragement that I have been receiving throughout 
the research work. 

Very special thanks to Prof. Dr hab. Marek Szyd{\l}owski for his enthusiasm and opportunities to work on exciting projects.  
I would like to thank my colleagues: Dr Andronikos Paliathanasis, Dr Mariafelicia De Laurentis, Dr Francisco Lobo, 
Dr Hermano Velten, Mariacristina Paolella, Leonidas Karpathopoulos, Aleksander Stachowski, and Adriano Oliveira for inspiring ideas and enjoyable atmosphere during the work.
\\

\noindent I am also indebted to my wonderful Family. Without your unconditional support and love I would never be here. 
Many thanks to my friends met in Wroc{\l}aw, Napoli and other places who have been supporting me since the very beginning. 

\bigskip


\endgroup 

\pagestyle{scrheadings} 

\cleardoublepage

\refstepcounter{dummy}

\pdfbookmark[1]{\contentsname}{tableofcontents} 

\setcounter{tocdepth}{2} 

\setcounter{secnumdepth}{3} 

\manualmark
\markboth{\spacedlowsmallcaps{\contentsname}}{\spacedlowsmallcaps{\contentsname}}
\tableofcontents 
\automark[section]{chapter}
\renewcommand{\chaptermark}[1]{\markboth{\spacedlowsmallcaps{#1}}{\spacedlowsmallcaps{#1}}}
\renewcommand{\sectionmark}[1]{\markright{\thesection\enspace\spacedlowsmallcaps{#1}}}

\clearpage

\begingroup 
\let\clearpage\relax
\let\cleardoublepage\relax
\let\cleardoublepage\relax

\endgroup 

\cleardoublepage

\pagenumbering{arabic} 

\cleardoublepage 





\chapter{Introduction}\label{introduction}


In November $2015$ we celebrated $100$ years of General Relativity introduced by Albert Einstein \cite{einstein1915field, einstein1916found}, a theory of gravitation 
which has changed our thinking about the Universe and about other phenomena related to gravitational interactions. Confirmed by many observations, such as the perihelion precession of 
Mercury \cite{einstein1916found}, the deflection of light by the Sun \cite{dyson1920determination}, the gravitational redshift of light 
\cite{pound1959gravitational, pound1960apparent} or the expanding universe \cite{hubble1929relation}, General Relativity is a basic theory that we use for any 
gravitational phenomena that we want to explain. This year has been also very special because of another anniversary: the first exact solution of
Einstein's field equations published in $1916$ by Karl Schwarzschild \cite{schwarzschild1916gravitationsfeld, schwarzschild1999gravitational}. Moreover,
on $11$th of February $2016$, we experienced a very exciting announcement about a merger of binary black hole whose observed effects were gravitational waves 
\cite{abbott2016observation}. This is the first direct detection of gravitational waves and the first observation of a binary black hole merger. One should mention
that the detected waveform matches the prediction of General Relativity for a gravitational wave emanating from such a particular binary system.
The success of Einstein's theory has made a lot of difficulties since there have appeared many problems in fundamental physics, astrophysics, and cosmology
that one is not able to explain with the help of General Relativity. However, abandoning it, seems to be something inappropriate. Instead of that one attempts to 
slightly modify the theory. The first modification was done by Einstein himself by introducing a cosmological constant to the gravitational action in order
to make the cosmological solutions static. He believed that the Universe does not expand. Nowadays, for cosmological purposes, one uses the equations with the 
cosmological constant in order to ensure that the field equations provide the scenario which is in agreement with observations indicating that the Universe undergoes 
the late-time acceleration \cite{copeland2006dynamics, huterer1999prospects}. The most popular approach to the cosmological constant problem is included in so-called 
$\Lambda$CDM model ($\Lambda$ Cold Dark Matter model) that is, the standard cosmological model. The model is described by Einstein's field equations considered in Friedmann - Robertson- 
Lemaitre - Walker (FRLW) metric background with the cosmological constant added to the gravitational Lagrangian. The effect of the acceleration is 
explained by exotic fluid called dark energy represented by $\Lambda$. It has negative pressure and contributes about $68.3\%$ of the total energy in the present-day 
observable Universe \cite{ade2014planck}. Another weird component of the standard cosmological model is dark matter \cite{report, capozziello2010beyond} which contributes $26.8\%$.
The dark matter interacts only gravitationally: there is no direct observational evidence that it exists since it does not emit electromagnetic radiation. Its existence and 
properties are given by gravitational effects such as for instance the motions of visible matter. The problem is mainly indicated by galaxies rotation which
is the discrepancy between observed 
galaxy rotation curves and the theoretical prediction based on the virial theorem. The amount of the ordinary baryonic matter that we are able to
detect is just $4.9\%$ of the total energy while neutrinos and photons contribute insignificantly so usually they are neglected in 
theoretical considerations. Besides the mentioned exotic ingredients which we do not understand, there are many other
unsolved problems as inflation \cite{starobinsky1980new, guth1981inflationary}, cosmological singularities (for instance Big Bang),
issues related to quantum field theories in curved spacetime, 
non-renormalization of Einstein's theory, unification of gravity with other interactions which has been already unified into the Standard Model, that 
is, electromagnetism, weak and strong interactions. 

Despite these problems, the success of the $\Lambda$CDM model makes alternative theories beyond the standard model unattractive to many physicists. It seems that it is 
easier to accept three unknown components which are required by the $\Lambda$CDM model, that is, dark matter, dark energy and the inflaton field as an agent of 
the inflation phase than to look for a new model or to modify the old one. Additionally, it is assumed 
that the Universe is isotropic and homogeneous on large scales (Cosmological Principle), inflation happened, and that it is based on General Relativity as a correct 
theory to describe the Universe aside from the quantum regime \cite{bull2015beyond}. The Cosmological Principle is deduced from Copernican Principle, it means that 
our galaxy worldline is not special so if we observe isotropy about our worldline, there is isotropy about other galaxies worldlines, too. That implies homogeneity and also 
leads to FRLW geometry of spacetime \cite{bull2015beyond, clarkson2010inhomogeneity}. However, it is not so sure that isotropic Cosmic Microwave 
Background (CMB) radiation means isotropic spacetime \cite{ellis1978expansion}. Such considerations indicate that we may not limit ourselves only to FRLW 
spacetime but also one should study anisotropic and/or inhomogeneous models.

It is believed that dark matter may correspond to weak interacting particles which have not been 
observed yet. There exist alternative models of gravity which explain dark matter effects without a need of the new particle, namely it is a purely gravitational 
effect. Similarly, there are proposals that also accelerating expansion can be described by some mechanism arising from modified theories of gravity instead of 
dark energy. There are two main ideas: dark energy as the cosmological constant or scalar fields, both with the feature of negative pressure. Introducing 
the $\Lambda$ component, that is, cosmological constant, just made the situation even worst since more problems appeared. Classically, it can be treated as a free 
parameter which can be fixed to any value that we want; particularly to the value indicated by cosmological observations. On the other hand, one usually considers
$\Lambda$ as vacuum energy coming from different matter fields and the theoretical predictions on its value can be done. The estimated value of the cosmological
constant is about (in reduced Planck units) $10^{-120}$ while from the Standard Model of particle physics one deals with the value $1$. It also seems to be very 
ambiguous that its observational
value is so small but it strongly dominates the Universe evolution today as well as it may contribute to the structure formation epoch. 

As already mentioned, the $\Lambda$CDM model requires the inflation epoch that happened after the Big Bang. The cosmological inflation is the early Universe accelerating 
expansion introduced by A. Guth \cite{guth1981inflationary}. The inflation theory turn out to explain a lot of compelling problems such as for example
cosmic size of the Universe, its large-scale structure, isotropy, homogeneity and flatness. Although its simplicity and explanations of the above issues, it also 
causes problems. One would like to understand what made the Universe evolution to start accelerating and then to slow down. The most popular idea is scalar field but 
immediately the question arises: what is a form of the potential of that field? There are also many other proposals coming from alternative theories of gravity. They are 
inspired by the Starobinsky proposal \cite{starobinsky1980new} which is in very good agreement with Planck data: he considers an extra quadratic 
term in the gravitational Lagrangian (see the Chapter \ref{chap1}).

Since the Einstein's gravity and $\Lambda$CDM model derived from it have passed positively Solar System tests (for review see 
for example \cite{will1981theory, de2010f}) and matched so far the observational data, one claims that Einstein's gravity is just an effective theory. Due to that fact 
one looks for different approaches in order to find a good theory which is able to answer the above problems and tells us more about 
the Universe that we live in. Moreover, we would like to have a theory which unifies all known interactions and describes quantum effects which had appeared after 
Big Bang. No satisfactory result has been obtained so far (string theory, supersymmetry) that 
could combine particle physics and gravitation. There has been many years of research since $1915$ but none of
proposed models was considered as satisfactory. Due to that fact, one needs new theories of gravity
which should be checked carefully with all possible tools that we possess. Each new theory should
pass many theoretical and observational tests and also agree with GR in the case of weak
gravitational limit.
Working on Extended Theories of Gravity, specially on Palatini theories can give clues to work on a
consistent theory merging gravitation and quantum physics. It should also get closer to the answer if
one needs to add scalar field in order to explain the inflation phenomena. Could the modified
geometry solve the problem of the accelerated epoch at the beginning of our Universe? Are there
different scenarios of the origin of the Universe indicating by the existence of singularities than the
one given by the $\Lambda$CDM model, that is, Big Bang? What are exactly cosmological singularities? Can we get to know more about them without Quantum Gravity theory?

The thesis is divided into two parts: the first one consists of some examples of Extended Theories of Gravity while in the Appendix we briefly introduce mathematical 
tools. The Appendix \ref{app_lie} describes Lie symmetry method and show how it can be used for solving ordinary and partial differential equations. We also 
represent a subclass of Lie symmetries, that is, the Noether symmetries having a serious consequences in physics. In order to see how Lie symmetry method works, 
illustrative examples are provided. We also discuss the connection to conformal algebra of Riemannian metric which is a phase space metric of a 
physical system. Moreover, in the Appendix \ref{apB} we show how the dynamical systems theory can be used for studies of cosmological models. As some of them can be
recast into two dimensional cases, we focus on phase portraits of linear systems in $2$-dimensional vector space. Later on, as an example we consider $\Lambda$CDM 
model as a dynamical system.

In the main part we examine three models of Extended Theories of Gravity. The Chapter \ref{chap1} includes a simple generalization of the $\Lambda$CDM model which 
is based on Palatini modified gravity with quadratic Starobinsky term and generalized Chaplygin gas as a matter source. We show that it provides inflation as well as 
the current accelerated expansion. The singularity which appears in the model turns out to be responsible for the inflationary epoch. We also perform statistical 
analysis in order to find values favored by astronomical data. Subsequently, we classify all evolutionary paths in the model phase space using dynamical system theory.
The another model that we study in the Chapter \ref{hg_chapter} also have something in common with the Palatini gravity. It is a recently proposed model
whose Lagrangian consists of the standard Einstein-Hilbert term considered in metric formalism, and an arbitrary function of the Palatini curvature scalar. For 
this investigation we use Lie and Noether symmetries in order to select $f(\mathcal{R})$ form. The symmetries also help us to solve the field 
equations for the selected model. Quantizing the model, we derive Wheeler-DeWitt equation: its invariant solution can be also given by the Lie symmetries which are 
determined by the methods discussed in the Appendix \ref{app_lie}. 

In the last Chapter \ref{chap3} we present the simplest representation of the scalar-tensor theory, it means, we focus on minimally coupled to 
gravity scalar field. Here, we are focused on cosmology provided by the Bianchi 
spacetimes. We examine anisotropic models in which the scalar field also contributes. We perform similar analysis using Lie symmetries methods as it was done 
for Hybrid Gravity. Additionally, we study WKB approximations in order to find classical solutions, that is, anisotropy parameters, which are given 
as functions of time. The further part of this chapter concerns
configurations of relativistic stars. We show that for a general (not specified) form of Extended Theories of Gravity one may write the equilibrium equations in the 
Tolman-Oppenheimer-Volkoff-like form with new definitions of energy density and pressure. Since the stability criterion of relativistic star system must be 
considered case by case, we study the problem for the minimally coupled scalar field.







\chapter{$f(\text{R})$ gravity in Palatini formalism} \label{chap1}
The most natural way (as we do not want to say the simplest one) to extend our considerations on gravity beyond General Relativity is to study some geometric modifications 
of the Einstein's theory. The geometric part of the gravitational action can be changed in many different ways. One may assume that constant of Nature are not really constant values. 
A scalar field might be added into Lagrangian and moreover, it can be minimally or non-minimally coupled to gravity (to Ricci scalar). One proposes much more 
complicated functionals than the simple linear one used in GR, for example $f(R)$ gravity. The latter approach has gained a lot of interest recently as the extra
geometric terms could explain not only the dark matter issue \cite{capozziello2006dark, capozziello2007low} but also the dark energy problem
because it produces the accelerated late-time effect at low cosmic 
densities (it means when the trace 
of the energy momentum tensor goes to zero). The field equations also differ from the Einstein's ones so they could provide different behavior of the early
Universe. 
The $f(R)$ gravity can be treated in two different ways: the metric approach and Palatini one \cite{sotiriou2010f}. The former arises to the fourth order differential equations 
which are difficult to handle and moreover, one believes that physical equations of motion should be of the second order. In contrast to the metric formalism,
the Palatini $f(\mathcal{R})$ gravity provides second order differential equations since the connection and metric are treated as independent objects. The Riemann and Ricci 
tensors are constructed with the connection while for building the Ricci scalar we also use the physical metric in order to contract the indices.  The Palatini 
approach is very important in cosmology because one may use the trace of the field equations derived with respect to the metric in order to obtain the $\mathcal{R}(a)$
dependence where $a=a(t)$ is a scale factor of the FRLW metric. Then we get the Friedmann equation which rules the dynamics of the Universe in a given model so 
the model might be compared with the observational data \cite{bor_kam}. Such dynamics has a form of Newtonian one so one deals with an effective potential term depending only on 
the scale factor $a(t)$. Furthermore, the theory coincides with GR (with a dynamical feature that the connection is a Levi-Civita connection of the metric in comparison
to GR where this is a priori assumption) if the functional is linear in $\mathcal{R}$.
There also exist disadvantages of such an approach: being in conflict with the Standard Model of particle physics
\cite{sotiriou2010f, flanagan2004palatini, iglesias2007not, olmo2008hydrogen}, surface singularities of static 
spherically symmetric objects in the case of polytropic EoS \cite{barausse2008curvature}, the algebraic dependence of the post-Newtonian metric on the 
density \cite{olmo2005gravity, sotiriou2006nearly}, and the
complications with the initial values problem in the presence of matter \cite{ferraris1994universality, sotiriou2006f}, although the problem was already solved 
in \cite{olmo2011hamiltonian}. Another one happens at 
microscopic scales, that is, the theory produces instabilities in atoms which disintegrate them. However, it was shown \cite{olmo2009dynamical} that high curvature
corrections do 
not cause such problem. What is also very promising, some of the Palatini Lagrangians avoid the Big Bang singularity. What should be also emphasized, the 
effective dynamics of Loop 
Quantum Gravity can be reproduced by the Palatini theory which gives the link to one of approaches to Quantum Gravity \cite{olmo2009covariant}. High curvature
correction of 
the form $f(\mathcal{R}^{\mu\nu}\mathcal{R}_{\mu\nu})$ changes the notion of the independent connection: in the simple Palatini f(R) gravity the connection is auxiliary field 
while in the more general 
Palatini theory it is dynamical without making the equations of motion second order in the fields. Furthermore, as already remarked in the Introduction, the 
squared curvature terms improve the renormalization \cite{stelle1977renormalization}.

Palatini theories seem to be very promising and hence more investigations should be performed. We will start with the simplest representant of the Palatini theories,
that is, we will study $f(\mathcal{R})$ gravity in the mentioned formalism. We will briefly introduce the main assumptions of the theory in order to construct 
our model which is examined from theoretical and observational points of view.

\section{Introduction of the model}
Palatini formalism of $f(\mathcal{R})$ theory of gravity is based on the gravitation action in which not only the standard Hilbert - Einstein action
$S=\frac{1}{2\kappa}\int\mathrm{d}^4x\sqrt{-g} R$ is replaced by an arbitrary function of the Ricci scalar $f(R)$ \cite{report, de2010f, capozziello2010beyond, sotiriou2010f}, but one uses Palatini scalar $\mathcal{R}$ instead 
of the metric one $R$:
\begin{equation}\label{action}
 S=\frac{1}{2\kappa}\int\mathrm{d}^4x\sqrt{-g}f(\mathcal{R})+ S_m(g_{\mu\nu},\psi).
\end{equation}
where $\kappa=8\pi G$ is as usually the Einstein constant.
The Palatini curvature scalar $\mathcal{R}=g^{\mu\nu}\mathcal{R}_{\mu\nu}(\hat{\Gamma})$ is constructed with the metric-independent connection $\hat{\Gamma}$.  
The metric $g_{\mu\nu}$ is used for raising and lowering indices. The action $S_m$ denotes a matter action which depends only on the metric $g_{\mu\nu}$ and 
matter fields but it is independent of the connection $\hat{\Gamma}$.
One varies the action with respect to two independent objects: the metric $g_{\mu\nu}$ and the connection $\hat{\Gamma}$. The first variation, after applying 
the Palatini formula 
\begin{equation}
 \delta\mathcal{R}_{\mu\nu}=\hat{\nabla}_\lambda\delta\hat{\Gamma}^\lambda_{\;\mu\nu}-\hat{\nabla}_\nu\delta\hat{\Gamma}^\lambda_{\;\mu\lambda},
\end{equation}
 gives rise to the equation
\begin{equation}\label{var_met}
 f'(\mathcal{R})\mathcal{R}_{\mu\nu}-\frac{1}{2}f(\mathcal{R})g_{\mu\nu}=\kappa T_{\mu\nu}.
\end{equation}
The prime denotes the differentiation with respect to $\mathcal{R}$ while $T_{\mu\nu}$ is the standard energy-momentum 
tensor given by the variation of the matter action with respect to $g_{\mu\nu}$:
\begin{equation}
 T_{\mu\nu}:=-\frac{2}{\sqrt{-g}}\frac{\delta(\sqrt{-g}\mathcal{L}_m)}{\delta(g^{\mu\nu})}.
\end{equation}
One should mention that the energy-momentum tensor is conserved \cite{sotiriou2010f, barraco1999conservation, koivisto2006note} by the covariant 
derivative which is defined with the Levi-Civita connection of the metric $g_{\mu\nu}$
\begin{equation}
 \nabla_\mu T^{\mu\nu}=0\;\;\;\text{but not }\;\;\bar{\nabla}_\mu T^{\mu\nu}=0.
\end{equation}
The consequence of the above condition is the motion of the test particles: they follow geodesics of the metric. Although, there exists another possibility, it means, 
that the particles follow the geodesics provided by the 
connection \cite{capozziello2009dark, capozziello2015extended, fatibene2013mathematical, capozziello2015equivalence, fatibene2012characterization}. 
But from now on, we will assume that they follow the metric 
ones so the theory satisfies the metric postulates \cite{sotiriou2010f, will1981theory} and Einstein Equivalence 
Principle \cite{einstein1989swiss, misner1973gravitation}.
The trace of (\ref{var_met}) with respect to $g^{\mu\nu}$ gives us the structural (master) equation of the spacetime which 
controls (\ref{var_met}) \cite{alle_bor1, alle_bor2, alle_bor3}:
\begin{equation}\label{struc}
  f'(\mathcal{R})\mathcal{R}-2f(\mathcal{R})=\kappa T.
\end{equation}
Assuming that one has a given function $f(\mathcal{R})$, we may solve (\ref{struc}) and express a solution as $\mathcal{R}(T)$. Hence, $f(\mathcal{R})$ is a 
function of $T$ being the trace of the energy-momentum tensor $T=g^{\mu\nu}T_{\mu\nu}$\cite{alle_bor1, alle_bor2, alle_bor3}.

Following the approach of \cite{alle_bor1, alle_bor2, alle_bor3}, the generalized Einstein's equations can be also written as
\begin{equation}\label{EinR}
 \hat{R}_{\mu\nu}(\Gamma)=g_{\mu\alpha}P^\alpha_\nu,
\end{equation}
where the operator $P^\alpha_\nu$ used above consists of two scalars $b$ and $c$ depending on $\hat{R}$
\begin{align}\label{operator}
 P^\alpha_\nu=&\frac{c}{b}\delta^\mu_\nu+\frac{1}{b}T^\mu_\nu,\\
 b=&b(\hat{R})=f'(\hat{R}),\;\;c=c(\hat{R})=\frac{1}{2} f(\hat{R}),
\end{align}
which will be useful later.

The variation with respect to the connection leads to the equation
\begin{equation}\label{var_con}
- \hat{\nabla}_\alpha(\sqrt{-g}f'(\mathcal{R})g^{\mu\nu})+
 \hat{\nabla}_\sigma(\sqrt{-g}f'(\mathcal{R})g^{\sigma(\mu})\delta^{\nu)}_\alpha=0,
\end{equation}
where $(\mu\nu)$ denotes a symmetrization over the indices $\mu$ and $\nu$. The trace with respect to $g$ allows us to write down the second equation of motion 
(\ref{var_con}) as
\begin{equation}\label{Lev_Civ}
 \hat{\nabla}_\alpha(\sqrt{-g}f'(\mathcal{R})g^{\mu\nu})=0
\end{equation}
which stands for the Levi-Civita connection of the metric 
\begin{equation}\label{Pal_met}
\tilde{g}_{\mu\nu}=f'(\mathcal{R})g_{\mu\nu}.
\end{equation}
One notices that choosing $f(\mathcal{R})=\mathcal{R}$ leads to 
General Relativity and from the equation (\ref{Lev_Civ}) we get the metric-independent connection $\hat{\Gamma}$ is a Levi-Civita connection of the metric $g_{\mu\nu}$. 
Immediately it follows that $\mathcal{R}_{\mu\nu}=R_{\mu\nu},\;\mathcal{R}=R$ so the equation (\ref{var_met}) becomes Einstein's equation. It should be noticed that 
in 
the case of General Gravity derived by Palatini formalism we deal with two equations of motion which one of them indicates that the connection $\hat{\Gamma}$ 
is the Levi-Civita one of the metric $g$, it means $\tilde{g}_{\mu\nu}=g_{\mu\nu}$. As a contrary to GR, this is a dynamical feature, not the assumption.

Moreover, the Palatini equations of motion may be written as ones depending only on the metric and matter field \cite{sotiriou2010f}. 
As one has (\ref{Pal_met}), we may use the conformal relations between Ricci tensors and scalars (recall that one uses the metric $g_{\mu\nu}$ but 
not $\tilde{g}_{\mu\nu}$ for raising and lowering indices):
\begin{equation}\label{conf_rel}
  \mathcal{R}_{\mu\nu}=R_{\mu\nu}+\frac{3}{2}\frac{F(\mathcal{R})_{;\mu}F(\mathcal{R})_{;\nu}}{F^2(\mathcal{R})}-
 \frac{1}{F(\mathcal{R})}\nabla_\mu F(\mathcal{R})_{;\nu}-\frac{1}{2}\frac{g_{\mu\nu}\Box F(\mathcal{R})}{F(\mathcal{R})},
\end{equation}
in order to rewrite the equation (\ref{var_met}) as
\begin{align}
 G_{\mu\nu}=&\frac{\kappa}{f'}T_{\mu\nu}-\frac{1}{2}g_{\mu\nu}\left( \mathcal{R}-\frac{f}{f'} \right) +
 \frac{1}{f'}(\nabla_\mu\nabla_\nu-g_{\mu\nu}\Box)f'\nonumber\\
 -&\frac{3}{2}\frac{1}{f'^2}\left( (\nabla_\mu f')(\nabla_\nu f')-\frac{1}{2}g_{\mu\nu}(\nabla f')^2 \right)
\end{align}
which is the standard GR equation with the modified source term, where $G_{\mu\nu}$ is the Einstein tensor, it means $G_{\mu\nu}=R_{\mu\nu}-\frac{1}{2}Rg_{\mu\nu}$.

Let us just briefly discuss perfect fluid energy-momentum tensor which will stand for the energy-momentum tensor in (\ref{var_met}):
\begin{equation}\label{perfect}
 T_{\mu\nu}=\rho u_\mu u_\nu + p h_{\mu\nu},
\end{equation}
where $\rho$ and $p=p(\rho)$ are energy density and pressure of the fluid, respectively. The vector $u^\mu$ is an observer 
co-moving with the fluid satisfying $g_{\mu\nu}u^\mu u^\nu=-1$ and $h^\mu_\nu=\delta^\mu_\nu+u^\mu u_\nu$ is 
a 3-projector tensor projecting 4-dimensional object on 3-dimensional hypersurface in the case when the observer $u$ is rotation-free.
Hence, the trace of (\ref{perfect}) is
\begin{equation}\label{trace}
 T=3p-\rho.
\end{equation}

\subsection{$f(\mathcal{R})$ gravity as a scalar-tensor theory}
The theory under our consideration may be also transformed into a Brans-Dicke theory with a self-interacting potential of a scalar field. The theories (for a 
special choice of the parameter $\omega$ - see below) are mathematically equivalent but one should be careful when apply physics: that is,
the theories do not have to be physically equivalent (see 
e.g. \cite{capozziello2015extended, fatibene2013mathematical, capozziello2015equivalence, fatibene2012characterization}).
We are going to introduce an
auxiliary field $\chi$ in order to get a dynamically equivalent theory, it means, different representations of the same theory \cite{sotiriou2010f}.

Let us introduce the field $\chi$ and write the dynamically equivalent action:
\begin{equation}
 S=\frac{1}{2\kappa}\int\mathrm{d}^4x\sqrt{-g}[f(\chi)+f'(\chi)(\mathcal{R}-\chi)]+ S_m(g_{\mu\nu},\psi)
\end{equation}
whose variation with respect to the field $\chi$ gives
\begin{equation}
 f''(\chi)(\mathcal{R}-\chi)=0.
\end{equation}
From the above condition it turns out that $\chi=\mathcal{R}$ if $f''(\chi)\neq0$ which obviously gives rise to the action (\ref{action}). If we redefine the scalar 
field $\chi$ by $\phi=f'(\chi)$ and define the potential of the field $\phi$ as
\begin{equation}
 V(\phi)=\chi(\phi)\phi-f(\chi(\phi)),
\end{equation}
then the Palatini action will take the following form
\begin{equation}
 S=\frac{1}{2\kappa}\int\mathrm{d}^4x\sqrt{-g}[\phi\mathcal{R}-V(\phi)]+ S_m(g_{\mu\nu},\psi).
\end{equation}
It is important to make a comment here that the just obtained action is not an action of the Brans-Dicke theory since $\mathcal{R}$ is not the Ricci scalar of the metric
$g_{\mu\nu}$. But we may use the conformal relation (\ref{conf_rel}) and rewrite the action which is now (skipping the boundary term)
\begin{equation}
 S=\frac{1}{2\kappa}\int\mathrm{d}^4x\sqrt{-g}[\phi R+\frac{3}{2\phi}\nabla_\mu\phi\nabla^\mu\phi-V(\phi)]+ S_m(g_{\mu\nu},\psi).
\end{equation}
It is a Brans-Dicke action with B-D parameter $\omega_0=-\frac{3}{2}$. The variations, now taken with respect to the metric and the scalar field, are
\begin{align}
 G_{\mu\nu}=&\frac{\kappa}{\phi}T_{\mu\nu}-\frac{3}{2\phi^2}\left( \nabla_\mu\phi\nabla_\nu\phi-\frac{1}{2}g_{\mu\nu}\nabla^\alpha\phi\nabla_\alpha\phi \right)\nonumber\\
 +&\frac{1}{\phi} (\nabla_\mu\nabla_\nu\phi-g_{\mu\nu}\Box\phi)-\frac{V}{2\phi}g_{\mu\nu},\label{Pal_BD1}\\
 \Box\phi=&\frac{\phi}{3}(R-V')+\frac{1}{2\phi}\nabla^\mu\phi\nabla_\mu\phi.\label{Pal_BD2}
\end{align}
If we take the trace of the equation (\ref{Pal_BD1}) in order to eliminate the Ricci scalar $R$ in the equation (\ref{Pal_BD2}), we will get
\begin{equation}\label{notdyn}
 2V-\phi V'=\kappa T.
\end{equation}
From the obtained relation we see that the scalar field $\phi$ is algebraically related to the matter source, in means, it is not a dynamical field. Due to that fact, the Palatini 
$f(\mathcal{R})$ gravity is in conflict with the Standard Model of particle physics when we consider it as a metric 
theory \cite{sotiriou2010f, flanagan2004palatini, iglesias2007not, barausse2008curvature}. 

It should be also mentioned that performing the conformal transformation (\ref{Pal_met}) (transferring the action into the 
Einstein frame) \cite{allemandi2006conformal, capozziello2010beyond, faraoni1999einstein}, but without rescaling 
the scalar field, the action is
\begin{equation}
 S'=\int\mathrm{d}^4x\sqrt{-g}\left[\frac{\tilde{R}}{2\kappa} - U(\phi)\right]+ S_m(\phi^{-1}h_{\mu\nu},\psi),
\end{equation}
where $U(\phi)=\frac{V(\phi)}{2\kappa\phi^2}$. The topic is very controversial hence we are not going to discuss it here as we will not work in Einstein frame.

\subsection{FRLW cosmology in Palatini formalism}
As the observations of the cosmic microwave background (CMB) indicate that our Universe is highly isotropic and homogeneous, it allows us to assume a perfect fluid 
description (\ref{perfect}) for the matter and the
Friedmann-Robertson-Lemaitre-Walker (FRLW) background metric
\begin{equation}\label{frlw}
 ds^2=-dt^2+a^2(t)\left[ \frac{1}{1-kr^2}dr^2+r^2(d\theta^2+\sin^2{\theta}d\phi^2) \right].
\end{equation}
The scalar $k=0,1,-1$ stands for the space curvature and $a(t)$ is a scale factor depending on cosmological time $t$. The energy-momentum
tensor (\ref{perfect}) satisfies the metric covariant conservation law $\nabla^\mu T_{\mu\nu}=0$ which gives arise to the continuity
equation
\begin{equation}\label{continuity}
 \dot{\rho}+3H(p+\rho)=0,
\end{equation}
where $H=\frac{\dot{a}}{a}$ is the Hubble constant. The relation between the pressure $p$ and energy density $\rho$, i.e.
equation of state $p=p(\rho)$, leads to a dependence of $\rho$ on the scale factor $a(t)$ (by using (\ref{frlw}) and (\ref{continuity})).
The generalized Einstein's field equations (\ref{EinR}) for the FRLW metric becomes the generalized Friedmann 
equation \cite{alle_bor1, alle_bor2}
\begin{equation}
 \left( \frac{\dot{a}}{a}+\frac{\dot{b}}{2b} \right)^2 + \frac{k}{a^2}=\frac{1}{2}P^1_1-\frac{1}{6}P^0_0,
\end{equation}
where $P^1_1$ and $P^0_0$ are the components of (\ref{operator}). Therefore, the modified Friedmann equation 
may be written as \cite{alle_bor1, alle_bor2, bor_kam, alle_bor3}:
\begin{equation}\label{friedCOS}
 \left( \frac{\dot{a}}{a}+\frac{\dot{b}}{2b} \right)^2+\frac{k}{a^2}=\frac{1}{2}\left( \frac{c}{b}+\frac{p}{b} \right) -
 \frac{1}{6}\left( \frac{c}{b}-\frac{\rho}{b} \right).
\end{equation}

\section{Palatini cosmology with Generalized Chaplygin Gas}

\subsection{Chaplygin Gas as a dark side of the Universe}

In the standard approach to cosmology one uses the barotropic equation of state (EoS) $p=\omega\rho$ for matter filling the homogeneous and isotropic 
universe. The matter is represented by the perfect fluid energy-momentum tensor 
\begin{equation}
 T_{\mu\nu}=\rho u_{\mu} u_{\nu} + ph_{\mu\nu}
\end{equation}
where $\rho$ is an energy density while $p$ is pressure of the fluid considered in the model. Together with the continuity equation (\ref{continuity}) the 
equation of state allows to write the energy density in terms of the scale factor $a(t)$. For the standard, barotropic equation one gets
\begin{equation}
 \rho=\rho(a)\propto a^{-3(1+\omega)}.
\end{equation}
Dark energy is represented by the fluid with negative pressure, that is, $\omega=-1$ in the barotropic EoS while the dark matter is supposed to behave 
as pressureless dust with barotropic EoS $\omega=0$.

There exists an idea that the dark side of the Universe can be unified into single exotic fluid, so-called Chaplygin Gas which recently has gained a lot 
of attention in the literature \cite{jackiw1999fluid, ogawa2000remark}. 
 It was introduced by Sergey Chaplygin in $1904$ in order to compute the lifting force on a wing of an airplane 
in aerodynamics \cite{chap} 
but it has also been used in cosmology
\cite{kamenshchik2001alternative, bento, lu2009cosmology, bilic2002unification, popov2010dark, naji2014effect, kremer2004palatini, 
gorini2003can, avelino2014nonlinear, kahya2015universe, fabris2011ruling} 
The interesting feature of Chaplygin gas is that it is the only fluid known up to now which has a super-symmetric generalization 
\cite{hoppe9311059supermembranes, jackiw2000supersymmetric}. Moreover, it has also a representation as tachyon 
field \cite{Benaoum, chimento2004extended}
and it is added to matter on branes in order to stabilize them in black hole bulks \cite{kamenshchik2000chaplygin}.
Chaplygin gas gives positive and bounded square of sound velocity $v_s^2=\frac{a}{\rho^2}$ that is very remarkable as it is a non-trivial problem for fluids 
with negative pressure \cite{kamenshchik2001alternative}.
Due to that interesting features it seems to be a very important object to study. Therefore, we would also like to apply that idea into our cosmological considerations.

The equation of state of the pure Chaplygin Gas is
\begin{equation}
  p=-\frac{A}{\rho},
\end{equation}
where $A$ is a positive constant. Applying that relation into the continuity equation (\ref{continuity}) one gets
\begin{equation}\label{chapl1}
 \rho=\sqrt{A+\frac{B}{a^6}}
\end{equation}
with $B$ being an integration constant. From that solution one sees immediately that assuming a positive value for $B$ the expression (\ref{chapl1}) 
for small $a(t)$ (i.e. $a^6\ll \frac{B}{A}$) gives rise to
\begin{equation}
 \rho\sim \frac{\sqrt{B}}{a^3}
\end{equation}
while for the large values of the scale factor 
\begin{equation}
 \rho \sim \sqrt{A},\;\;\;\;p\sim -\sqrt{A}.
\end{equation}
The first approximation corresponds to a universe dominated by dust-like matter whereas the second one to an empty universe with a cosmological
constant $\sqrt{A}$, that is, a de-Sitter universe \cite{kamenshik2001alternative, kamenshchik2000chaplygin}.
For the intermediate epoch between a dust dominated universe to a de Sitter universe we have that
\begin{equation}
 \rho\approx \sqrt{A}+\sqrt{\frac{B}{4A}}a^{-6},\;\;\;\;p\approx -\sqrt{A}+\sqrt{\frac{B}{4A}}a^{-6}
\end{equation}
which describes the mixture of a cosmological constant with stiff matter for which the EoS is $\omega=1$. It should be noticed that in a model with Chaplygin 
Gas one deals with a smooth evolution from the dust dominated phase to the nowadays accelerated expansion run by cosmological constant $\sqrt{A}$. That 
process is reached by using only one fluid.

There is a very natural generalization of the Chaplygin gas, so-called Generalized Chaplygin Gas (GCG) whose equation of state is written as
\begin{equation}
 p=-\frac{A}{\rho^\alpha},
\end{equation}
where constants $A$ and $\alpha$ satisfy $A>0$ and $0<\alpha\leq1$. For $\alpha=1$ one deals with the original Chaplygin Gas. 
Substituting GCG into the conservation law (\ref{continuity}) in FRLW spacetime we get
\begin{equation}
 \rho=\left(A+\frac{B}{a^{3(1+\alpha)}}\right)^{\frac{1}{1+\alpha}},
\end{equation}
where $B>0$ is an integration constant. One notices that in the GCG model, similarly like in pure Chaplygin Gas, the early stage of the Universe is dominated by dust
($\rho\propto a^{-3}$) while at late times by cosmological constant (vacuum energy, $\rho\simeq\text{const}$).

\subsection{Starobinsky's model $f(\mathcal{R})=\mathcal{R}+\gamma \mathcal{R}^2$}

In the early $1980$s A. Guth \cite{guth1981inflationary} and A. Starobinsky \cite{starobinsky1980new} proposed models that introduced inflation into cosmic evolution of our 
Universe. Guth studied magnetic monopoles which do not exist in our Universe; in order to explain that, he proposed inflation.
In turn, Starobinsky showed that there was a link bitween curvature-squared corrections to the Einstein-Hilbert action and quantum corrections which were supposed to play
an important role during the early Universe. That provided solutions to cosmological problems and has been showed to have a very good agreement with observational 
data (for example Planck \cite{planck2015planck}). The model, apart the standard gravitational Lagrangian, contains an additional term consisting 
of the quadratic term of the Ricci scalar $R^2$, that is,
\begin{equation}
  S=\frac{1}{2\kappa}\int\mathrm{d}^4x\sqrt{-g}(R+\gamma R^2)+ S_m(g_{\mu\nu},\psi),
\end{equation}
where $\gamma$ is a small parameter with the dimension of mass. Such a choice of the Lagrangian preserves the GR effects in weak gravitational field, 
for example in our Solar System. The quadratic term gains an importance in the case of strong gravity (such as neutron stars or black holes) or the very 
early stage of the Universe. The last one will be a subject of our discussion. 

Let us consider the Starobinsky ansatz $f(\mathcal{R})=\mathcal{R}+\gamma\mathcal{R}^2$ but with the Palatini curvature 
scalar $\mathcal{R}=g^{\mu\nu}\mathcal{R}_{\mu\nu}$ instead of metric Ricci scalar. Applying it to the structural equation (\ref{struc}) and using the 
trace of the perfect fluid energy-momentum tensor (\ref{trace}) to the right part of the (\ref{struc}) we are able to find the 
relation $\mathcal{R}=\mathcal{R}(a)$:
\begin{equation}
 \mathcal{R}=\left( A+Ba^{-3(1+\alpha)} \right)^{\frac{-\alpha}{1+\alpha}}\left(4A+Ba^{-3(1+\alpha)}   \right) \label{palatinir}
\end{equation}
for the Universe filled with Generalized Chaplygin Gas \cite{chap, lu2009cosmology, bento}.
It allows us to write the Friedmman equation (\ref{friedCOS}) as a function of the scale factor $a(t)$
\begin{equation}
  H^2\equiv \left( \frac{\dot{a}}{a} \right)^2=M^2(a)\cdot\left[N(a)-P(a)-\frac{K}{a^2}\right]
\end{equation}
where
{\small
\begin{align*}
 M(a)&=\frac{\rho^{  \alpha}  \left[2B \gamma a^{-3(1+\alpha)} +
 (8 A \gamma + \rho^\alpha) \right] }
 { \left[-\frac{B^2}{\rho} \gamma\ a^{-6(1 + \alpha)}+ 
  \frac{B}{\rho} (A (7 + 9 \alpha) \gamma + \rho^\alpha) a^{-3(1 + \alpha)} + 
  \frac{A}{\rho} (8 A \gamma + \rho^\alpha)\right] }\\
  N(a)&=\frac{
   A + 6 A \gamma \rho + 
    9 A^2 \gamma \rho^{-\alpha} + \rho^{
     1 + \alpha} (1 + \gamma \rho)}{
   4 \left[6 A \gamma + \rho^\alpha (1 + 2 \gamma \rho)\right]}  \\
  P(a)&= \frac{ a^{-3(1 + \alpha)}B\left[B \gamma\rho^{-\alpha} a^2 + 
       (8 A \gamma\rho^{-\alpha} + 1) \right] - 
      2 (-8 A^2 \gamma\rho^{-\alpha} - 2 A  + \rho^{
         1 +  \alpha}) }  
         {12 
    \left[2 B \gamma a^{ -3(1 + \alpha)} + (8A \gamma + \rho^\alpha) \right]}.
\end{align*}
}
Since we would like to examine the model with respect to observational constraints, the Friedmann equation needs to be properly parametrized. The above 
form seems to be difficult to handle and for that reasons one may try to parametrize the quantities appearing in (\ref{friedCOS}) 
(see \cite{borowiec2016inflationary}).  Due to that 
fact we will introduce parameters $A_{\text{s}}$ and $\rho_{\text{ch},0}$ related to physics \cite{bento, borowiec2016inflationary} 
 instead of the theoretical ones,
that is, $A$ and $B$. The new parameters of Generalized Chaplygin Gas are defined as
\begin{equation}
	\rho_{\text{ch}}=\rho_{\text{ch},0}\left(A_{\text{s}}+
	\frac{1-A_{\text{s}}}{a^{3(1+\alpha)}}\right)^{\frac{1}{1+\alpha}}=
	3H_0^2\Omega_{\text{ch},0}\left(A_{\text{s}}+\frac{1-A_{\text{s}}}{a^{3(1+\alpha)}}\right)^{\frac{1}{1+\alpha}},
\end{equation}
where $A_{\text{s}} \rho_{\text{ch},0}^{1+\alpha}=A$,\; $\rho_{\text{ch},0}^{1+\alpha}(1-A_{\text{s}})=B$. The quantity $ \rho_{\text{ch},0}$ corresponds 
to the present epoch's value while $H_0$ is a present value of the Hubble constant which is 
 $H_0=67.27\frac{\text{km}}{\text{sMpc}}$ (Planck
mission \cite{planck2015planck}). One also defines a new dimensionless parameter that is related 
to the parameter $\gamma$
\begin{equation}
 \Omega_{\gamma}=3\gamma H_0^2.
\end{equation}
Before we will parametrize the rest of the quantities appearing in (\ref{friedCOS}), let us introduce the quantity $K$:
\begin{equation}\label{defK}
 K(a)=\frac{3A_{\text{s}}}{A_{\text{s}}+(1-A_{\text{s}})a^{-3(1+\alpha)}}.
\end{equation}
The function $K(a)$ takes the values from the interval $[0,\;3)$ as the values of the scale factor lies in $a\in [0,\text{ }+\infty)$. We will 
use that quantity not only to write the Friedmann equation in a nicer form but it will be helpful during the further analysis of singularities that 
the model possesses. Moreover, it is also associated with the squared velocity of sound, 
that is, $c^2_\text{s}=\frac{\partial p}{\partial \rho}=\frac{\alpha A_{\text{s}}}{A_{\text{s}}+(1-A_{\text{s}})a^{-3(1+\alpha)}}=\frac{1}{3}\alpha K$.
Similarly, like for the parameter $\gamma$, we will need dimensionaless functions instead of $\mathcal{R}$, $b$, $c$, $\rho_{\text{ch}}$ in (\ref{friedCOS}).
The new ones are defined with the parameters $A_\text{s}$, $\rho_{\text{ch},0}$ and the function $K(a)$ as follows:
\begin{align}
\Omega_{\text{ch}}=&\Omega_{\text{ch},0}\left(A_{\text{s}}+\frac{1-A_{\text{s}}}{a^{3(1+\alpha)}}\right)^{\frac{1}{1+\alpha}}=
\Omega_{\text{ch},0}\left(\frac{3A_\text{s}}{K}\right)^{\frac{1}{1+\alpha}},\\
 \Omega_R=&\frac{\mathcal{R}}{3 H_0^2}=\Omega_{\text{ch},0}\left(A_{\text{s}}+
 \frac{1-A_{\text{s}}}{a^{3(1+\alpha)}}\right)^{\frac{1}{1+\alpha}}\frac{4+ \frac{(1-A_{\text{s}})}{A_{\text{s}}}a^{-3(1+\alpha)}}{1+\frac{(1-A_{\text{s}})}{A_{\text{s}}}a^{-3(1+\alpha)}}
 \nonumber\\=&\Omega_{\text{ch}}(K+1),\\
\Omega_c=&\frac{c}{3H_0^2}=\frac{1}{6H_0^2}f(\mathcal{R})=\frac{\mathcal{R}}{6H_0^2}(1+\gamma\mathcal{R})=\frac{\Omega_{R}}{2}(1+\Omega_{\gamma}\Omega_{R})\nonumber\\
=&\frac{\Omega_{\text{ch}}(K+1)}{2}(1+\Omega_{\gamma}\Omega_{\text{ch}} (K+1)),\\
\Omega_k=&-\frac{k}{H_0^2 a^2},\\
b=&f'(\mathcal{R})=1+2\Omega_{\gamma} \Omega_R=1+2\Omega_{\gamma}\Omega_{\text{ch}}(K+1).
\end{align}
Let us also define the another function $d(t):=\frac{\dot{b}}{H}$, where $\dot{b}=\frac{db}{dt}$, which one may rewrite as a function 
of $K,\;\Omega_\gamma\;\text{and}\;\Omega_\text{ch}$:
\begin{equation}
 d=2\Omega_{\gamma}\Omega_{\text{ch}}(3-K)[\alpha(1-K)-1].
\end{equation}
With the just defined dimensionless parameters, we can write the normalized Friedmann equation (\ref{friedCOS}) as follows
\begin{equation}
 \frac{H^2}{H_0^2}=\frac{b^2}{\left(b+\frac{d}{2}\right)^2}\left(\Omega_{\gamma}\Omega_{\text{ch}}^2\frac{(K-3)(K+1)}{2b}+\Omega_{\text{ch}}
+\Omega_k\right).\label{hubble}
\end{equation}
Since the radiation, whose equation of state is $p_r=\frac{1}{3}\rho_r$, does not contribute to the trace of the energy-momentum tensor, the structural 
equation (\ref{struc}) is the same, that is, the solution $\mathcal{R}=\mathcal{R}(a)$ (\ref{palatinir}) does not change. That property allows as to add to
the normalized Friedmann equation (\ref{hubble}) the radiation term in a form of a dimensionless 
parameter $\Omega_{\text{r}}=\Omega_{\text{r,0}} a^{-4}=\frac{\rho_{\text{r}}}{3H_0^2}a^{-4}$. In that case (\ref{hubble}) takes the form
\begin{equation}
 \frac{H^2}{H_0^2}=\frac{b^2}{\left(b+\frac{d}{2}\right)^2}\left(\Omega_{\gamma}\Omega_{\text{ch}}^2\frac{(K-3)(K+1)}{2b}+\Omega_{\text{ch}}+
 \frac{\Omega_{\text{r}}}{b}+\Omega_k\right).\label{hubble2}
 \end{equation}
One may consider the coordinate transformation $t\rightarrow \tau \colon \frac{|b| dt}{|b+\frac{d}{2}|}=d\tau$ and apply it to the equation (\ref{hubble})
if it is non-singular:
\begin{equation}\label{lagr_newton}
\frac{H^2(\tau)}{H_0^2}=\Omega_{\gamma}\Omega_{\text{ch}}^2\frac{(K-3)(K+1)}{2b}+\Omega_{\text{ch}}
+\Omega_k.
\end{equation}
The new Hubble parameter is defined as $H(\tau)=a(\tau)^{-1}\frac{da(\tau)}{d\tau}$. One should also notice that the new time $\tau$ is a growing 
function of the original cosmological time $t$.
The re-parametrization of time taken under an examination whether it is a diffeomorphism or not will allow us to determine the position of the singularity 
$a_\text{sing}$. Let us define a function $f(K,\alpha,A_{\text{s}}, \Omega_{\gamma})=2b+d$ (which is just a denominator of the re-parametrization) whose the
zero value indicates the singularity $a_{\text{sing}}\text{: } f(K(a_{\text{sing}}))=0$
\begin{equation}\label{eq:13}
 \alpha K^2-3(1+\alpha)K-\frac{K^{\frac{1}{1+\alpha}}}{\Omega_{\gamma}\Omega_{\text{ch,0}}\left(3A_{\text{s}}\right)^{\frac{1}{1+\alpha}}}+1=0.
\end{equation}
As the above equation cannot be solved algebraically, let us
 consider less complex case, that is, the case of the original Chaplygin Gas (for which the parameter $\alpha=1$):
\begin{equation}
K^4-12K^3+38K^2-\chi K+1=0,
\end{equation}
where we have defined
\begin{equation}
\chi =\left(12+\frac{1}{3 A_{\text{s}}\Omega _{\gamma }^2\Omega _{\text{ch},0}^2}\right).
\end{equation}
The quantity $\chi$ belongs to the interval $\chi\in [12,\text{ }\infty)$.
Let us recall that we are interested in the real solutions of the above algebraic equation in the interval $[0,\;3)$. Hence, one finds that
\begin{equation}
K_{\text{sing}} = 3- \frac{\zeta}{\sqrt{6}} -
 \sqrt{\left( \frac{16}{3}+\frac{(9 \chi-364 )}{3\xi }-\frac{1}{12} \xi
 -\frac{\sqrt{6} (\chi-12 )}{4\zeta}\right)},
\end{equation}
where
{\small
\begin{align}
\zeta(\chi)&=\sqrt{16+\frac{2(364-9 \chi )}{\xi }  +\frac{\xi}{4} },\\
\xi(\chi) &= \left(55448-2052 \chi +\frac{27 \chi ^2}{2}+\frac{3}{2} \sqrt{3(\chi-12)^2 \left(27 \chi ^2-12496-648 \chi \right)}\right)^{1/3}.
\end{align}
}
One notices that the position of the singularity depends on the $\chi$ parameter 
which for $\Omega_{\gamma}\ll 1$ is $\chi=\left(3A_{\text{s}}\Omega_{\gamma }^2\Omega_{\text{ch},0}^2\right)^{-1}$.
Using the definition (\ref{defK}), we are able to express the scale factor for this case (remember that $\alpha=1$) as
{\footnotesize
\begin{equation}
a_{\text{sing}}=\left[\frac{A_{\text{s}}}{1-A_{\text{s}}}
\left(\frac{3}{3- \frac{\zeta}{\sqrt{6}}- \sqrt{\left(\frac{16}{3}+\frac{(9 \chi-364 )}{3
			\xi }-\frac{1}{12} \xi -\frac{\sqrt{6} (\chi-12 )}{4\zeta}\right)}}-1\right)\right]^{\frac{-1}{3(1+\alpha)}}  .
\end{equation}
}

Let us consider now the case $\alpha=0$. This implies that the matter content of our universe is the same as in $\Lambda$CDM model. It was already mentioned 
that the case $\alpha=\gamma=0$ reconstructs $\Lambda$CDM model completely. However, we are considering the model with the presence of the
quadratic term, that is, $\gamma\neq 0$. For $\gamma=0$ one gets $b=1, d=0$ and the equation (\ref{eq:13}) has no
solutions at all. Moreover, one remembers that the value $\gamma<<1$ should be very small in order to locate the singularity in an appropriate epoch and
also to have a model which reproduces GR equations for weak gravitational field (e.g. our Solar System). 
The case $\alpha=0$ significantly simplifies (\ref{eq:13}) and hence we are able to find singular solutions
\begin{equation}
K_{\text{sing}}=\frac{1}{3+\frac{1}{3\Omega_{\gamma}\Omega_{\text{ch,0}}A_{\text{s}}}}\;\;
\;\;\text{and}\;\;\;\;
a_{\text{sing}}=\left(\frac{1-A_{\text{s}}}{8A_{\text{s}}+\frac{1}{\Omega_{\gamma}\Omega_{\text{ch,0}}}}\right)^\frac{1}{{3}}.
\end{equation}

In the general case ($\alpha\neq\{0,1\}$) the re-parametrization function is
\begin{equation}\label{repar}
\frac{b^2}{\left(b+\frac{d}{2}\right)^2}=
\frac{\left(1+2\Omega_{\gamma}\Omega_{\text{ch,0}}(K+1)\right)^2}{\left(1+\Omega_{\gamma}\Omega_{\text{ch,0}}(3K+\alpha K(3-K)-1)\right)^2}.
\end{equation}
We should also mention that the density parameters $\alpha$, $\Omega_{\text{ch},0}$,
$\Omega_{\text{k},0}$, $A_{\text{s}}$, $\Omega_{\gamma}$ are not independent as they satisfy the constraint condition
\begin{equation}\label{constraint}
\begin{array}{l}
1 - \Omega_{\text{ch},0} - \Omega_{k,0} =\frac{\Omega_{\gamma}\Omega_{\text{ch,0}}}{2+4\Omega_{\gamma}\Omega_{\text{ch,0}}(3A_{\text{s}}+1)}\times\\ \\
\times(1-A_{\text{s}})(1-3\alpha A_{\text{s}})\left(12-3\Omega_{\text{ch,0}}+
\frac{6\Omega_{\gamma}\Omega_{\text{ch,0}}}{1+2\Omega_{\gamma}\Omega_{\text{ch,0}}(3A_{\text{s}}+1)}\right)  .
\end{array}
\end{equation}

\subsection{Statistical analysis of the model}

From that point of our consideration we will assume that the model is flat, it means $\Omega_{\text{k},0}=0$. There are only three parameters to
estimate, that is, $A_s, \alpha$ and $\Omega_\gamma$ since the value of the parameter $\Omega_\text{ch}$ will be derived from the constraint condition 
(\ref{constraint}). Moreover, we will also assume that the value of the today's Hubble constant is $H_0=67.27\frac{\text{km}}{\text{s Mpc}}$ according to 
the Planck mission \cite{planck2015planck}. As it was
already mentioned, the value of 
$\Omega_{\gamma}$ is small; we have also found an upper bound $\Omega_{\gamma}<10^{-9}$. If the value of $\Omega_{\gamma}$ had been bigger than the boundary, 
then the epoch of the singularity would have been shifted to the epoch of recombination or later \cite{borowiec2016inflationary, szyd_stach}. 

Two models have been taken under statistical analysis and estimation procedure in \cite{borowiec2016inflationary}: the model with radiation (\ref{hubble2}) and with baryonic matter whose 
Friedmann equation is
\begin{equation}
\frac{H^2}{H_0^2}=\frac{b^2}{\left(b+\frac{d}{2}\right)^2}\left(\Omega_{\gamma}\Omega_{\text{ch}}^2\frac{(K-3)(K+1)}{2b}+\Omega_{\text{ch}}+
\Omega_{\text{bm}}+\Omega_k\right).
\end{equation}
The parameter $\Omega_{\text{bm}}=\Omega_{\text{bm},0}a^{-3}$ is related to the presence of baryonic visible matter 
for which the value $\Omega_{\text{bm},0}=0.04917$ is assumed following the Planck estimation \cite{planck2015planck}. In the following part we will consider 
only the model with radiation.

For the statistical analysis we have used a large set of data such as the SNIa, BAO, CMB and lensing observations,
measurements of $H(z)$ for galaxies and the Alcock-Paczy{\'n}ski test, Union 2.1, that is, the sample of $580$ supernovae \cite{suzuki2012hubble}
(see details and likelihood functions in \cite{borowiec2016inflationary}). For the estimation procedure 
of the model parameters the code CosmoDarkBox \cite{borowiec2016inflationary} has been used. This code uses the Metropolis-Hastings 
algorithm \cite{metropolis1953equation, hastings1970monte}.

\begin{figure}
	\centering
	\includegraphics[scale=1]{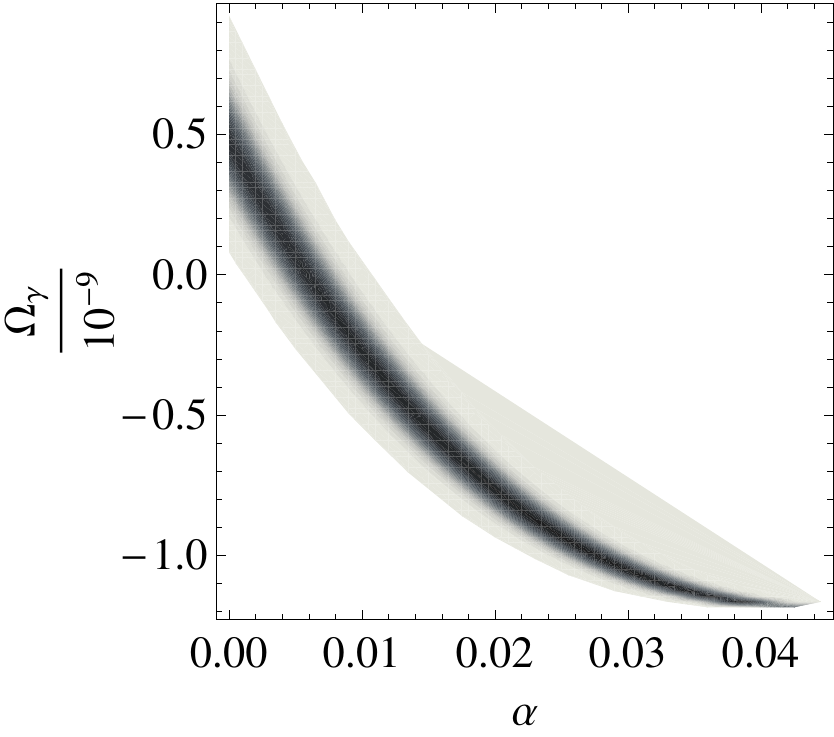}
	\caption{The likelihood function of two model parameters ($\alpha$, $\Omega_{\gamma}$) with the marked $68\%$ and $95\%$ confidence
	levels for model with the GCG and radiation. We assume $H_0=67.27\frac{\text{km}}{\text{s Mpc}}$, $A_\text{s}=0.6908$.}
	\label{fig:6a}
\end{figure}

\begin{figure}
	\centering
	\includegraphics[scale=1]{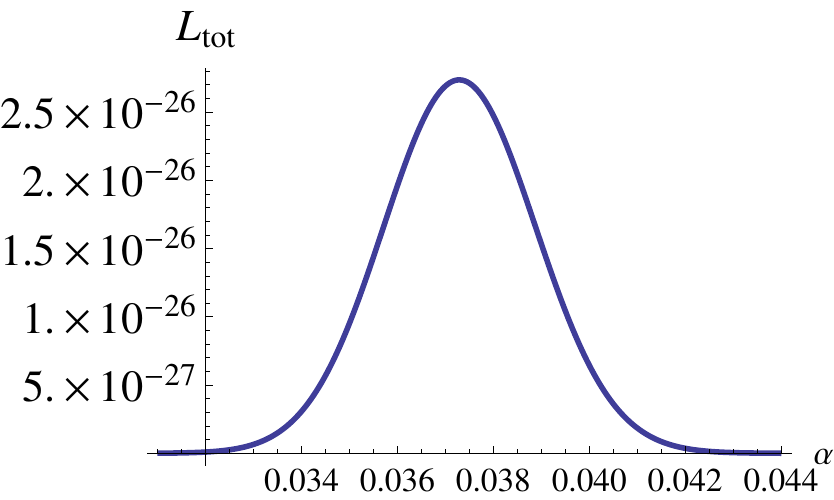}
	\caption{Diagram of PDF for parameter
		$\alpha$ obtained as an intersection of a likelihood function for model with the GCG and radiation. 
		Two planes of intersection likelihood function are
		$\Omega_{\gamma}=-1.15\times 10^{-9}$ and $A_{\text{s}}=0.6908$.}
	\label{fig:7}
\end{figure}

The results of our statistical analysis for the Generalized Chaplygin Gas with radiation are represented in the tables \ref{table:1} and \ref{table:3} 
as well as in the figure \ref{fig:6a}. On the picture \ref{fig:6a} there is a likelihood function with $68\%$ and $95\%$ confidence level. The diagram of 
probability density function (PDF) is presented in the figure \ref{fig:7}.

The value of $\chi^2$  for the best fit for the model with the Generalized Chaplygin Gas and radiation is $117.722$ 
while the value of reduced $\chi^2$ is equaled to $0.1892$. Moreover, we have used the Bayesian Information Criterion (BIC) whic is defined in the following
way \cite{schwarz1978estimating, kass1995bayes}
\begin{equation}
\text{BIC}= \chi^2 \,+\, j\, \ln(n),
\end{equation}
where $j$ is a number of parameters and $n$ is a number of data points. In our statistical analysis we have used $625$ data points, hence $n=625$. 
Although the number of parameters for our model is
$5$ ($H_0$, $\Omega_{\text{r},0}$, $\Omega_{\gamma}$, $A_{\text{s}}$, $\alpha$) we took $j=3$ in 
computation of BIC because values for $H_0$ and $\Omega_{\text{r},0}$ are assumed in the estimation.
The value of BIC for our
model with radiation is $137.036$. For comparison: BIC of $\Lambda$CDM model is $125.303$ (the value of $\chi^2$ is $118.866$ while reduced $\chi^2$ is $0.1908$). 
For computation of BIC
of $\Lambda$CDM model we took $j=1$ because, as previously, we assumed that the values of $H_0$,
and $\Omega_{\text{r},0}$ are already known.
In consequence, the only free parameter is $\Omega_{\text{m},0}$ representing matter. The difference between BIC of our model
and $\Lambda$CDM model is $\Delta\text{BIC}=11.733$. If the value $\Delta$BIC is
between the numbers $2$ and $6$ then the evidence against the model is positive in comparison to the model of the null hypothesis. If that
value is more than $6$, the evidence against the model is strong \cite{kass1995bayes}. Consequently, the evidence in favor $\Lambda$CDM model is 
strong in comparison to our model. But one should notice that all models which posses more than one parameter to be estimated will have a poor evidence in 
comparison to $\Lambda$CDM model.

\begin{table}
	\caption{The best fit and errors for model with the GCG and radiation for the case where we assume the value of $\Omega_{\gamma}$ from the
		interval $(-1.2\times10^{-9}, 10^{-9})$. We assume also $A_{\text{s}}$ from the interval $(0.67, 0.72)$, and $\alpha$ from
		the interval $(0, 0.06)$. The value of $\chi^2$ for the best fit is equaled $117.722$.}
	\label{table:1}
	\begin{center}
		\begin{tabular}{llll} \hline
			parameter & best fit & $ 68\% $ CL & $ 95\% $ CL\\
			\hline \hline
			$A_{\text{s}}$ & $0.6908$ & $\begin{array}{c} +0.0066
			\\ -0.0069
			\end{array}$ & $\begin{array}{c} +0.0104
			 \\ -0.0098
			\end{array}$
			\\ \hline
			$\alpha$ & $0.0373$ & $\begin{array}{c} +0.0083
			\\ $-0.0373$
			\end{array}$ & $\begin{array}{c} +0.0131
			\\ -0.0373
			\end{array}$ \\ \hline
			$\Omega_{\gamma}$ & $-1.156\times 10^{-9}$ & $\begin{array}{c} +2.156\times 10^{-9}
			\\ -0.010\times 10^{-9}
			\end{array}$ & $\begin{array}{c} +2.156\times 10^{-9}
			\\ -0.015\times 10^{-9}
			\end{array}$
			\\ \hline
		\end{tabular}
	\end{center}
\end{table}

\begin{table}
	\caption{The best fit and errors for model with the GCG and radiation for the case where we assume the value of $\Omega_{\gamma}$ from the
		interval $(-1.2\times10^{-9}, 0)$. We assume also $A_{\text{s}}$ from the interval $(0.67, 0.72)$, and $\alpha$ from the interval $(0, 0.06)$. 
		The value of $\chi^2$ for the best fit is equaled $117.722$.}
	\label{table:3}
	\begin{center}
		\begin{tabular}{lllll} \hline
			parameter & best fit & $ 68\% $ CL & $ 95\% $ CL\\
			\hline \hline
			$A_{\text{s}}$ & $0.6908$
			& $\begin{array}{c} +0.0065
			\\ -0.0068
			\end{array}$ & $\begin{array}{c} +0.0103
			\\ -0.0098
			\end{array}$
			\\ \hline
			$\alpha$ & $0.0373$
			& $\begin{array}{c} +0.0080
			\\ -0.0373
			\end{array}$ & $\begin{array}{c} +0.0129
			\\ -0.0373
			\end{array}$\\ \hline
			$\Omega_{\gamma}$ & $-1.156\times 10^{-9}$
			& $\begin{array}{c} +1.156\times 10^{-9}
			\\ -0.008\times 10^{-9}
			\end{array}$ & $\begin{array}{c} +1.156\times 10^{-9}
			\\ -0.014\times 10^{-9}
			\end{array}$
			\\ \hline
		\end{tabular}
	\end{center}
\end{table}

\subsection{Cosmological singularities}\label{sec_sing}
\begin{figure}
  \centering
  \includegraphics[width=1\linewidth]{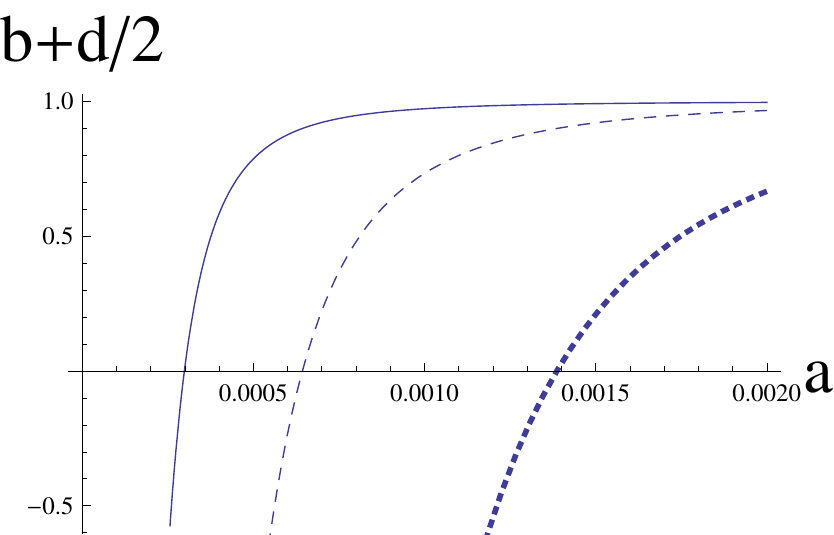}
  \caption{The diagram represents function $b(a)+d(a)/2$ for different values of the positive $\Omega_{\gamma}$ and shows that it is growing
  function of scale factor. Zero of this function represents a value of the scale factor for the freeze singularity $a_{\text{fsing}}$. The continuous
  line is for $\Omega_{\gamma}=10^{-10}$, the dashed line for $\Omega_{\gamma}=10^{-9}$ and the dotted line is for $\Omega_{\gamma}=10^{-8}$. Is is
  assumed that $A_{\text{s}}=0.7264$ and $\alpha=0.0194$.}
  \label{fig:1}
\end{figure}

Beyond the initial singularity, there also appears another singularity in our model. It arises as the denominator of the Hubble function (\ref{hubble2}) may equal to zero 
(that is, the re-parametrization of time (\ref{repar}) will not be a diffeomorphism). Let us examine it.
The picture of the function $b+\frac{d}{2}=f(A_\text{s},\Omega_\text{ch,0},\Omega_\gamma,\alpha)$ as a function of the scale factor is given in the figure \ref{fig:1}.

\begin{figure}
  \centering
  \includegraphics[width=1\linewidth]{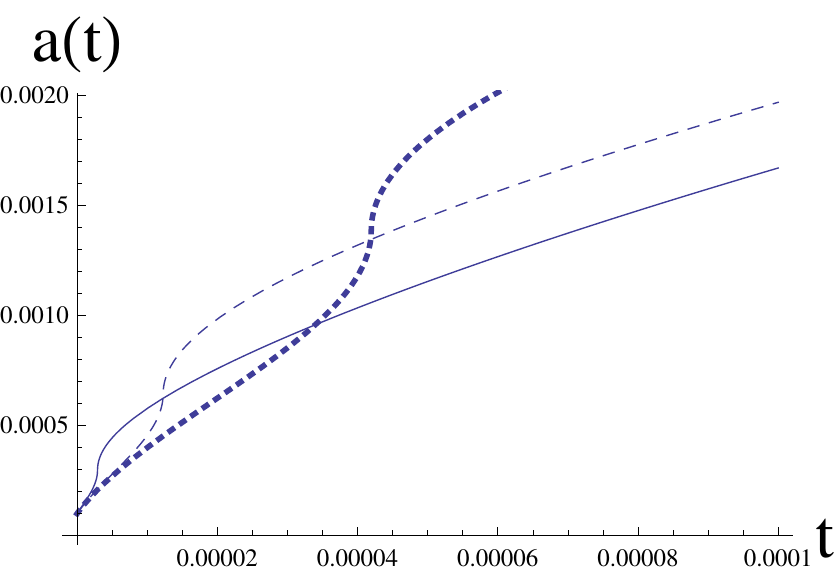}
  \caption{The diagram represents function $a(t)$ for positive $\Omega_{\gamma}$. For the scale factor of the freeze singularity,
  the function $a(t)$ has a vertical inflection point. The continuous line is for $\Omega_{\gamma}=10^{-10}$, the dashed line is
  for $\Omega_{\gamma}=10^{-9}$ and the dotted line is for $\Omega_{\gamma}=10^{-8}$. Is is assumed that $A_{\text{s}}=0.7264$ and $\alpha=0.0194$. We
  assume that $8\pi G=1$ and we chose $\frac{\text{s Mpc}}{\text{100 km}}$ as a unit of time $t$.}
  \label{fig:6}
\end{figure}

\begin{figure}
  \centering
  \includegraphics[width=1\linewidth]{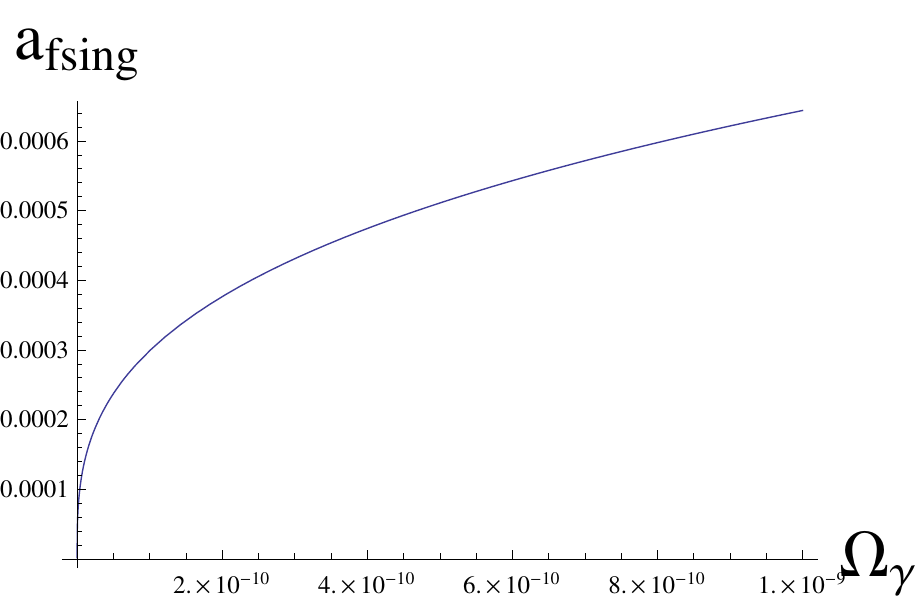}
  \caption{The diagram shows the relation between positive $\Omega_\gamma$ and $a_{\text{fs}}$ obtained for $A_s=0.7264$ and $\alpha=0.0194$. We see
  that this relation is a monotonic function. If $\Omega_\gamma\longrightarrow 0$ then $a_{\text{fsing}}\longrightarrow 0$.}
  \label{fig:2}
\end{figure}

As we consider singularities of FLRW models filled with perfect fluid of effective energy density $\rho_{eff}$ and pressure $p_{eff}$, we may classify them 
due to a well-known classification \cite{Nojiri:2005sx, Singh:2010qa} of finite-time, future singularities. The classification consists of four
groups with respect to the behaviors of effective energy density and pressure as well as scale factors and Hubble rates. We will briefly remind their definitions.
\begin{itemize}
 \item Type I (Big Rip): Energy density, pressure, and scale factor diverge.
 \item Type II (Sudden Singularity): Energy density and scale factor are finite values while pressure diverges.
 \item Type III (Big Freeze): Energy density and pressure diverge at a finite value of a scale factor.
 \item Type IV (Big Brake): Energy density and pressure go to zero at a finite value of a scale factor while higher derivatives of Hubble rate diverge.
\end{itemize}
Let us mention that the cosmological singularities classification can be enriched by adding subclasses of the above 
types \cite{dabrowski2014singularities, dkabrowski2009barotropic, frampton2011little, frampton2012pseudo}.   
In our model we may express the effective quantities and equation of state $w_{\text{eff}}=\frac{ p_{\text{eff}}}{\rho_{\text{eff}}}$ in terms of the potential
\begin{align}
 \rho_{\text{eff}}&= -\frac{6V}{a^2},\\
 p_{\text{eff}}&=-\rho_{\text{eff}}-\frac{1}{3}\frac{d(\rho_{\text{eff}})}{d(\ln a)},\\
 w_{\text{eff}}&=-1-\frac{1}{3}\frac{d(\ln \rho_{\text{eff}})}{d(\ln a)}
\end{align}
which diverges while the scale factor is finite. The pressure diverges too as well as $\dot{a}$. We notice that the singularity is a singularity of acceleration because the derivative of the 
potential diverges. The crucial observation is that it goes to plus infinity on the left from the singular point while from the right hand side it goes to minus infinity.
That behavior is represented on the picture \ref{fig:6} of the scale factor as a function of time: we observe the inflection point $t=t_\text{sing}$. One may also show 
the relation of the singularity as a function of the parameter $\Omega_\gamma$ which is depicted in \ref{fig:2}. Numerical simulations showed that the singularity 
is sensitive on value changes of the parameter $\Omega_\gamma$ while the dependence on the parameter $\alpha$ is very weak. Therefore, for the values of $\alpha$ from the 
interval $(0,1)$ the singularity does not differ.

\subsection{Dynamical system analysis}
Let us briefly discuss dynamical system analysis of the model considered in the previous sections \cite{szyd_stach}. Similarly as for $\Lambda CDM$ model (see 
the Appendix \ref{apB}), our model can be investigated as a two-dimensional dynamical system of a Newtonian type \cite{bor_kam, zoo_marek}. Since one deals with 
a degenerate singularity of the type III \cite{nojiri2015singular, odintsov2015singular}, the phase space structure is complicated: it divides the evolutionary 
paths on two $\Lambda CDM$ types of 
 evolution. The full trajectories should be sewn along the singularity \cite{Bautin:1976mt}.
 That singularity has an intermediate character \cite{herrera2013intermediate} and its presence in the early evolution of the universe provides the 
 inflationary behavior (so-called singular inflation introduced in \cite{barrow2015singular}).
 
 The equation (\ref{lagr_newton}) can be seen as a Hamiltonian of the considered model with the one-dimensional potential 
 \begin{equation}
  \tilde{V}(a)=-\frac{1}{2}a^2\left(\Omega_{\gamma}\Omega_{\text{ch}}^2\frac{(K-3)(K+1)}{2b}+\Omega_{\text{ch}}+\Omega_k\right)
 \end{equation}
whose motion is along the energy levels $\mathcal{H}=E=\text{const}$. Here, the scale factor is a function of the rescaled cosmological time, that is, 
$H_0t=\tau$. In cosmology, the scale factor $a(t)$ play a role of positional variable while
a localization critical points and their type is determined by a shape of a potential. The considered dynamics is dynamics of a particle with unit mass in the potential
$V(a)$ over the energy level. Before going further, we will write down necessary notions that will be used later 
in the section \cite{zoo_marek, szyd_stach}.

Let us consider a potential $V(x)$ of a cosmological model. One deals with the system
\begin{align}\label{dyn_sys}
 \frac{dx}{dt}=&y,\nonumber\\
 \frac{dy}{dt}=&-\frac{\partial V}{\partial x},\\
 E=&\frac{y^2}{2}+V(x).\nonumber
\end{align}

\begin{figure}
  \centering
  \includegraphics[width=1\linewidth]{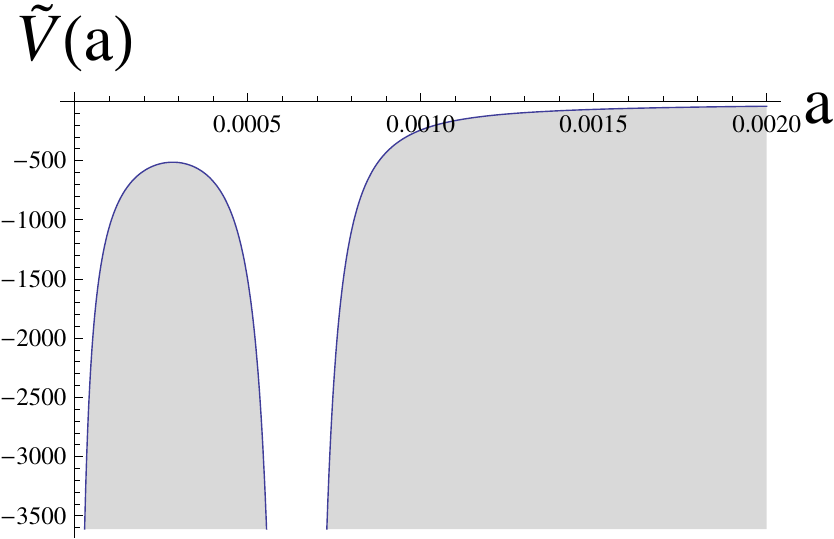}
  \caption{The diagram presents the potential $\tilde{V}(a)$ for $A_s=0.7264$, $\alpha=0.0194$ and $\Omega_{\gamma}=10^{-9}$. The shaded region represents a non-physical domain forbidden for motion of a classical system for which $\dot{a}^2\geq 0$.}
  \label{fig:3}
\end{figure}

\begin{itemize}
\item A static universe is represented by a critical point of
the system (\ref{dyn_sys}). It always lies on the $x$-axis, that is, $y=y_0 = 0$, $x = x _0$.
\item The point ($x_0$, 0) is a critical point of a Newtonian system if that is a critical point of the function of the potential
$V(x)$, it means: $V (x) = E$, where $E= \frac{y^2}{2} + V (x)$ is total
energy of the system. Spatially flat models refer to $y=\dot{x}$; $E = 0$ while the ones with the spatial curvature $k\neq0$ (constant) have
$E=-\frac{k}{2}$.
\item A critical point ($x_0$ , 0) is saddle one if it is a strict local maximum of the potential $V(x)$.
\item If ($x_0$ , 0) is a strict local minimum of the analytic function $V (x)$ then one deals with
a center.
\item ($x_0$, 0) is a cusp if it is a horizontal inflection point of the $V(x)$.
\end{itemize}
With the above definitions one agrees that critical points of a system and their stabilities are determined by the potential which is showed in the figure 
\ref{fig:3}.
 The configuration space is $\{a:\;a\geq0\}$ over the energy level $E=0$ as the Hamiltonian system has a form $\mathcal{H}(p,a)=\frac{1}{2}p^2_a+\tilde{V}(a)=0$. The domain which 
 is admissible for the universe motion is $\{a:\;\tilde{V}\leq0\}$ with a boundary $\{a:\;\tilde{V}=0\}$. It should be noticed that the domain $E-\tilde{V}<0$ is forbidden for classical motion.
 
 One may also consider different energy levels than the above one. They will correspond to different types of evolution providing different scenarios of the Universe. 
 We classify them in the following way (see the picture \ref{fig:4}):
\begin{figure}
  \centering
  \includegraphics[width=1\linewidth]{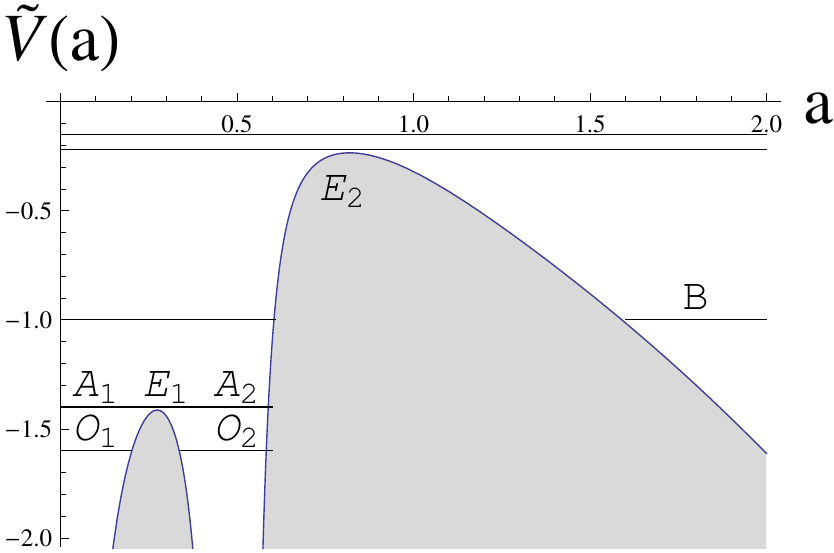}
  \caption{Diagram of the potential of dynamical system of a Newtonian type. The classification of trajectories are presented the configuration
  space. The shaded domain $E-\tilde{V}<0$ is forbidden for motion of classical systems.}
  \label{fig:4}
\end{figure}
\begin{enumerate}
\item $O_1$ --- oscillating universes with initial singularities;
\item $O_2$ --- `oscillatory solutions' without the initial and final singularity but with the freeze singularity;
\item $B$ --- bouncing solutions;
\item $E_1$, $E_2$ --- solutions representing the static Einstein universe;
\item $A_1$ --- the Einstein-de Sitter universe starting from the initial singularity and approaching asymptotically static Einstein universe;
\item $A_2$ --- a universe starting asymptotically from the Einstein universe, next it undergoes the freeze singularity and approaches to a maximum
size. After approaching this state it collapses to the Einstein solution $E_1$ through the freeze singularity;
\item $A_1$ --- expanding universe from the initial singularity toward to the Einstein universe $E_2$ with an intermediate state of the freeze singularity;
\item $EM$ --- an expanding and emerging universe from a static $E_2$ solution (Lemaitre-Eddington type of solution);
\item $IM$ --- an inflectional model: the relation $a(t)$ possesses an inflection point. That is an expanding universe from the initial singularity
undergoing the freeze type of singularity.
\end{enumerate}
The last two solutions $EM$ and $IM$ lie above the maximum of the potential $\tilde{V}$.
\begin{figure}
  \centering
  \includegraphics[width=1\linewidth]{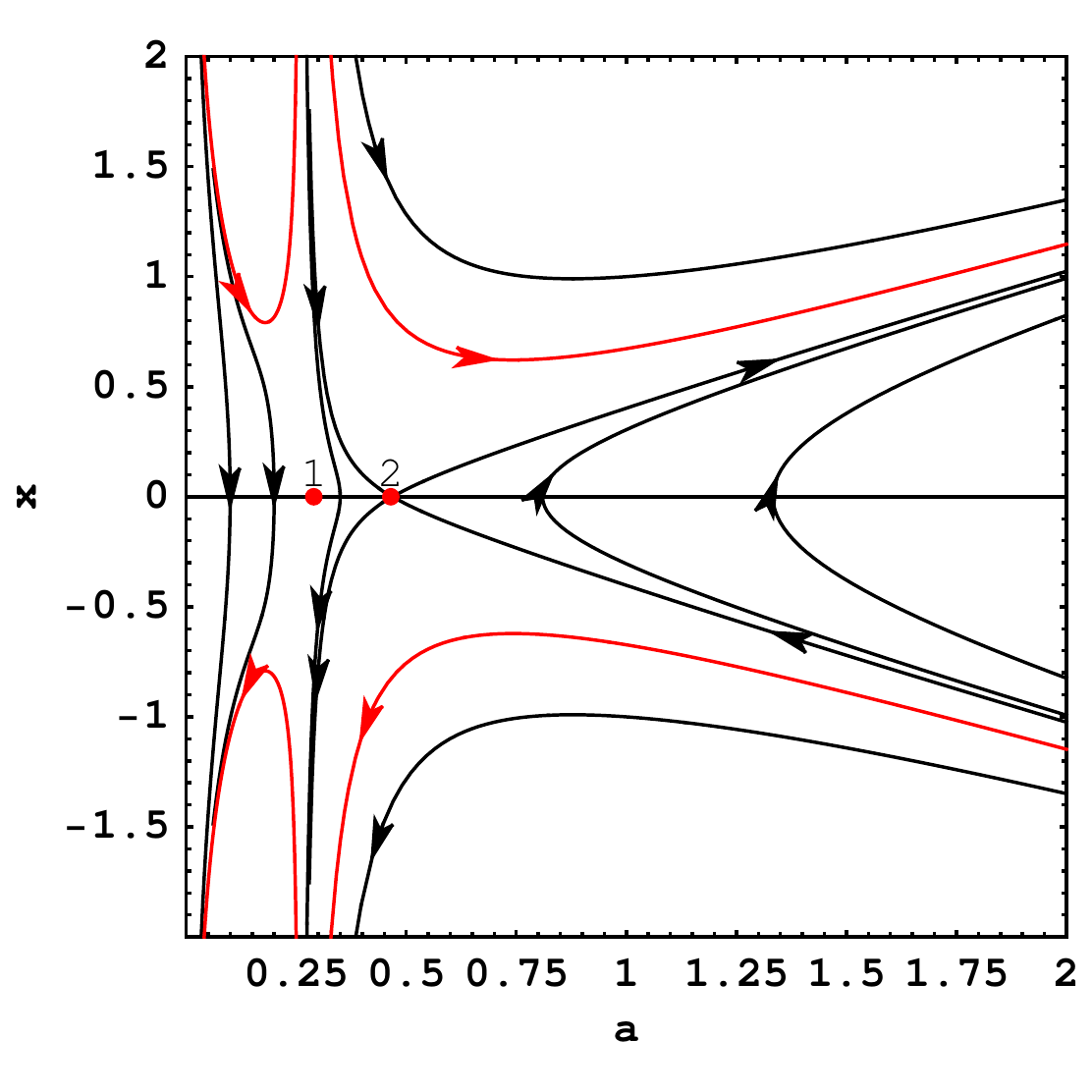}
  \caption{The diagram represents the phase portrait of the system (\ref{eq:ds1}-\ref{eq:ds2}) for positive $\Omega_{\gamma}$. The red
  trajectories represent the spatially flat universe. Trajectories under the top red trajectory and below the bottom red trajectory represent
  models with negative curvature. Positive curvature models are represented by the trajectories between the top and bottom red ones.}
  \label{fig:5}
\end{figure}

It should be mentioned that the singularity appearing just after the Big Bang one which is beyond the standard classification (see \cite{Nojiri:2005sx,Singh:2010qa} and 
the discussion in the (\ref{sec_sing}). The acceleration $\ddot{a}$ at the singularity point is undefined: left-hand side limit of the derivative 
of the potential is positive while the right-hand side limit has a negative value. We treat it as two sewn singularities: the first one is of the type III while the 
second one also belongs to that type but with reverse type. It will be better understood if we construct the phase portrait of the system which will also allow us to
classify all evolution path of the phase space. Firstly, let us defined a new potential $V(a)$
\begin{equation}
 V(a)=-\frac{ a^2}{2}\left(\Omega_{\gamma}\Omega_{\text{ch}}^2\frac{(K-3)(K+1)}{2b}+\Omega_{\text{ch}}+\Omega_k\right)
\end{equation}
such that $a'^2=-2V(a)$ where we have defined the new re-parametrization of time as $'\equiv\frac{d}{d\sigma}=\frac{b+\frac{d}{2}}{b}\frac{d}{d\tau}$. The system is now 
\begin{align}
p &= \dot a =x \label{eq:ds1}\\
\ddot a &= \dot x= -\frac{\partial \tilde{V}(a)}{\partial a} = \frac{x^2}{ m}\frac{\partial m}{\partial a}
-m^2\frac{\partial V(a)}{\partial a}, \label{eq:ds2}
\end{align}
with the quantity $m=\frac{b}{b+\frac{d}{2}}$ for simplification. We will divide dynamics into two parts: for the scale factor $a<a_{\text{fsing}}$ and
$a>a_{\text{fsing}}$ so it belongs to the class of the sewn dynamical systems \cite{Bautin:1976mt}. The configuration space is glued along the singularity $a_{\text{fsing}}$. The dynamical system (after the re-parametrization) for the first case
is 
\begin{align}
\dot a &= x,\\
\dot x &=\frac{x^2}{m}\frac{\partial m}{\partial a}-m^2\frac{\partial V_1(a)}{\partial a},
\end{align}
where $V_1=V(-\eta (a-a_s)+1)$ with respect to the new time $\sigma$ while $\eta(a)$ notes the Heaviside function while the second one $a>a_{\text{fsing}}$
is analogously given by the function $V_2=V \eta (a-a_s)$. 

Now on, we are able to draw the phase portrait of system (\ref{eq:ds1}-\ref{eq:ds2}) for positive $\Omega_{\gamma}$. It is presented on the picture \ref{fig:5}. 
The case of negative values of $\Omega_{\gamma}$ can be found in \cite{szyd_stach} and will not be considered here. 
The red trajectories concern evolution of spatially flat models: the Universe expands, starting from the initial singularity. It goes through the freeze 
singularity and after the accelerating phase it goes toward the de Sitter attractor. Trajectories under the top red trajectory and below the bottom red trajectory
are models with negative curvature while the ones with positive spatial curvature lie between the red trajectories. 
 The point of sewing is located at infinity ($a=a_{\text{fsing}}$, $\dot{a}\equiv x=\infty$). Note that all trajectories of open models are passing 
 through the freeze singularity.
  The phase portrait possesses the reflectional symmetry $x\rightarrow -x$. Trajectories from the domain $x<0$ continue their evolution into domain $x>0$.
  Due to this symmetry one can identify the corresponding point on the line $\{ b=0 \}$ and make from the line $\{ b=0 \}$ a circle $S^1$.
  Therefore the phase space is a cylinder. The line $\{ b=0 \}$ is not shown on the picture \ref{fig:5}.
  
  \subsection{Conclusions}
We have been discussing a cosmological model based on the modified Lagrangian $f(\mathcal{R})=\mathcal{R}+\gamma \mathcal{R}^2$ considered in Palatini formalism. As a 
source we have used Generalized Chaplygin Gas. After the physical parametrization we were able to find the best fits for the density
parameters $\Omega_\gamma,\,A_\text{s},\;\alpha,\,H_0$. As expected, the value of the parameter $\Omega_\gamma$ must be very small and due to that fact it has an 
insignificant influence on the physics in our Solar System. It was shown that the model provides inflationary scenario as well as it reaches a good agreement with the 
present day experimental data. The inflation is given by a singularity of the type III (freeze singularity) which appeared naturally as a pole in the Newtonian 
potential since we had treated the dynamics of the Universe as a dynamics of a unit-mass particle in a potential. The pictures \ref{fig:6} and \ref{fig:2} show that 
the value of the scale factor depends strictly on the values of the density parameter $\Omega_\gamma$ while the dependence on the other parameters was very weak and hence 
we could neglect their influences. With these properties, we could restrict the value of $\Omega_\gamma$ to the interval $[0;10^{-9})$ in order to have the 
singularity before the recombination epoch. That guaranteed the preservation of post-recombination physics of the Universe. Such construction provides the four
phases of the cosmic evolution:  the decelerating phase dominated by matter, an intermediate
inflationary phase corresponding to III type singularity, a phase of matter domination (decelerating phase) and finally, the phase of acceleration of current universe.
Moreover, the Big Bang singularity is also preserved as we incorporated the radiation term into the Friedmann equation. We call the singularity, which provides the 
inflation, a singularity of the type III but some comment should be given here. A freeze singularity, as well as Big Rip and Sudden singularities, has not a well-defined 
acceleration $\ddot{a}$; it diverges. Moreover, the left and right limits at the singular point of the acceleration differ in our case. The scale factor is 
finite whilst the effective energy density, pressure and the Hubble rate diverge. It 
suggests that we deal with a freeze singularity which are characterized by the generalized Chaplygin equation of state \cite{kiefer2010avoidance}. A new feature 
is the one already mentioned: the acceleration is infinite, that is, it goes to the plus infinity on the left from the singular point and to the minus infinity
 from the right hand side. Due to that fact (see the picture \ref{fig:5}) one may glue the past and future trajectories in the singular point. Such type of a weak
 singularity has been called a "degenerate freeze singularity". The degenerate freeze singularity vanishes if 
$\Omega_\gamma=0$ and therefore one deals with the standard $\Lambda$CDM model. It allows us to suppose that the presence of such singularity is related to Palatini 
formalism. Further studies on that topic are urgent and will be performed in the nearest future.

Furthermore, using the dynamical system approach, we were able to examine all evolutionary paths of our cosmological model. The phase space structure of dynamics is 
organized through the two saddle points which represent static Einstein universes. The degenerate freeze singularity lies between the saddle points on the $x$-axis. 
The set of sewn trajectories is a critical point located at the infinity $(\dot{a}=\infty,a=a_{\text{fsing}})$ and they pass through that point. Let us notice
that on the phase portrait \ref{fig:5} we have drawn a red trajectory which represents the flat model ($k=0$). It divides the rest of the trajectories into 
domains occupied by closed ($k=+1$) and open ($k=-1$) models. The closed model's trajectories are located between the top and the bottom red ones. We see that all 
open models possess the degenerate freeze singularity. 

One may ask a question whether the existence of singularities disqualifies Palatini $f(\mathcal{R})$ theories for it seems that they can be artifacts of the 
formalism. It could limit its application. However, we would like to notice that we have just considered a very simple modification of General Relativity. Other 
forms of $f(\mathcal{R})$ functionals, as for example higher order terms in $\mathcal{R}$ or its inverse, may happen not to possess singularities which we do not know 
how to treat on the classical level. One believes that Quantum Gravity will solve the problem of singularities.


\chapter{Hybrid metric - Palatini gravity}\label{hg_chapter} 



Hybrid metric - Palatini gravity was introduced for the first time by T. Harko et al. \cite{Harko:2011nh} in order to avoid disadventages of $f(R)$ gravity, both in metric 
and in Palatini formalisms, while its modified dynamics provides self-accelerated cosmic expansion without inserting exotic dark energy into the equations. 
The presented approach includes corrections to General Relativity which have a form of an arbritrary function $f(\mathcal{R})$ considered in Palatini formalism. It 
means, similarly to Palatini $f(\mathcal{R})$ gravity, one deals with two independent objects, that is, a metric $g_{\mu\nu}$ and a connection $\hat{\Gamma}$ which turns 
out to be a Levi-Civita connection of a metric conformally related to the metric $g_{\mu\nu}$. Moreover, one may express the theory as a scalar-tensor one which makes 
the studies on the hybrid gravity easier to handle. Post-Newtonian analysis (weak-field and slow-motion limits) \cite{Harko:2011nh, capozziello2015hybrid} shows that
the theory passes the Solar system constraints even for light scalar fields. The coupling of the scalar field to the local system depends on the current cosmic 
amplitude which is supposed to be sufficiently small. That affects the dynamics of galaxes and evolution of the Universe: the modified gravitational potential 
\cite{Harko:2011nh} seems to describe the flatness of rotational curves of galaxies very well. The Cauchy problem was also considered \cite{capozziello2014cauchy}.
Moreover, some specific example of the hybrid gravity 
model \cite{capozziello2012wormholes} supports the stability of wormholes solutions. There are also speculations that there is no need to introduce an
exotic matter violating null energy condition in order to satisfy the wormhole throat condition because such geometry can be obtained by the higher order curvature terms.

Cosmology provided by hybrid gravity has been also extensively examinated \cite{Capoz}. By introducing a field $X=\kappa T+R$ (showing the deviation from the GR trace 
equation $R=-\kappa T$) one may parametrize the astrophysical scales according to the cosmic densities because it is considered as the chameleon mechanism. At short 
scales, as Solar System one, the variable $X\rightarrow0$ reduces the model to Einstein one while its value grows at larges scales in order to have the effects of 
dark energy and dark matter. Any cosmological behavior, similarly to other ETG's, can be achieve by an effective potential of scalar field which affects in the 
form of the function $f(\mathcal{R})$. The examination of the stability of Einstein static Universe was also done \cite{bohmer2013einstein}. For that purpose they 
considered linear homogeneous perturbations of the model and showed that there exist stable solutions. Dynamics of scalar perturbations was also studied for that 
framework and expressed in Newtonian and synchronous gauges \cite{lima2014dynamics}. What is interesting, they found a family of functions $f(\mathcal{R})$ which 
reproduce a cosmology that is not distinguisable from the one provided by $\Lambda$CDM model while the meaningful modifications from GR appear in the distant past what 
suggests a different scenario for the early Universe. Further analysis on that topic is found in \cite{lima2016constraints} where they put observational data 
constraints on the scalar field value being in an agreement with allowed early-time deviations from GR. Dynamical system analysis was also applied to the hybrid gravity 
\cite{carloni2015dynamical} where they examined the model of the form $f(\mathcal{R})\sim\mathcal{R}^n$. They showed that for this case the points characterizing GR limits
are unstable, it means, there is no value of parameters and initial conditions that lead to cosmic evolutions indistinguishable from GR. 
Furthermore, thermodynamics properties have been also studied 
at the apparent horizon in the FRLW background \cite{azizi2015thermodynamics}. In the non-equilibrium description (continuity equation of the 
dark fluid coming from the modifications is not satisfied) they obtained that a new entropy production is generated violating the first law of thermodynamics. Assuming 
the continuity equation for the dark fluid, the law is not violated (equilibrium approach). On the other hand, the second law of thermodynamics for hybrid gravity is 
satisfied but with extra restrictions imposed on the cosmological even horizon.

The theory was also investigated from astrophysical point of view. The extension of the virial theorem due to hybrid gravity \cite{capozziello2013virial} might be a tool 
for observational test of that theory. One applies the theorem in order to obtain the mean density of galaxies, clusters etc., to determind the total mass of such 
objects as well as the stability of gravitationally bounded systems. They showed that most of the mass in a cluster is in the form of the geometric mass, that is, 
it is included in the extra geometric terms (scalar field) modifiyng Einstein's gravity. That scalar field contribution to the gravitational energy play a role of dark 
matter at the galactic level. In \cite{capozziello2013galactic} all quantities including the scalar field were expressed in terms of observable parameters opening a 
possibility to direct tests of hybrid gravity. Their theoretical considerations indicate that galactic rotation curves and the mass discrepancy in galaxies can be 
explained by the hybrid gravity field equations being perfectly consistent with observational data. Similar conclusions are given by \cite{borka2015probing}, 
where they used the observational data of stars moving around the centre of our Galaxy.

Hybrid metric - Palatini gravity survived long enough to see its generalisation: as a brane system, that is five-dimensional hybrid gravity \cite{fu2016hybrid} and 
four-dimensional one \cite{tamanini2013generalized} which was done by considering arbitrary functions both 
the metric and Palatini curvature scalars in the gravitational action. The authors proved that such a generalisation leads to GR with two scalar fields coupled to 
each other when one uses the conformal transformation procedure. The model also provides the late time acceleration. Unfortunately, in 
\cite{koivisto2013ghosts} one shows that this gravity model possesses propagating degrees of freedom which are ghosts or tachyons. The only hybrid theory (they have 
also considered $f(R,\mathcal{R}_{\mu\nu}\mathcal{R}^{\mu\nu})$ case) that is viable and reduces to GR is hybrid metric-Palatini one which we are also going to 
study. Our examinations will concern a selection procedure for viable models given by Lie and Noether symmetries. It will also allow us to solve cosmic evolutionary equations 
as well as to find exact solutions of Wheeler-DeWitt equations for the previously obtained models. The results of the analysis were published in \cite{wojnar} while 
the similar Lie symmetries analysis for Bianchi spacetime in the hybrid gravity framework was investigated in \cite{wojnar2}.


\section{Introduction of the model}

We are going to consider a special form of the action introduced in \cite{Harko:2011nh} which consists of two parts: one deals with the 
standard Hilbert - Einstein action and an additional term which is constructed with arbitrary function of the Palatini curvature scalar $\mathcal{R}$: 
\begin{equation}
S=\frac{1}{2\kappa}\int\mathrm{d}^{4}x\sqrt{-g}[R+f(\mathcal{R})]+S_{m}.
\label{action_hyb}%
\end{equation}
The Palatini scalar $\mathcal{R}$ is built with two independent object which are the metric $g$ (of Lorentzian signature) and the connection $\hat{\Gamma}$:
\begin{equation}
 \mathcal{R}=g^{\mu\nu}\mathcal{R}_{\mu\nu},\;\;
 \mathcal{R}_{\mu\nu}=\hat{\Gamma}^\alpha_{\mu\nu,\alpha}-\hat{\Gamma}^\alpha_{\mu\alpha,\nu}+\hat{\Gamma}^\alpha_{\alpha\lambda}\hat{\Gamma}^\lambda_{\mu\nu}-
 \hat{\Gamma}^\alpha_{\mu\lambda}\hat{\Gamma}^\lambda_{\alpha\nu}.
\end{equation}
The third term, $S_{m}=\int\mathrm{d}^{4}x\sqrt{-g}\mathcal{L}_m$, stands for the standard matter action and in this approach we assume that is connection 
independent. The Ricci scalar $R$ is obtained from 
the metric $g_{\mu\nu}$.

The variation of the above action with respect to the metric gives the gravitational field equations (Einstein's modified equations)
\begin{equation}
\label{eq_hyb}G_{\mu\nu}+f^{\prime}(\mathcal{R})\mathcal{R}_{\mu\nu}-\frac
{1}{2}f(\mathcal{R})g_{\mu\nu}=\kappa T_{\mu\nu},
\end{equation}
where $G_{\mu\nu}$ is the Einstein tensor of the metric $g$ and $f'(\mathcal{R})=df(\mathcal{R})/d \mathcal{R}$. The matter energy - momentum tensor
was defined as usual, it means
\begin{equation}
 T_{\mu\nu}:=-\frac{2}{\sqrt{-g}}\frac{\delta(\sqrt{-g}\mathcal{L}_m)}{\delta(g^{\mu\nu})}.
\end{equation}
The variation with respect 
to the connection $\hat{\Gamma}$ provides (as in the pure $f(\mathcal{R})$ Palatini gravity) the following equation
\begin{equation}
 \tilde{\nabla}_\alpha(\sqrt{-g}f'(\mathcal{R})g^{\mu\nu})=0
\end{equation}
which reveals that the independent connection is the Levi-Civita connection of the metric $\tilde{g}_{\mu\nu}=f'(\mathcal{R})g_{\mu\nu}$. It means that the 
metrics $g$ and $\tilde{g}$ are conformally related and the function $f'(\mathcal{R})$ is a conformal factor. It requires that $f'(\mathcal{R})$ is
a positive defined function. There is a well-known relation between the two curvatures \cite{dabrowski}
\begin{equation}\label{conf}
 \mathcal{R}_{\mu\nu}=R_{\mu\nu}+\frac{3}{2}\frac{F(\mathcal{R})_{;\mu}F(\mathcal{R})_{;\nu}}{F^2(\mathcal{R})}-
 \frac{1}{F(\mathcal{R})}\nabla_\mu F(\mathcal{R})_{;\nu}-\frac{1}{2}\frac{g_{\mu\nu}\Box F(\mathcal{R})}{F(\mathcal{R})},
\end{equation}
where $_{;\mu}\equiv\nabla_\mu$ is the metric connection, $\Box\equiv\nabla^\mu\nabla_\mu$ d'Alembertian operator, and we have defined $F(\mathcal{R}):=f'(\mathcal{R})$.

The trace of the modified Einstein equations (\ref{eq_hyb}) is called the hybrid
structural equation or hybrid master equation. Assuming that $f(\mathcal{R})$ has analytic solutions, one may obtain that the Palatini curvature
$\mathcal{R}$ is expressed in terms of the variable $X$ from the algebraic relation 
\begin{equation}
F(\mathcal{R})\mathcal{R}-2f(\mathcal{R})=\kappa T+R\equiv X.
\label{master}%
\end{equation}
The variable $X$ measures the deviation from the General Relativity trace equation $R=-\kappa T$ and with that definition one 
has $F(\mathcal{R}(X))\equiv F(X)$. It is possible to reformulate the previous field equations in terms of the variable $X$ and the function $F(X)$ \cite{Capoz}.

The action of the hybrid gravity theory can be also transformed into a scalar-tensor theory action in the similar manner as for the pure metric and Palatini 
case \cite{report,olmo}. One needs to introduce an auxiliary field $E$ such that \cite{Harko:2011nh,Capoz}
\begin{equation}
S=\frac{1}{2\kappa^{2}}\int\mathrm{d}^{4}x\sqrt{-g}[R+f(E)+f_E(\mathcal{R}-E)]+S_m,
\end{equation}
where $f_E\equiv\frac{df(E)}{dE}$. The field $E$ is dynamically equivalent to the Palatini scalar $\mathcal{R}$ if
$f^{\prime\prime}(\mathcal{R})\neq0$. Let us define a scalar field and its potential 
\begin{equation}
\phi\equiv f^{\prime}(E), \qquad   V(\phi)=Ef^{\prime}(E)-f(E). \label{lagr}%
\end{equation}
Applying them to the above action it becomes
\begin{equation}
S=\frac{1}{2\kappa^{2}}\int\mathrm{d}^{4}x\sqrt{-g}[R+\phi\mathcal{R}%
-V(\phi)]+S_m.\label{ac2}
\end{equation}
The variation of (\ref{ac2}) with respect to the metric, the scalar field $\phi$ and the connection provides the field equations
\begin{align}
 G_{\mu\nu}+\phi\mathcal{R}_{\mu\nu}-\frac{1}{2}(\phi\mathcal{R}-V)g_{\mu\nu}&=\kappa T,\label{row1}\\
 \mathcal{R}-V_{,\phi}&=0,\label{row2}\\
 \hat{\nabla}_\alpha(\sqrt{-g}f'(\mathcal{R})g^{\mu\nu})&=0.
\end{align}
Using the conformal relation (\ref{conf}) between $R$ and $\mathcal{R}$ (let us remind that we are using the metric $g_{\mu\nu}$ for lowering and raising indices),
that is
\begin{equation} \label{eq:conformal_R}
\mathcal{R}=R+\frac{3}{2\phi^2}\nabla_\mu \phi \nabla^\mu \phi-\frac{3}{\phi}\Box \phi,
\end{equation}
the standard scalar-tensor form can be obtained:
\begin{equation}
S=\frac{1}{2\kappa^{2}}\int\mathrm{d}^{4}x\sqrt{-g}\left[(1+\phi)R+\frac{3}{2\phi
}\nabla^{\mu}\phi\nabla_{\mu}\phi-V(\phi)\right]. \label{lagr0}%
\end{equation}
Applying the conformal relations to the equation (\ref{row1}) and (\ref{row2}) together with the trace of (\ref{row1}) we have
\begin{align}
 (1+\phi)G_{\mu\nu}-\frac{3}{4\phi}g_{\mu\nu}\nabla^\alpha\phi\nabla_\alpha\phi+
 \frac{3}{2\phi}\nabla_\mu\phi\nabla_\nu\phi \nonumber\\ 
 +g_{\mu\nu}\Box\phi-\nabla_\mu\nabla_\nu\phi+g_{\mu\nu}V=\kappa T_{\mu\nu},\label{ein1}\\
 -\Box\phi+\frac{1}{2\phi}\nabla^\alpha\phi\nabla_\alpha\phi+\frac{\phi}{3}(2V-(1+\phi)V_{,\phi})=\frac{\kappa\phi}{3}T.\label{kg11}
\end{align}
The equation (\ref{kg11}) is the second-order evolution equation for the scalar field $\phi$ and can be interpreted as an effective Klein-Gordon equation. It is 
important to notice that in hybrid gravity theory, unlike in the Palatini case, the scalar field is dynamical (\ref{notdyn}).


\subsection{Hybrid gravity cosmology}

In order to examine the hybrid gravity model let us consider Friedmann - Robertson - Lemaitre - Walker (FRLW) spatially flat metric:
\begin{equation}
ds^{2}_{FRLW}=-dt^{2}+a^{2}\left(  t\right)  \left(  dx^{2}+dy^{2}+dz^{2}\right),
\label{FRW}%
\end{equation}
where $a(t)$ is the scale factor. One easily finds that the Ricci scalar is $R=6(2H^2+\dot{H})$ where one defines the Hubble 
rate function $H=\frac{\dot{a}(t)}{a(t)}$. The dot denotes the cosmic time derivative, it means $\frac{\mathrm{d}}{\mathrm{d}t}$. The field 
equations in the scalar-tensor representation (\ref{ein1}), (\ref{kg11}) are
\begin{align}
 3H^2=\frac{1}{1+\phi}\left[ \kappa\rho + \frac{V}{2} -3\dot{\phi}\left( H+\frac{\dot{\phi}}{4\phi}  \right) \right],\label{fri1}\\ 
 2\dot{H}=\frac{1}{1+\phi}\left[ -\kappa(\rho+p)+H\dot{\phi}+\frac{3}{2}\frac{\dot{\phi}^2}{\phi}-\ddot{\phi}  \right],\label{fri2}\\ 
 \ddot{\phi}+3H\dot{\phi}-\frac{\dot{\phi}^2}{2\phi}+\frac{\phi}{3}[2V-(1+\phi)V_{,\phi}]=-\frac{\kappa\phi}{3}(\rho-3p).\label{KG1}
\end{align}
$\rho$ and $p$ are the energy density and pressure of the cosmic fluid coming from the trace of perfect fluid energy - momentum tensor:
\begin{align}
 T_{\mu\nu}=&(\rho+p) u_\mu u_\nu + p g_{\mu\nu},\\
 T=&g^{\mu\nu}T_{\mu\nu}=3p-\rho,
\end{align}
where $u^\mu=(1,0,0,0)$ is a co-moving observer with the normalization condition $u^\alpha u_\alpha=-1$. The conservation of the matter energy - momentum 
tensor is (assuming that it is a metric theory $\nabla_\mu T^{\mu\nu}=0$)
\begin{equation}
 \dot{\rho}+3H(\rho+p)=0.
\end{equation}
The above Klein - Gordon equation (\ref{KG1}) can be written in the following form \cite{Harko:2011nh}:
\begin{equation}\label{kg}
 \ddot{\phi}+3H\dot{\phi}-\frac{\dot{\phi}^2}{2\phi}+M^2_\phi(T)\phi=0,
\end{equation}
where $M^2_\phi(T):=m^2_\phi-\frac{1}{3}\kappa T=\frac{1}{3}[2V-(1+\phi)V_{,\phi}-\kappa T]$. Assuming that the scalar field $\phi$ is not rapidly changing, then 
$\frac{\dot{\phi}^2}{2\phi}\sim0$ and (\ref{kg}) represents a massive scalar field on a FRLW background. The dynamical behavior of the scalar field at late 
times depends on a form of the potential appearing in the Klein - Gordon equation. In \cite{Harko:2011nh} the authors consider two models which are consistent at 
Solar System and cosmological scale with asymptotically de Sitter behavior. One may also examine the deceleration
parameter \cite{misner1973gravitation, Capoz} defined by 
\begin{equation}\label{dece}
 q=\frac{d}{dt}\frac{1}{H}-1=-\frac{\dot{H}}{H^2}-1.
\end{equation}
One deals with accelerated expansion of the Universe when $q<0$. There is also an important class of cosmological models for which the deceleration parameter
equals zero (so-called marginally accelerating models). Being a monotonically decreasing function the deceleration parameter whose the value reaches
$q=0$ indicates models which start from 
decelerating states ($q>0$) and end in accelerated ones. Such models can be interpreted as the turning point between a structure formation epoch 
and dark energy \cite{Capoz}.

Starting from now on, we will skip the matter action $S_m$ in (\ref{action_hyb}) so we will always consider vacuum case ($T_{\mu\nu}=0$) unless one indicates directly 
an introduction of standard matter. In the absence of matter, we may rewrite the modified Friedmann equations (\ref{fri1}) and (\ref{fri2}) as:
\begin{align}
3H^{2}  &  = \kappa \rho_{\rm eff} ,\label{fr1} \\
2\dot{H}   &  = -\kappa (\rho_{eff}+p_{\rm eff}),\label{fr2}
\end{align}
where $\rho_{\rm eff}$ and $p_{\rm eff}$ are effective energy density and pressure given by
\begin{align}
(1+\phi)\kappa\rho_{\rm eff}  &  =-\frac{3}{4\phi}\dot{\phi}^2+\kappa V-3H\dot{\phi}\label{fe.03a}\\
(1+\phi)\kappa p_{\rm eff}  &  = -\frac{3}{4\phi}\dot{\phi}^2-\kappa V+2H\dot{\phi}+\ddot{\phi} .\label{fe.03b}%
\end{align}
Using the formulas (\ref{fr1}) and (\ref{fr2}) the deceleration parameter is
\begin{equation}
 q=\frac{3}{2}\frac{\rho_{eff}+p_{eff}}{\rho_{eff}}-1
\end{equation}
so with the definitions (\ref{fe.03a}) and (\ref{fe.03b}) with the Klein - Gordon equation (\ref{kg}) in vacuum one gets the dependence on the scalar field:
\begin{equation}
 q=-\frac{3}{2}\left(\frac{\frac{\dot{\phi}^2}{\phi}+4H\dot{\phi}+\frac{2}{3}\kappa\phi[2V-(1+\phi)V_{,\phi}]}
 {-\frac{3\dot{\phi}^2}{4\phi}+\kappa V-3H\dot{\phi}}\right)-1.
\end{equation}
As mentioned above, dynamics of cosmological models may be studied with respect to the deceleration parameter. In \cite{Capoz} the authors showed the 
existence of a few viable accelerating solutions (recall that $\phi\equiv\frac{df(\mathcal{R})}{d\mathcal{R}}$) for the hybrid gravity model among which
there are power-law accelerating models.

From the cosmological point of view, hybrid gravity is a promising and interesting model which should be further investigated. Solutions, it means forms of the 
scalar field potential or functions $f(\mathcal{R})$, should satisfy restricted requirements. Since the field equations are quite difficult and demand a form
of $V(\phi)$ (when for example one wants to solve Wheeler-DeWitt equation), we need special tools to deal with them. Due to that fact, we would like to 
study FRLW cosmology of the hybrid gravity model with respect to symmetries. For the further investigations we will need a point-like Lagrangian deduced from
(\ref{lagr0})
\begin{equation}
\mathcal{L}=6a\dot{a}^{2}(1+\phi)+6a^{2}\dot{a}\dot{\phi}+\frac{3}{2\phi}%
a^{3}\dot{\phi}^{2}+a^{3}V(\phi). \label{lagr1}%
\end{equation}


\section{Noether symmetries in cosmology}\label{sec1symetr}
Using Noether symmetries approach to cosmology is nothing new but the huge number of works considering the theorem in cosmological applications (see for example
\cite{capozziello2008f, capozziello2009noether, kucukakca2012noether, roshan2008palatini, zhang2010noether, vakili2008noether, vakili2008noether2})
just shows that it is a powerful tool in theoretical physics. Noether symmetries do not only allow to solve differential equations by finding integrals of 
motion but also, in the case of Extended Theories of Gravity, they select models which are viable due to Noether symmetries of the system. 
They are a physical criterion as each symmetry is related to a conserved quantity which has a physical meaning \cite{capozziello2009noether}. Moreover, as 
it was shown in \cite{capozziello2000selection}, Noether symmetries are a selection rule to recover classical behaviors in cosmic evolution:
\begin{theorem}
 In the semi-classical limit of quantum cosmology and in the framework of minisuperspace approximation, the reduction procedure of dynamics, due to existence of
 Noether symmetries, allows to select a subset of the solution of Wheeler-DeWitt equation where oscillatory behaviors are found. As consequence, correlations between
 coordinates and canonical conjugate momenta emerge so that classical cosmological solutions can be recovered.  Vice-versa,  if  a  subset  of  
 the  solution  of  WDW  equation  has  an  oscillatory  behavior,  due to 
 \begin{equation}
  -i\partial_j |\Psi>=\Sigma_j |\Psi>,\;\;\;j=1,...,k,\;\;\;k\;\text{- number of symmetries},
 \end{equation}
 where $|\Psi>$ is a wave function of the Universe while $\Sigma$ is a constant of motion (first integral), 
 conserved momenta have to exist and Noether symmetries are present.
 In other words, Noether symmetries select classical universes.
\end{theorem}
From the above conclusion of Capozziello and Lambiase \cite{capozziello2000selection} the importance of Noether symmetries in (quantum) cosmology is evident. Let us
briefly discuss it.

In the canonical quantization approach (based on $3+1$ decomposition of Einstein's field equations, also called ADM formalism \cite{arnowitt1959dynamical}) to
quantum gravity (see for example \cite{Kiefer2007ria}) one deals with so-called superspace of 
geometrodynamics which is an 
infinite-dimensional configuration space where the classical dynamics takes place. It is a space of all $3$-metric and matter field configurations which are 
defined on $3$-manifold. The superspace metric is constructed with $3$-metrics and matter fields, and it appears in the kinetic term of Hamiltonian 
constraint (see for example \cite{gourgoulhon}). A quantum state of the Universe is represented by a wave functional not depending explicitly on the coordinate 
time $t$. Wheeler-DeWitt (WDW) equation is, roughly speaking, a quantized version of Einstein's field equations (together with so-called momentum constraint 
\cite{Kiefer2007ria, gourgoulhon, capozziello2012hamiltonian}). It is a second-order hyperbolic functional differential equation describing the dynamical 
evolution of the wave function in superspace, that is, the wave function of the Universe. There are many difficulties arising from this approach. One of them is
the infinite dimension of the superspace that makes WDW equations impossible to fully integrate. In order to deal with that problem one consider a toy model of quantum 
cosmology by imposing restrictions on the metric and matter fields engaged to a game. Such a simplified superspace is called minisuperspace. The simplest one 
possesses homogeneous and isotropic metrics, and matter fields. Corresponding supermetric is called minisupermetric. With finite dimension of configuration 
spaces one may solve WDW equations and try to interpret obtained results. The popular interpretation is given by Hartle; so-called Hartle 
criterion \cite{hartle1986ingravitation, capozziello2000selection} says that strong peaks of the wave function of the Universe indicates classical 
trajectories (it means universes). It corresponds to the oscillatory behavior of the wave function (see the analogical situation in
the non-relativistic quantum mechanics) for which the system behaves like a classical one while exponential regime of the wave function is classically forbidden. 

Coming back to Noether symmetries of a minisuperspace cosmological model under consideration, one shows 
\cite{capozziello2000selection} that conserved momenta are connected to oscillatory parts of the wave function in the directions of corresponding symmetries.
Finding Noether symmetries and corresponding
constants of motion allows not only to reduce and solve differential equations but also to specify oscillatory parts of the wave function which is an exact solution 
of the WDW equation.

\subsection{Noether symmetries of hybrid gravity model}

We are going to use Noether symmetries in order to solve classical equations of motion derived from the hybrid gravity Lagrangian (\ref{lagr1}). We will follow 
the approach developed in \cite{TsamAnd} (see notes in (\ref{ap_noether}) for details). One observes that the Lagrangian (\ref{lagr1})
\begin{equation}
\mathcal{L}=6a\dot{a}^{2}(1+\phi)+6a^{2}\dot{a}\dot{\phi}+\frac{3}{2\phi}%
a^{3}\dot{\phi}^{2}+a^{3}V(\phi)
\end{equation}
has the
standard form ${\cal L}=\mathcal{T}-\mathcal{V}_{\rm eff}$. The kinetic energy $\mathcal{T}=\frac{1}{2}g^{(2)}_{\mu\nu}\dot{x}^{\mu}\dot{x}^{\nu}$ indicates 
the minisuperspace metric
\begin{equation}
ds_{\left(  2\right)  }^{2}=12a\left(  1+\phi\right)  da^{2}+12a^{2}%
dad\phi+\frac{3}{\phi}a^{3}d\phi^{2} , \label{HG.02}%
\end{equation}
while $\mathcal{V}_{\rm eff}=-a^{3}V\left(  \phi\right)$ stands for effective potential.
Applying the theorem \ref{theo_noether} to the Lagrangian (\ref{lagr1}) we are able to find such forms of potential $V\left(\phi\right)$ that 
the hybrid gravity Lagrangian will admit Noether point symmetries. 

Since the Lagrangian is time-independent, it admits the Noether symmetry $\partial_{t}$ with 
the Hamiltonian (it means the first Friedmann equation (\ref{fr1})) as a conservation law, that is
\begin{equation}
E_{H}=6a\left(  1+\phi\right)  \dot{a}^{2}+6a^{2}\dot{a}\dot{\phi}+\frac
{3}{2\phi}a^{3}\dot{\phi}^{2}-a^{3}V\left(  \phi\right).
\end{equation}
Let us recall that we are considering a vacuum case so the Einstein equation $G_{0}^{0}=0$ being a constraint gives that $E_{H}=0$. The Noether 
 condition (\ref{noe_cond}) gives us, 
together with $\xi=1$, the following system of partial differential equations:
\begin{align}
 \eta(1+\phi)+a^2\rho_{,a}+a(\rho+2(1+\phi)\eta_{,a})=&0,\label{noe1}\\
 \phi(3\eta+4\phi\eta_{,\phi}-a(\rho-2\phi\rho_{,\phi}))=&0,\label{noe2}\\
 4\phi^2\eta_{,\phi}+a^2\rho_{,a}+2a\phi(24\eta\eta_{,\phi}+\rho_{,\phi}+\eta_{,a})=&0,\\
 2(1+\phi)\eta_{,t}+a\rho_{,t}=&0,\label{noe4}\\
 2\eta_{,t}+\frac{a}{\phi}\rho_{,t}=&0,\\
 a\rho V'(\phi)+3V(\phi)\eta=&0.
\end{align}
Since the potential $V(\phi)$ does not depend on the scale factor, $\eta=0$ and hence the potential is a constant value. Applying that result into
the equations (\ref{noe1}), (\ref{noe2}) and (\ref{noe4}) we found that $\rho=\frac{\sqrt{\phi}}{a}$ and therefore we 
have just shown that there also exists an extra Noether symmetry for a constant potential $V(\phi)=V_{0}$:
\begin{equation}
X_{1}=\frac{\sqrt{\phi}}{a}\partial_{\phi}
\end{equation}
while the corresponding Noether integral is of the form%
\begin{equation}
I_{1}=3\frac{a}{\sqrt{\phi}}\left(  2\phi\dot{a}+a\dot{\phi}\right)  .
\end{equation}
Instead of looking for solutions of field equations in $(a,\phi)$ coordinates (which would be tough), let us perform the following coordinate transformation%
\begin{equation}
a=u^{\frac{2}{3}}, \qquad \phi=v^{2}u^{-\frac{4}{3}},
\end{equation}
under which the Lagrangian of the field equations became%
\begin{equation}
{\cal L}\left(  u,v,\dot{u},\dot{v}\right)  =\frac{8}{3}\dot{u}^{2}+6u^{\frac{2}{3}%
}\dot{v}^{2}+V_{0}u^{2} .
\end{equation}
In the $(u,v)$ variables the field equations have the forms
\begin{align}
\frac{8}{3}\dot{u}^{2}+6u^{\frac{2}{3}}\dot{v}^{2}-V_{0}u^{2} &  =0 , \label{HG.04a}\\
\ddot{u}-\frac{3}{4}u^{-\frac{1}{3}}\dot{v}^{2}-\frac{3}{8}V_{0}u  & =0, \label{HG.05a}\\
\ddot{v}+\frac{2}{3u}\dot{u}\dot{v}  &  =0,
\end{align}
while we wrote the Noether integral as $\bar{I}_{1}=u^{\frac{2}{3}}\dot{v}$. Replacing one of the velocities by
$\dot{v}=\bar{I}_{1}u^{-\frac{2}{3}}$, we are able to write the general solution of the above system as
\begin{equation}
\int\frac{du}{\sqrt{\frac{3}{8}V_{0}u^{2}-\frac{9}{4}\bar{I}_{1}^{2}%
u^{-\frac{2}{3}}}}=\int dt. \label{HG.04ab}%
\end{equation}
The solution of (\ref{HG.04ab}) exists but the scale factor cannot be expressed in terms of the cosmic time $t$ in an algebraic and easy way.
But we can rewrite the first Friedmann equation (\ref{HG.04ab}) in terms of the scale factor $a(t)$ as the Hubble function (recall that $H=\frac{\dot{a}}{a}$):
\begin{equation}
\frac{H^{2}}{H_{0}^{2}}=\left(  \Omega_{\Lambda}+\Omega_{r}a^{-4}\right),
\label{HG.05aa}%
\end{equation}
where we have defined the density parameters of the cosmological constant and radiation as
\begin{equation}
\Omega_{\Lambda}=\frac{1}{6}\frac{V_{0}}{H_{0}^{2}}, \qquad {\rm and} \qquad
\Omega_{r}=-\frac{\bar{I}_{1}^{2}}{H_{0}^{2}},
\end{equation}
respectively. Let us just notice that in order to have a physical solution, the integral of motion has to be a complex value. Comparing the result to the 
$\Lambda$CDM model one sees that the hybrid gravity model introduces radiation term - let us underline that we are considering vacuum case from the very beginning.
It means that such a geometric modification provides "matter fluids" which are a cosmological constant responsible for the late time acceleration (as 
the $\Lambda$CDM model does) and radiation. In the contrary to $\Lambda$CDM model, the radiation term appears "naturally" in the Friedmann equation derived 
from hybrid gravity while in the first case one needs to put it by hand since radiation fluid does not contribute to the trace of an energy momentum tensor.

One may also perform a similar analysis after introducing dust to the model, it means introducing $\rho_{D}=\rho_{m,0}a^{-3}$ in (\ref{fe.03a}).
Hence, the equation (\ref{HG.04a}) becomes
\[
\frac{8}{3}\dot{u}^{2}+6u^{\frac{2}{3}}\dot{v}^{2}-V_{0}u^{2}=\rho_{m0}.
\]
The analytical solution is written as
\[
\int\frac{du}{\sqrt{\frac{3}{8}V_{0}u^{2}+\frac{3}{8}\rho_{m0}-\frac{9}{4}%
\bar{I}_{1}^{2}u^{-\frac{2}{3}}}}=\int dt.
\]
Simply, we may also write the Hubble function as
\[
\frac{H^{2}}{H_{0}^{2}}=\left(  \Omega_{\Lambda}+\Omega_{m}a^{-3}+\Omega
_{r}a^{-4}\right)  ,
\]
where we have defined another density parameter corresponding to the dust ${\displaystyle \Omega_{m}=\frac{\rho_{m0}}{6H_{0}^{2}}}$.

\subsection{Noether symmetries of conformal hybrid gravity Lagrangian}
\label{ConT}
The examination of hybrid gravity with respect to Noether symmetries gave us a trivial solution. However, one may try to perform a conformal transformation of
our Lagrangian (\ref{lagr1}) and apply results of \cite{TsamC02,TsamC01,AnIJGMP} (see the theorem \ref{theo_confor} and the section (\ref{useful_theo})).
As the dynamical system provided 
by the Lagrangian (\ref{lagr1}) is conformally 
invariant ($E_H=0$) one is going to look for new solutions in conformal frames. We will focus on the case when 
the lapse function is a function of the scale factor, it means $d\tau=N(a)dt$
\begin{equation}
ds^{2}=-N^{-2}\left(  a\left(  \tau\right)  \right)  d\tau^{2}+a^{2}\left(
\tau\right)  \left(  dx^{2}+dy^{2}+dz^{2}\right)  . \label{FRW1}%
\end{equation}
One may also consider the lapse function which depends on a scalar field 
 and show that hybrid gravity is conformally related to a Brans-Dicke-like scalar-tensor theory \cite{wojnar}. 
 
The already mentioned Lagrangian for the conformal FRLW spacetime
(\ref{FRW1}) is given as:
\begin{eqnarray}
&& {\cal L}\left(  a,\phi,a^{\prime},\phi^{\prime}\right)  =  \frac{a^{3}V\left(  \phi\right)
}{N\left(  a\right)  } 
\nonumber \\
&&+N\left(  a\right)  \left[
6a\left(  1+\phi\right)  a^{\prime2}+6a^{2}a^{\prime}\phi^{\prime}
     +\frac{3}{2\phi}a^{3}\phi^{\prime2}\right] , \label{FRW.02}%
\end{eqnarray}
where the prime denotes $d/d\tau$. 
The conformal kinetic metric and Ricci scalar of that metric are given by
\begin{equation}
d\bar{s}_{\left(  2\right)  }^{2}=N\left(  a\right)  \left(  12a\left(
1+\phi\right)  da^{2}+12a^{2}dad\phi+\frac{3}{\phi}a^{3}d\phi^{2}\right) ,
\label{FRW.03}%
\end{equation}
and
\[
R_{\left(  2\right)  }=-\frac{a^{2}NN_{,aa}-a^{2}N_{,a}^{2}-N^{2}}%
{12a^{3}N^{3}},
\]
respectively. We will consider the case when $R_{\left(  2\right)  }=0$
so the problem reduces to the dynamics of
Newtonian physics \cite{TsamC02}. Now the lapse function is of the form
\begin{equation}
N\left(  a\right)  =a^{-1}e^{N_{0}a}. \label{FRW.04}%
\end{equation}
Applying the lapse function (\ref{FRW.04}) into the Lagrangian (\ref{FRW.02}) we may use the geometric approach developed in \cite{TsamAnd}. 
The set of differential equations, that we need to solve, coming from the Noether symmetry approach is
\begin{align}
 \xi_{,a}=&0,\\
 \xi_{,\phi}=&0\\
 \rho+2(1+\phi)\eta_{,a}+a\rho_{,a}=&0,\label{k1}\\
 2\phi(\eta+2\phi \eta_{,\phi})-a(\rho-2\phi\rho_{,\phi})=&0,\label{k2}\\
 4\phi(1+\phi)+(a^2\rho_{,a}+2\phi[\eta+a(\rho_{,\phi+\eta_{,a}})])=&0,\label{k3}\\
 \rho a V'(\phi)+4\eta V(\phi)=&0,\label{k4}
\end{align}
where we have already used that $\xi_{,t}=0$. The case when $\xi_{,t}\neq0$ was considered in \cite{wojnar} where computer algebra was used. Subtracting the 
equation (\ref{k3}) from (\ref{k2}) and using (\ref{k1}) one finds that $a\eta_{,a}=2\phi\eta_{,\phi}$. Let us examine the case for the constant $\eta=-\frac{1}{2}$. 
That ansatz provides the equation (\ref{k1}) as $\rho=\frac{A(\phi)}{a}$, where $A(\phi)$ is an unknown function of the scalar field. Applying the result 
into (\ref{k2}) or (\ref{k3}) one gets the equation determining the function $A(\phi)$: 
\begin{equation*}
 \phi+A(\phi)+A'(\phi)=0
\end{equation*}
with the solution $A(\phi)=\phi+V_1\sqrt{\phi}$, where $V_1$ is a constant. Using the obtained solutions to (\ref{k4}) we find that
\begin{equation}
 V\left(  \phi\right)  =V_{0}\left(  \sqrt{\phi}+V_{1}\right)  ^{4}.
\label{FRW.07}%
\end{equation}
and the extra Noether symmetry which is admitted by the conformal system has the form:
\begin{equation}
X_{1}=-\frac{1}{2}\partial_{a}+\frac{\phi+V_{1}\sqrt{\phi}}{a}\partial_{\phi}.
\label{FRW.05}%
\end{equation}
The corresponding conservation law is
\begin{equation}
I_{X_{1}}=6\left(  V_{1}\sqrt{\phi}-1\right)  \dot{a}+3\frac{a}{\sqrt{\phi}%
}V_{1}\dot{\phi}. \label{FRW.06}%
\end{equation}
There exists also the second symmetry vector \cite{wojnar} with the corresponding conservation law:
\begin{equation}
X_{2}=2\tau\partial_{\tau}+a\left(  \sqrt{\phi}V_{1}+1\right)  \partial
_{a}-2V_{1}\sqrt{\phi}\left(  \phi+1\right)  \partial_{\phi}  , \label{FRW.08}%
\end{equation}
and
\begin{equation}
I_{X_{2}}=12a\left(  1+\phi\right)  \dot{a}+6a^{2}\left(  1-\frac{V_{1}}%
{\sqrt{\phi}}\right)  \dot{\phi},  \label{FRW.09}%
\end{equation}
respectively, with the potential given by
\begin{equation}
V\left(  \phi\right)  =V_{0}\left(  1+\phi\right)  ^{2}\exp\left(  \frac
{6}{V_{1}}\arctan\sqrt{\phi}\right)  . \label{FRW.10}%
\end{equation}
We have chosen $N_{0}=0$ for both cases because in the case of $N_{0}\neq0$, one finds
that the Lagrangian (\ref{FRW.02}) admits extra Noether symmetries only in the
case of the trivial potential $V\left(  \phi\right)  =0$.


From the definition of the potential (\ref{lagr}) one may try to find a form of the function $f(\mathcal{R})$. For the first potential
(\ref{FRW.07}) the equation ($\ref{lagr}$) is
\begin{equation}
Ef^{\prime}(E)-f(E)=V_{0}\left(  \sqrt{f^{\prime}\left(  E\right)  }%
+V_{1}\right)  ^{4}.
\end{equation}
Let us differentiate it with respect to $f'({E})$:
\begin{equation}
\frac{2V_{0}}{\sqrt{f^{\prime}\left(  E\right)  }}\left(  \sqrt{f^{\prime
}\left(  E\right)  }+V_{1}\right)  ^{3}-E=0 ,
\end{equation}
and defining $y=\sqrt{f^{\prime}\left(  E\right)  }$, one has the polynomial equation%
\begin{equation}
\left(  y+V_{1}\right)  ^{3}-\frac{E}{2V_{0}}y=0 . \label{HG.051g1}%
\end{equation}
If we put $V_{1}=0$, the solution of the equation (\ref{HG.051g1}) is the ``Starobinsky-like'' ansatz: 
\begin{equation}
f\left(  E\right)  =\frac{E^{2}}{4V_{0}}\text{.} \label{GR_sol}%
\end{equation}
One notices that the solution (\ref{GR_sol}) after applying to the hybrid master equation (\ref{master}) reproduces the General Relativity trace equation, it 
means $X\equiv \kappa T+R=0$.

Similarly, for the potential (\ref{FRW.10}) one has the following equation to solve:
\begin{equation}
Ef^{\prime}(E)-f(E)=V_{0}\left[1+f^{\prime}\left(  E\right)  \right]
^{2}
 \exp\left(  \frac{6}{V_{1}}\arctan\sqrt{f^{\prime}\left(  E\right)
}\right).
\end{equation}

\section{Exact and invariant solutions}\label{hamil_class}
As we have found the symmetries and potentials appearing in the conformal Lagrangian we may look for the exact solution of the field equations.
We will focus on the potential (\ref{FRW.07}) while the case of the potential (\ref{FRW.10}) is considered in \cite{wojnar}.

Symmetries are very helpful if one wants to find a suitable transformation of variables appearing in field equations. Using the symmetry (\ref{FRW.05}) one may 
find Lie invariants by
\begin{equation}
 -2da=\frac{ad\phi}{\phi+V_1\sqrt{\phi}}.
\end{equation}
The relation between the scalar field $\phi$ and a new variable $v$ is
\begin{equation}
 \phi=\left( \frac{v}{a}-V_1 \right)^2.
\end{equation}
Inserting the above result into the conservation law (\ref{FRW.06}) and performing some simple algebra one finds the following coordinate transformation:
\begin{equation}
a=Cv+u,~\phi=\left(  \frac{v}{Cv+u}-V_{1}\right)  ^{2},
\end{equation}
where $C=V_{1}/(1+V_{1}^{2})$ and the new variable $u$ is constructed with the Noether integral (\ref{FRW.06}). In the new coordinates, the Lagrangian
(\ref{FRW.02})  becomes
\begin{equation}
{\cal L}\left(  u,v,u^{\prime},v^{\prime}\right)  =6\left(  V_{1}^{2}+1\right)
u^{\prime2}+\frac{6}{\left(  V_{1}^{2}+1\right)  }v^{\prime2}+V_{0}v^{4}.
\label{An.01}%
\end{equation}
Let us perform a second transformation of the form
\begin{eqnarray}
x&=&\sqrt{12\left(  V_{1}^{2}+1\right)  }u,
    \\    
y&=&\sqrt{\frac{12}{\left(  V_{1}%
^{2}+1\right)  }}v , \label{An.02}%
\end{eqnarray}
which transforms the Lagrangian (\ref{An.01}) into a much easier form
\begin{equation}
{\cal L}\left(  x, y, x^{\prime}, y^{\prime}\right)  =\frac
{1}{2}x^{\prime2}+\frac{1}{2}y^{\prime2}+\bar{V}_{0}y^{4}  ,\label{An.03}%
\end{equation}
where $\bar{V}_{0}=\frac{V_{0}}{144}\left(  V_{1}^{2}+1\right)  ^{2}$. The Hamiltonian of the field equations is given by
\begin{equation}
\mathcal{H}=\frac{1}{2}p_{x}^{2}+\frac{1}{2}p_{y}^{2}-\bar{V}_{0}y^{4}, \label{An.04}%
\end{equation}
with the momenta $p_{x},p_{y}$ defined below. The Hamilton equations of the 
system ($p'_{q_i}=-\frac{\partial \tilde{H}}{\partial q_i},\;q'_i=\frac{\partial \tilde{H}}{\partial p_{q_i}},\;q_i=\{x,y\}$) are
\begin{equation}
x^{\prime}=p_{x},  \qquad   y^{\prime}=p_{y}%
\end{equation}
\begin{equation}
p_{x}^{\prime}=0, \qquad p_{y}^{\prime}=4\bar{V}_{0}y^{3}.
\end{equation}
Since the Hamiltonian does not depend on time explicitly we may write the Hamilton-Jacobi equation as $S=\bar{S}(x,y,\tilde{H})-\tilde{H}t$. Let us recall that
the Hamiltonian constraint is $\tilde{H}=0$ so the equation is
\begin{equation}
 \left(\frac{\partial\bar{S}}{\partial x}\right)^2+\left(\frac{\partial\bar{S}}{\partial y}\right)^2-2\bar{V}_0y^4=0,
\end{equation}
where $\frac{\partial S}{\partial q_i}=p_{q_i}$. From the Hamilton equations and Hamilton-Jacobi equation one gets that
\begin{align}
 x=c_1\tau + c_2,\;\; y'=\varepsilon\sqrt{2V_0 y^4-c_1^2},\;\;\varepsilon=\pm1,\label{xy_roz}\\
 S=c_1x+\bar{S}(y)+S_0,\;\;\;\bar{S}(y)=\varepsilon\int{\sqrt{2V_0 y^4-c_1^2}dy}.
\end{align}
Let us write (\ref{xy_roz}) as
\begin{align}
x\left(  \tau\right)  &=c_{1}\tau+c_{2}, \label{An.04a}\\
\int\frac{dy}{\sqrt{2\bar{V}_{0}y^{4}-c_{1}^{2}}}&=\varepsilon\left(  \tau
-\tau_{0}\right). \label{An.05}
\end{align}
and consider firstly a simply case when $V_1=c_2=0$. For these assumptions one gets that $a=u=(12)^{-\frac{1}{2}}x=\tilde{c}_1\tau$. From the conformal 
transformation of the time coordinate $dt=a(\tau)d\tau$ the radiation solution can be obtained ($a_0=\sqrt{\frac{2}{\tilde{c}_1}}$):
\begin{equation}
 a(t)=a_0\sqrt{t}.
\end{equation}
We may also perform the another simplification in order to get a more interesting solution. Let us assume that $V_1\neq0$ and $c_1=0$. Then we can solve (\ref{An.05})
\begin{equation}
 y\left(  \tau\right)  =-\varepsilon\frac{1}{\sqrt{2V_{0}}}\frac{1}{\left(
\tau-\tau_{0}\right)}
\end{equation}
so now the scale factor can be expressed as 
\begin{equation}
 a(\tau)=a_1-a_2\frac{1}{\tau-\tau_0}\label{sfact}
\end{equation}
where $a_1=\frac{c_2}{\sqrt{12(V_1^2+1)}}$ and $a_2=\frac{12\varepsilon}{\sqrt{24V_0(V_1^2+1)}}$.
In order to write the Hubble function\footnote{Recall that $H=\frac{1}{a}\frac{da}
{dt}=\frac{1}{a^{2}}\frac{da}{d\tau}$} as a function of the scale factor
\begin{equation}
 H\left(  \tau\right)  = \frac{a'}{a^{2}}=\frac{a_2}{(a_1(\tau-\tau_0)-a_2)^2}
\end{equation}
we need to have $\tau-\tau_0=\frac{a_2}{a_1-a}$ from (\ref{sfact}) and now
\begin{equation}
H\left(  a\right)  =a_{2}^{-1}\left(  a_{1}a^{-1}-1\right)  ^{2}.
\label{Hg.00}%
\end{equation}
The normalized Friedmann equation is defined as $\frac{H^2(a)}{H^2_0}=\mathcal{F}(a)$, where $H_0=H(a(t)=1)$ is a present value of the Hubble constant while $\mathcal{F}(a)$ is 
a function containing density parameters which are comparable with astrophysical data. For the present time we set that $ a(t)=1 $
and from the equation (\ref{Hg.00}) we deduce $a_{2}^{-1}%
=H_{0}/\left(  \left\vert a_{1}\right\vert +1\right)  ^{2}$. The Hubble function is
\begin{equation}
 H(a)=H_0\left( \frac{a_1-a}{a(a_1-1)} \right)^2
\end{equation}
and finally, the normalized Friedman equation can be written in the following form
\begin{equation}
\frac{H^{2}\left(  a\right)  }{H_{0}^{2}}=\Omega_{r}a^{-4}+\Omega_{m}%
a^{-3}+\Omega_{k}a^{-2}+\Omega_{f}a^{-1}+\Omega_{\Lambda} , \label{Hg.01}%
\end{equation}
where
\begin{eqnarray}
\Omega_{f}=\frac{\left\vert 4a_{1}\right\vert }{\left(  \left\vert
a_{1}\right\vert +1\right)  ^{4}}, && \qquad \Omega_{\Lambda}=\frac{1}{\left(
\left\vert a_{1}\right\vert +1\right)  ^{4}} , \label{Hg.02a}%
    \\
\Omega_{r}=\frac{\left\vert a_{1}\right\vert ^{4}}{\left(  \left\vert
a_{1}\right\vert +1\right)  ^{4}}, && \qquad \Omega_{m}=\frac{\left\vert 4a_{1}%
\right\vert ^{3}}{\left(  \left\vert a_{1}\right\vert +1\right)  ^{4}%
} ,  
    \\
 \Omega_{k}&=&\frac{\left\vert 6a_{1}\right\vert ^{2}}{\left(  \left\vert
a_{1}\right\vert +1\right)  ^{4}}\,.\label{Hg.02}%
\end{eqnarray}

\begin{figure}[ptb]
\includegraphics[height=8.25cm]{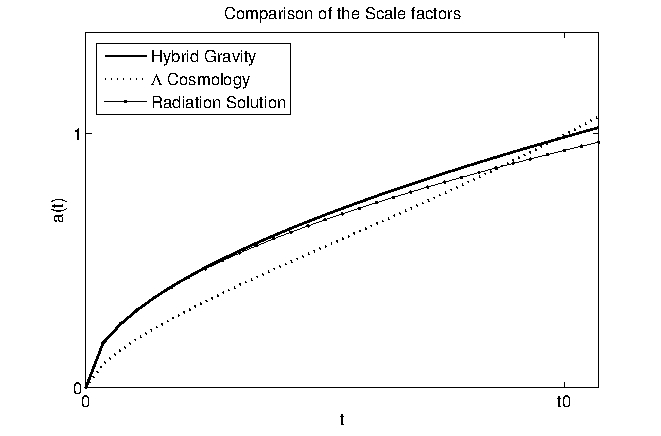}
\caption{Comparison of the hybrid gravity scale factor with that of 
$\Lambda$CDM-cosmology~$a_{\Lambda}\left(  t\right)  $ and the radiation
solution $a_{r}\left(  t\right)  =a_{0r}\sqrt{t}$ where $t_{0}$ is the present
time, $a_{\Lambda}\left(  t_{0}\right)  =1$. For the hybrid gravity solution, we set $\left \vert a_{1}\right\vert >1$. }%
\label{fig1}%
\end{figure}

From the above analysis arises a conclusion that each power of $\sqrt{\phi}$ appearing in the potential
(\ref{FRW.07}), that is $V\left(  \phi\right)  =V_{0}\left(  \sqrt{\phi}+V_{1}\right)^{4},$ introduces into the Friedmann equation
a power term of the scale factor which has a cosmological meaning. Each term represents fluid filling the Universe and they are: radiation, 
dust, curvature-like fluid,  a dark energy fluid
with equation of state $p_{f}=-\frac{2}{3}\rho_{f}$ and a cosmological constant, respectively. Let us again recall that we have been 
considering empty spacetime, that is, vacuum case of the model. Moreover, we have assumed that 
the spacetime is spatially flat: the curvature term which appeared in the Friedmann equation comes from hybrid gravity. We would also like to notice that
for large redshift $1+z=a^{-1}$ the Friedmann equation (\ref{Hg.01}) behaves like the radiation solution which is presented in the picture \ref{fig1}:
the hybrid gravity coincides with radiation solution in the early Universe. There 
are also drawn the scale factor of the standard $\Lambda$CDM cosmology and the radiation one.

\subsection{Wheeler-DeWitt equation of hybrid gravity model}\label{WDW_hg}
Roughly speaking, WDW equation is a quantized version of a Hamiltonian of a
considered system. In the case of hybrid gravity applied to FRLW cosmology we deal with the $2$-dimensional minisuperspace 
described by the minisuperspace metric $G_{ij}=\text{diag}(1,1)$, the WDW it has the following form 
\begin{equation}
\Box\Psi -a^{3}V(\phi )\Psi =0,
\end{equation}%
where $\Box =\frac{1}{\sqrt{|G|}}\frac{\partial }{\partial x^{i}}\left(
\sqrt{|G|}G^{ij}\frac{\partial }{\partial x^{i}}\right) $ is the Laplace
operator, $\Psi $ is a wave function of the Universe and $x^{i}=\{a,\phi \}$. Let us notice that the WDW equation is a Klein-Gordon equation with the d'Alembertian
defined by the minisuperspace metric $G_{ij}$.%

From the Hamiltonian (\ref{An.04}), following canonical quantisation procedure \cite{dirac1996general, Kiefer2007ria, matschull1996dirac}, one may
determine the WDW equation (recall that
the dimension of the minisuperspace is two and the minisuperspace is flat), which has the form
\begin{equation}
\label{WDW.01}\Psi_{,xx}+\Psi_{,yy}-2V_{0}y^{4}\Psi=0\,,
\end{equation}
where $\Psi$ is the Wave Function of the Universe \cite{halliwell}.
We would like to solve the above equation and again we can use symmetries to do that. We could use Lie symmetries 
method \cite{StephaniB,hydon} in order to find a generic symmetry vector and bring it to play to lower the number of independent variables.
In our case the linearized symmetry condition by which we look for symmetries is lengthy but fortunately one may follow the theorem \ref{KG_theo} formulated in
 \cite{AnIJGMP} which combine conformal algebras of minisuperspace with symmetries of Klein-Gordon equation. Unfortunately,
the two-dimensional Riemannian space has an infinite conformal algebra but we need at least one conformal Killing vector satisfying the condition 
(\ref{pot_theo}) to solve the WDW equation (\ref{WDW.01}). 

It is easy to check that the vectors $X_1=\partial_x$ and $X_2=\partial_y$ are Killing vectors of the minisuperspace metric $G_{ij}$, it 
means they satisfy the condition
\begin{equation}
 X_{i;j}+X_{j;i}=0.
\end{equation}
There exists also the homothetic vector $X_3=x\partial_x+y\partial_y$ with the conformal factor $\psi=1$:
\begin{equation}
  X_{i;j}+X_{j;i}=2\psi G_{ij}
\end{equation}
but it can be checked that only the vector $X_1$ satisfies 
the condition (\ref{pot_theo}). That means that the generic Lie symmetry vector is
\begin{equation}
 X=b_1\partial_x+(b_2\Psi+b(x,y))\partial_\Psi,\label{sym}
\end{equation}
where $b(x,y)$ is a function that satisfies WDW equation (\ref{WDW.01}). Let us now reduce the number of variables of the equation by the zeroth 
order invariants $\{Y,Z\}$. One gets that
\begin{equation}\label{sym_hg}
 \frac{dx}{b_1}=\frac{d\Psi}{b_2\Psi},\;Z=y
\end{equation}
so the invariant functions coming from (\ref{sym_hg}) are $\{\Psi=Ye^{\mu x},y\}$, where $\mu\in \mathbb{C}$ \cite{StephaniB}. Hence the considered WDW equation reduces to
the second order ordinary differential equation:
\begin{equation}
Y_{,yy}+\left(  \mu^{2}-2\bar{V}_{0}y^{4}\right)  Y=0.
\end{equation}
One recognizes the one-dimensional time-dependent oscillator. The solutions due to Lie point symmetries of such a system were considered in \cite{LeachOSc, Abraham, TypeII}. The
solution of the above equation is
\[
Y\left(  y\right)  =y_{1}e^{w\left(  y\right)  }+y_{2}e^{-w\left(  y\right)  },
\]
where 
\begin{equation}
w\left(  y\right)  =\frac{\sqrt{2}}{2}\int\sqrt{\left(  2\bar{V}%
_{0}y^{4}-\mu^{2}\right)  }dy.
\end{equation}
We are able finally to write the invariant solution of the WDW equation (\ref{WDW.01}) as
\begin{equation}
\Psi\left(  x,y\right)  =\sum_{\mu}\left[  y_{1}e^{\mu x+w\left(  y\right)
}+y_{2}e^{\mu x-w\left(  y\right)  }\right]\,.
\end{equation}

\section{Remarks}

Let us briefly conclude our considerations on hybrid gravity model applied to FRLW cosmology. We have used Noether and Lie point symmetries approaches which 
allowed us to solve classical field equations arising from Lagrangian in the first case and to find a Wave Function of the Universe which is a solution of 
the Wheeler-DeWitt equation being a quantized version of a cosmological Hamiltonian. The analysis was performed in the scalar-field representation of the hybrid 
gravity model what resulted into a system with two independent variables: the scale factor of the Universe $a(t)$ and the scalar field $\phi$. The field 
equations as well as WDW equations required a choice of the potential $V(\phi)$ in order to be solved, that is, coming back to the original representation of the theory, one needs to 
specify the gravitational Lagrangian $f(\mathcal{R})$. We examined two cases: first one gave as the solution with cosmological constant and radiation fluid for 
the constant potential $V_0$ (from the Noether symmetries procedure) while considering the conformal frame resulted in the much richer model: for the power-law 
potential we obtained the Universe filled with five fluids: radiation, dust, curvature-like fluid, dark energy one with EoS $p_f=-\frac{2}{3}\rho_f$ and 
cosmological constant. It is important to underline that we have considered spatially flat metrics ($k=0$) and vacuum equations in both models, that is, the right-hand side of the modified 
Einstein's equations is zero.

Following DeWitt \cite{halliwell}, the solution of the WDW equation is called a Wave Function of the Universe which is related to a probability that an
observed universe emerges with some initial conditions which might be specified by for example, "no boundary condition" \cite{hartle1983wave} 
or "tunneling from nothing" \cite{vilenkin1982creation, vilenkin1984quantum}. The oscillatory or exponential behavior depends on the signs of the variables as
$x,y$ are functions of $\phi$ which can be positive or negative defined. The positivity of the scalar field can be interpreted as quintessence while negative 
scalar field corresponds to phantom field. For the case $\phi=0$ one recovers GR, that is, the function $f(\mathcal{R})$ turns out to be a cosmological constant. 
It should be also mentioned that the only power-law Lagrangian of hybrid gravity that admits Noether symmetries has a form $f(\mathcal{R})\sim\mathcal{R}^2$
while in $f(R)$ gravity in metric approach one deals with $f(R)=R^n$, and similarly in pure Palatini one the power-law function is $f(\mathcal{R})=\mathcal{R}^n$.

Hybrid gravity, according to up-to-now studies, seems to be a theory worth of further considerations, especially for astrophysical objects such as neutron stars 
or black holes because no such examination has been performed so far. It is capable to recover the various cosmological epochs 
as shown in \cite{capozziello2013hybrid} and because of the scalar-tensor representation, it passes solar system 
tests\cite{de2010f, will1981theory, borka2015probing}.




\chapter{Other theories of gravity} \label{chap3}
Besides the cosmological constant $\Lambda$ introduced in order to explain late time accelerating expansion, a minimally coupled scalar 
field is one of the simplest modification of the Einstein's field equations. Similarly as the
cosmological constant, it can be treated both as the geometric modification as well as exotic fluid inserted on the right hand side of the equations. 
There are much more interesting but also more difficult to handle models which concern a non-minimally coupled scalar field (for example Brans-Dicke theory \cite{brans1961mach}) since their field
equations have similar forms as $4+1$ decomposition 
of Kaluza-Klein field equations \cite{kaluza1921unitatsproblem, klein1926quantentheorie, klein1926atomicity, williams2015field}: the $5$-dimensional
Kaluza-Klein theory unifies gravitation and electromagnetism \cite{faraoni2004cosmology}. So far, we have investigated two non-minimally coupled scalar field models:
Palatini $f(\mathcal{R})$ can be considered as scalar-tensor theory (although we do not treat it in this way) as well as Hybrid Gravity which we studied in 
scalar-tensor representation. Now on, we are going to examine minimally coupled scalar field in both cosmology and astrophysics.

In the presented chapter the first part will concern
cosmological considerations. We will focus on anisotropic models in the framework of scalar-tensor theory of gravity. Mainly we will focus on Bianchi $I$ and Bianchi
$II$ models. Later on, we will examine configurations of relativistic 
stars in the first part. We will briefly recall main results coming from General Relativity and then turn to Extended Theories of Gravity. Since the stability 
criterion must be investigated case by case, we will present, as the simplest example of ETGs, a minimally coupled scalar field.




\section{Bianchi cosmology in scalar - tensor theory of gravity}
We have already mentioned that anisotropic models of Universe can be also very important since we do not really know if our Universe is isotropic.
The simplest generalization providing anisotropy is the assumption that instead of one scale factor of the Universe one deals with three, each one for one spatial direction.
Such a model is so-called Bianchi $I$. There are more of Bianchi spacetimes and we will briefly introduce them. After that, we will focus on Lie symmetries in Bianchi 
scalar-field cosmology. Similarly as it was done for the Hybrid Gravity, we will look for invariant and exact solutions of Wheeler-DeWitt equations.  
To find classical solutions, we will use WKB approximation. More detailed discussion can be find in \cite{wojnar2}. 

Despite the fact that on the large scale the observed Universe is homogeneous and isotropic there are visible anisotropies in the cosmic microwave background.
The existence of the anisotropies means that the Universe does not expanse in the same way in all directions as standard cosmological model assumes. If our 
Universe is considered on the large scale, the simple model described by FRLW metric is sufficient and agrees with astronomical observations (so-called LCDM standard 
cosmological model). As the anisotropies do not increase and are very small one supposes that anisotropic models isotropize as time approaches our epoch 
\cite{capozziello1997nother,barrow1995universe}. This makes it important to study models which are not isotropic at early times and therefore the dynamics 
of anisotropies should be understood.
There are also considerations \cite{capozziello1997nother, rothman1986can} that anisotropies before the inflation could be a reason for the coupling 
between the gravitational field and the inflaton 
field (scalar field minimally or non-minimally coupled to gravity). A lot of attention has been given to a scalar field in inflationary 
models \cite{demianski1992scalar, linde1982new} but also because of a
possibility that it 
could explain dark matter problem and the damping of cosmological constant \cite{ford1987cosmological}. Unfortunately, the presence of scalar fields in 
cosmology arises to another problem which is an unknown form of their potentials.

Anisotropic but homogeneous models of universes are described by Bianchi models. Bianchi spacetime manifolds are foliated along the time axis 
with 3-dimensional homogeneous hypersurfaces admitting a group of motion $G3$. There are nine possible groups \cite{ryan2015homogeneous, tsamparlis2011geometric}
which gives nine possibles models
which can be taken under consideration. All physical variables appearing in the models depend on time only which reduce the Einstein and other governing 
equations to ordinary differential equations. The line element of the Bianchi models in $3+1$ decomposition has a following
form \cite{ryan2015homogeneous,misner1969quantum}
\begin{equation}
ds^{2}=-\frac{1}{N(t)^{2}}dt^{2}+\bar{g}_{ij}(t)\omega^{i}\otimes\omega^{j},
\label{B2}%
\end{equation}
where $N(t)$ is the lapse function and $\{\omega^{i}\}$ denotes the canonical basis of $1$-forms satisfying the Lie algebra
\begin{equation}
d\omega^{i}=C_{jk}^{i}\omega^{j}\wedge\omega^{k} \label{B2.2}%
\end{equation}
where $C_{jk}^{i}$ are the structure constants of the algebra. The spatial metric $\bar{g}_{ij}$ is diagonal and can be factorized as follows
\begin{equation}
g_{ij}(t)=e^{2\lambda(t)}e^{-2\beta_{ij}(t)} \label{B2.3}%
\end{equation}
where $e^{\lambda\left(  t\right)  }$ is the scale factor of the Universe and
the matrix $\beta_{ij}$ is diagonal and traceless. The matrix $\beta_{ij}$
depends on two independent quantities $\beta_{1},$ $\beta_{2}~$\ which are
called the anisotropy parameters \cite{misner1969quantum}
\begin{equation}
\beta_{ij}=diag\left(  \beta_{1},-\frac{1}{2}\beta_{1}+\frac{\sqrt{3}}{2}%
\beta_{2},-\frac{1}{2}\beta_{1}-\frac{\sqrt{3}}{2}\beta_{2}\right)  
\label{B2.4}%
\end{equation}
and, in these variables, it is $\sqrt{\bar{g}}=e^{3\lambda}$. We will consider only a subclass of the Bianchi models, so-called class A, since there exist Lagrangians 
of field equations for them (for details, see for example \cite{ryan2015homogeneous, tsamparlis2011geometric, wojnar2}). Let us additionally mention that 
for the line element (\ref{B2}) together with the definitions (\ref{B2.3}) and (\ref{B2.4}), the Ricci scalar of the Bianchi class A spacetimes is
\begin{equation}
R=R_{\left(  4\right)  }+R^{\ast} \label{B4a}%
\end{equation}
where (the dot denotes the differentiation with respect to the time $t$) 
\begin{equation}
R_{\left(  4\right)  }=\frac{3}{2}N\left(  4N\ddot{\lambda}+4\dot{N}%
\dot{\lambda}+8N\dot{\lambda}^{2}+N\dot{\beta}_{1}^{2}+N\dot{\beta}_{2}%
^{2}\right)  \label{B4b}%
\end{equation}
and \ $R^{\ast}=R^{\ast}\left(  \lambda,\beta_{1},\beta_{2}\right)  $ is the
component of the three dimensional hypersurface. The exact forms of $R^{\ast}$ for some of the Bianchi models are given in the table \ref{Bianchi3dR}.
\begin{table}[tbp] \label{Bianchi3dR}\centering
\caption{The Ricci scalar of the $3$d hypersurfaces of the class A Bianchi
spacetimes.}%
\begin{tabular}
[c]{cc}\hline\hline
\textbf{Model} & $R^{\ast}\left(  \lambda,\beta_{1},\beta_{2}\right)
$\\\hline
Bianchi I & $0$\\
Bianchi$~$II & $-\frac{1}{2}e^{\left(  4\beta_{1}-2\lambda\right)  }$\\
Bianchi VI$_{0}$/VII$_{0}$ & $-\frac{1}{2}e^{-2\lambda}\left(  e^{4\beta_{1}%
}+e^{-2\left(  \beta_{1}-\sqrt{3}\beta_{2}\right)  }\pm2e^{\beta_{1}+\sqrt
{3}\beta_{2}}\right)  $\\
Bianchi VIII & $-\frac{1}{2}e^{-2\lambda}\left(
\begin{array}
[c]{c}%
e^{4\beta_{1}}+e^{-2\beta_{1}}\left(  e^{\sqrt{3}\beta_{2}}+e^{-\sqrt{3}%
\beta_{2}}\right)  ^{2}+\\
-2e^{-\beta_{1}}\left(  e^{\sqrt{3}\beta_{2}}-e^{-\sqrt{3}\beta_{2}}\right)
^{2}%
\end{array}
\right)  $\\
Bianchi IX & $-\frac{1}{2}e^{-2\lambda}\left(
\begin{array}
[c]{c}%
e^{4\beta_{1}}+e^{-2\beta_{1}}\left(  e^{\sqrt{3}\beta_{2}}-e^{-\sqrt{3}%
\beta_{2}}\right)  ^{2}+\\
-2e^{-\beta_{1}}\left(  e^{\sqrt{3}\beta_{2}}+e^{-\sqrt{3}\beta_{2}}\right)
^{2}%
\end{array}
\right)  +1$\\\hline\hline
\end{tabular}
\label{Bianchi3dR}%
\end{table}%

Now on we are ready to study Bianchi models in scalar-tensor cosmology with minimally coupled scalar field. The action has a well-known form
\begin{equation}
S=\int dx^{4}\sqrt{-g}\left(  R-\frac{1}{2}g^{\mu\nu}\phi_{,\mu}\phi_{\nu
}+V\left(  \phi\right)  \right)  \label{B15}%
\end{equation}
while the Lagrangian $L\equiv L\left(  N,\lambda,\beta_{1},\beta_{2},\phi,\dot{\lambda},\dot{\beta}_{1}%
,\dot{\beta}_{2},\dot{\phi}\right)$ is obtained from (\ref{B2}), (\ref{B4a}), and (\ref{B4b}) \cite{demianski1992scalar}
\begin{equation}
L =N(t)e^{3\lambda}\left(  6\dot{\lambda
}^{2}-\frac{3}{2}\left(  \dot{\beta}_{1}^{2}+\dot{\beta}_{2}^{2}\right)
-\frac{1}{2}\dot{\phi}^{2}\right)  +\frac{e^{3\lambda}}{N(t)}\left(  V\left(
\phi\right)  +R^{\ast}\right). \label{B16}%
\end{equation}
The field equations with respect to the variables $\lambda,\,\beta_1,\,\beta_2,\,\phi$ are
\begin{equation}
4\ddot{\lambda}+\left(  6\dot{\lambda}^{2}+\frac{3}{2}(\dot{\beta_{1}}%
^{2}+\dot{\beta_{2}}^{2})+\frac{1}{2}\dot{\phi}^{2}\right)  +\frac{\dot{N}}%
{N}\dot{\lambda}-\frac{1}{N^{2}}\left(  V+R^{\ast}+\frac{1}{3}\frac{\partial
R^{\ast}}{\partial\lambda}\right)  =0, \label{B17}%
\end{equation}
\begin{equation}
\ddot{\beta}_{\left(  1,2\right)  }+\frac{\dot{N}}{N}\dot{\beta}_{\left(
1,2\right)  }+3\dot{\lambda}\dot{\beta}_{\left(  1,2\right)  }+\frac{1}%
{3N^{2}}R_{,\left(  1,2\right)  }^{\ast}=0, \label{B9}%
\end{equation}
\begin{equation}
\ddot{\phi}+3\dot{\lambda}\dot{\phi}+\frac{\dot{N}}{N}\dot{\phi}+\frac
{1}{N^{2}}\frac{\partial V}{\partial\phi}=0, \label{B19}%
\end{equation}
whilst the $00$ modified Einstein's equation is
\begin{equation}
Ne^{3\lambda}\left(  6\dot{\lambda}^{2}-\frac{3}{2}\left(  \dot{\beta}_{1}%
^{2}+\dot{\beta}_{2}^{2}\right)  -\frac{1}{2}\dot{\phi}^{2}\right)
-\frac{e^{3\lambda}}{N}\left(  V+R^{\ast}\right)  =0. \label{B18}%
\end{equation}
Under coordinate transformations
\begin{equation}
 \left(
\lambda,\beta_{1},\beta_{2}\right)  =\left(  \frac{\sqrt{3}}{6}x,\frac
{\sqrt{3}}{3}y,\frac{\sqrt{3}}{3}z\right)\;\;\text{and}\;\;N\left(  t\right)  =\bar
{N}\left(  t\right)  e^{-3\lambda},
\end{equation}
the equation (\ref{B18}) becomes%
\begin{equation}
\frac{1}{2}\bar{N}\left(  \dot{x}^{2}-\dot{y}^{2}-\dot{z}^{2}-\dot{\phi}%
^{2}\right)  -\frac{1}{\bar{N}}e^{\sqrt{3}x}\left(  V\left(  \phi\right).
+R^{\ast}\right)  =0 \label{B20}%
\end{equation}
From the kinetic part of the Lagrangian (\ref{B16}) we notice that one deals with a flat $4$-dimensional minisuperspace. Above equation, after the quantization 
procedure (see the sections (\ref{sec1symetr}), (\ref{WDW_hg}) as well as \cite{Kiefer2007ria}) can be transformed into WDW equation 
\begin{equation}
\Psi_{,xx}-\Psi_{,yy}-\Psi_{,zz}-\Psi_{,\phi\phi}-2e^{\sqrt{3}x}\left(
V\left(  \phi\right)  +R^{\ast}\right)  \Psi=0 \label{B22}%
\end{equation}
which has a form of Klein-Gordon equation in the $4$-dimensional flat space $M^4$. As we want to apply the procedure of \cite{AnIJGMP, TsamAnd}, which was briefly described
in the section (\ref{useful_theo}), we will need the conformal algebra of the $M^4$ spacetime. Its algebra is $15$-dimensional: one may show that 
$C(D)\simeq O(D+2)$, where the dimension of the orthogonal algebra $ O(D+2)$ is $\frac{(D+1)(D+2)}{2}$
\cite{barut1986theory}. The 
considered spacetime admits ten Killing vectors:
\begin{align}
 K_{(x)}=\partial_x,\;\;K_{(y)}=\partial_y\;\;K_{(z)}=\partial_z\;\;K_{(\phi)}=\partial_\phi,\\
 R_{\left(  xy\right)  }=y\partial_{x}+x\partial_{y},\;\;~R_{\left(  xz\right)  }=z\partial_{x}+x\partial_{z},\\
 R_{\left(  yz\right)}=z\partial_{y}-y\partial_{z},\;\;~R_{\left(  x\phi\right)  }=\phi\partial_{x}+x\partial_{\phi}\mathbf{,}\\
R_{\left(  y\phi\right)  }=\phi\partial_{y}-y\partial_{\phi}~,~R_{\left(
z\phi\right)  }=\phi\partial_{z}-z\partial_{\phi},
\end{align}
one gradient homothetic Killing vector 
\begin{equation}
 H=x\partial_{x}+y\partial_{y}+z\partial_{z}+\phi\partial_{\phi},
\end{equation}
and four special conformal Killing vectors
\begin{align*}
C_{\left(  x\right)  }  &  =\frac{1}{2}\left(  x^{2}+y^{2}+z^{2}\right)
\partial_{x}+xy\partial_{y}+xz\partial_{z}+\frac{1}{2}\phi
^{2}\partial_{x}+\phi x\partial_{\phi},\\
C_{\left(  y\right)  }  &  =xy\partial_{x}+\frac{1}{2}\left(  x^{2}%
+y^{2}-z^{2}\right)  \partial_{y}+zy\partial_{z}-\frac{1}{2}\phi
^{2}\partial_{y}+\phi y\partial_{\phi},\\
C_{\left(  z\right)  }  &  =xz\partial_{x}+yz\partial_{y}+\frac{1}{2}\left(
x^{2}+z^{2}-y^{2}\right)  \partial_{z}-\frac{1}{2}\phi
^{2}\partial_{z}+\phi z\partial_{\phi},\\
C_{\left(  \phi\right)  }&=x\phi\partial_{x}+y\phi\partial_{y}%
+z\phi\partial_{z}+\frac{1}{2}\left(  x^{2}+\phi^{2}-y^{2}-z^{2}\right)
\partial_{\phi},
\end{align*}%
for which the conformal factors are $\psi_{\left(  x\right)  }=x,\;\psi_{\left(  y\right)  }=y$,$\psi_{\left(  z\right)  }=z,\;\psi_{\left(  \phi\right)  }=\phi$,
respectively.

Now on, we are ready to look for Lie symmetries of the WDW equation (\ref{B22}). Using the theorem \ref{KG_theo} from the Appendix \ref{app_lie} we have found that 
the WDW equation under consideration admits Lie symmetries not only for special forms of the potential $V(\phi)$ but also for arbitrary one. The special forms of the 
potential are $V(\phi)=0$ for which the scalar field $\phi$ behaves like stiff matter, $V(\phi)=V_0$ with $V_0\neq0$ and exponencial one $V(\phi)=V_0 e^{\mu\phi}$. 
The mentioned results are presented in the tables \ref{BianchiSFV0} and \ref{BianchiSFVphi}.%

\begin{table}[tbp] \centering
\caption{Lie symmetries of the WDW equation of the Class A Bianchi models in
scalar field cosmology for  $V(\phi)=0$.}%
\begin{tabular}
[c]{ccc}\hline\hline
\textbf{Model~}$V\left(  \phi\right)  =0$ & $\#~$ & \textbf{Lie Symmetries}%
\\\hline
Bianchi I & $16$ & $\Psi\partial_{\Psi},~K_{\left(  x\right)  },~K_{\left(
y\right)  },~K_{\left(  z\right)  },~K_{\left(  \phi\right)  },~R_{\left(
xy\right)  },~R_{\left(  xz\right)  }$\\
&  & $~R_{\left(  yz\right)  },~R_{\left(  x\phi\right)  },~R_{\left(
y\phi\right)  },~R_{\left(  z\phi\right)  },~\bar{H},~\left(  \bar{C}_{\left(
x\right)  }-x\Psi\partial_{\Psi}\right)  ,~$\\
&  & $\left(  \bar{C}_{\left(  y\right)  }-y\Psi\partial_{\Psi}\right)
,~\left(  \bar{C}_{\left(  z\right)  }-z\Psi\partial_{\Psi}\right)  ,~\left(
\bar{C}_{\left(  \phi\right)  }-\phi\Psi\partial_{\Psi}\right)  $\\
Bianchi$~$II & $7$ & $\Psi\partial_{\Psi},~K_{\left(  z\right)  },~K_{\left(
\phi\right)  },~K_{\left(  x\right)  }-\frac{1}{2}K_{\left(  y\right)
},~R_{\left(  z\phi\right)  }$\\
&  & $R_{\left(  xz\right)  }-\frac{1}{2}R_{\left(  yz\right)  },~R_{\left(
x\phi\right)  }-\frac{1}{2}R_{\left(  y\phi\right)  }$\\
Bianchi VI$_{0}$/VII$_{0}$ & $3$ & $\Psi\partial_{\Psi},~K_{\left(
\phi\right)  },~K_{\left(  x\right)  }+\frac{1}{4}K_{\left(  y\right)  }%
+\frac{\sqrt{3}}{4}K_{\left(  z\right)  }$\\
&  & $R_{\left(  x\phi\right)  }+\frac{1}{4}R_{\left(  y\phi\right)  }%
+\frac{\sqrt{3}}{4}R_{\left(  z\phi\right)  }$\\
Bianchi VIII/IX & $2$ & $\Psi\partial_{\Psi},~K_{\left(  \phi\right)  }%
$\\\hline\hline
\end{tabular}
\label{BianchiSFV0}%
\end{table}%

\begin{table}[tbp] \centering
\caption{Lie symmetries of the WDW equation of the Class A Bianchi models in
scalar field cosmology for non-zero potentials.}%
\begin{tabular}
[c]{ccc}\hline\hline
\textbf{Model}~$V\left(  \phi\right)  =V_{0}$ & $\#~$ & \textbf{Lie
Symmetries}\\\hline
Bianchi I & $7$ & $\Psi\partial_{\Psi},~K_{\left(  y\right)  },~K_{\left(
z\right)  },~K_{\left(  \phi\right)  },~R_{\left(  yz\right)  },~R_{\left(
y\phi\right)  },~R_{\left(  z\phi\right)  }~$\\
Bianchi$~$II & $4$ & $\Psi\partial_{\Psi},~K_{\left(  z\right)  },~K_{\left(
\phi\right)  },~R_{\left(  z\phi\right)  }~$\\
Bianchi VI$_{0}$/VII$_{0}$ & $2$ & $\Psi\partial_{\Psi},~K_{\left(
\phi\right)  }$\\
Bianchi VIII/IX & $2$ & $\Psi\partial_{\Psi},~K_{\left(  \phi\right)  }$\\
&  & \\
\textbf{Model}~$V\left(  \phi\right)  =V_{0}e^{\mu\phi}$ & $\#~$ & \textbf{Lie
Symmetries}\\
Bianchi I & $7$ & $\Psi\partial_{\Psi},K_{\left(  y\right)  },~K_{\left(
z\right)  },~R_{\left(  yz\right)  },~\frac{\sqrt{3}}{3}\mu K_{(x)}-K_{(\phi
)},~~$\\
&  & $R_{\left(  y\phi\right)  }+\frac{\sqrt{3}}{3}\mu R_{\left(  xy\right)
},~R_{\left(  z\phi\right)  }+\frac{\sqrt{3}}{3}\mu R_{\left(  xz\right)  }$\\
Bianchi$~$II & $4$ & $\Psi\partial_{\Psi},~K_{\left(  z\right)  },~K_{\left(
x\right)  }-\frac{1}{2}K_{\left(  y\right)  }-\frac{\sqrt{3}}{\mu}K_{\left(
\phi\right)  }$\\
&  & $R_{\left(  z\phi\right)  }+\frac{\sqrt{3}}{3}\mu\left(  R_{\left(
xz\right)  }-\frac{1}{2}R_{\left(  yz\right)  }\right)  $\\
Bianchi VI$_{0}$/VII$_{0}$ & $2$ & $\Psi\partial_{\Psi},~K_{\left(  x\right)
}-\frac{1}{2}K_{\left(  y\right)  }-\frac{\sqrt{3}}{2}K_{\left(  z\right)
}-\frac{\sqrt{3}}{\mu}K_{\left(  \phi\right)  }$\\
Bianchi VIII/IX & $1$ & $\Psi\partial_{\Psi}$\\
&  & \\
\textbf{Model}~$V\left(  \phi\right)  =V\left(  \phi\right)  $ & $\#~$ &
\textbf{Lie Symmetries}\\
Bianchi I & $4$ & $\Psi\partial_{\Psi},~K_{\left(  y\right)  },~K_{\left(
z\right)  },~R_{\left(  yz\right)  }~$\\
Bianchi$~$II & $2$ & $\Psi\partial_{\Psi},~K_{\left(  z\right)  }~$\\
Bianchi VI$_{0}$/VII$_{0}$ & $1$ & $\Psi\partial_{\Psi}$\\
Bianchi VIII/IX & $1$ & $\Psi\partial_{\Psi}$\\\hline\hline
\end{tabular}
\label{BianchiSFVphi}%
\end{table}%

\subsection{Invariant solutions of WDW equation and WKB approximation}
As the first step let us examine Bianchi I spacetime. Here and further, we will assume that $\bar{N}\left(  t\right)  =1$ which allows us to write the 
equation (\ref{B22}) in the form
\begin{equation}
\Psi_{,xx}-\Psi_{,yy}-\Psi_{,zz}-\Psi_{,\phi\phi}-2e^{\sqrt{3}x}V\left(
\phi\right)  \Psi=0\,. \label{BSF.01}%
\end{equation}
The case of zero potential gives rise to $(1+3)$ wave equation in $E^3$ which was considered in \cite{Abraham}. The field equations with the constant potential 
 turn out to have a form of the ones coming from General Relativity with stiff matter and cosmological constant. Applying to the equation (\ref{BSF.01}) 
 the zeroth-order invariants of the Lie symmetries 
 \begin{equation}
\bar{X}_{\left(  i\right)  }=K_{\left(  i\right)  }+\mu_{\left(  i\right)
}\Psi\partial_{\Psi}~,~i=y,z,\phi\label{BSF.02}%
\end{equation}
which form a closed Lie algebra, the WDW equation may be reduced to the linear second-order ordinary differential equation
\begin{equation}
\Phi^{\prime\prime}-\left(  \mu_{\left(  y\right)  }+\mu_{\left(  z\right)
}+\mu_{\left(  \phi\right)  }+2V_{0}e^{\sqrt{3}x}\right)  \Phi=0\,. \label{BSF.03}%
\end{equation}
The Wave Function of the Universe is now $$\Psi=\Phi\left(  x\right)  \exp\left(  \mu_{\left(  y\right)  }%
y+\mu_{\left(  z\right)  }z+\mu_{\left(  \phi\right)  }\phi\right).$$ The prime in (\ref{BSF.03}) denotes the differentiation with respect to the variable $x$. The 
solution of (\ref{BSF.03}) exists 
\begin{equation}
\Phi\left(  x\right)  =\Phi_{1}J_{c}\left(  i\frac{2\sqrt{6V_{0}}}{3}%
e^{\frac{\sqrt{3}}{2}x}\right)  +\Phi_{2}Y_{c}\left(  i\frac{2\sqrt{6V_{0}}%
}{3}e^{\frac{\sqrt{3}}{2}x}\right)  \label{BSF.04}%
\end{equation}
where $J_{c},Y_{c}$ are the Bessel functions of the first and second kind while
the constant $c=\frac{2\sqrt{3}}{3}\left(  \sqrt{\mu_{\left(  y\right)
}^{2}+\mu_{\left(  z\right)  }^{2}+\mu_{\left(  \phi\right)  }^{2}}\right)  $.

When one deals with the exponential potential in (\ref{BSF.01}), it is convenient to apply the Lie invariants
\begin{equation}
\bar{X}_{\left(  y\right)  },\bar{X}_{\left(  z\right)  },~\frac{\sqrt{3}}%
{3}\mu K_{(x)}-K_{(\phi)}+\mathbf{\nu}\Psi\partial_{\Psi}.%
\end{equation}
The procedure gave us the WDW equation (\ref{BSF.01}) reduced to
\begin{equation}
\left(  3-\mu^{2}\right)  \Phi^{\prime\prime}\left(  w\right)  +6\nu
\Phi^{\prime}-\left(  \left(  \mu_{\left(  y\right)  }^{2}+\mu_{\left(
z\right)  }^{2}\right)  \mu^{2}-3\mathbf{\nu}^{2}+2V_{0}\mu^{2}e^{\mu
w}\right)  \Phi=0 \label{BSF.05}%
\end{equation}
where $\Phi^{\prime}=\frac{d\Phi\left(  w\right)  }{dw}$ and $w=\frac{\sqrt{3}}{\mu}x+\phi$. The wave function became
$$\Psi\left(  x,y,z,\phi\right)  =\Phi\left(  w\right)  \exp\left(
\frac{\sqrt{3}\mathbf{\nu}}{\mu}x+\mu_{\left(  y\right)  }y+\mu_{\left(
z\right)  }z\right).  $$
The solution of (\ref{BSF.05}) depends on the value of the  constant $\mu$. For the $\mu\neq\sqrt{3}$ one gets
\begin{equation}
\Phi\left(  w\right)  =\exp\left(  \frac{3\mu w}{\mu^{2}-3}\right)  \left[
\Phi_{1}J_{\bar{c}}\left(  2\sqrt{\frac{2V_{0}}{\mu^{2}-3}}e^{\frac{\mu}{2}%
w}\right)  +\Phi_{2}Y_{\bar{c}}\left(  2\sqrt{\frac{2V_{0}}{\mu^{2}-3}%
}e^{\frac{\mu}{2}w}\right)  \right],
\end{equation}
where $\bar{c}=\frac{2}{\left\vert \mu^{2}-3\right\vert }\sqrt{3\mathbf{\nu
}^{2}-\left(  \mu^{2}-3\right)  \left(  \mu_{\left(  y\right)  }^{2}%
+\mu_{\left(  z\right)  }^{2}\right)  }$  while for the constants $\left\vert \mu\right\vert =\sqrt{3},\;\;\nu\neq0$ we have
\begin{equation}
\Phi\left(  w\right)  =\Phi_{0}\exp\left[  \frac{1}{2\mathbf{\nu}}\left(
\mu_{\left(  y\right)  }^{2}+\mu_{\left(  z\right)  }^{2}\right)
-\frac{\mathbf{\nu}}{2}w+\frac{\sqrt{3}}{3}\frac{V_{0}}{\mathbf{\nu}}%
e^{\sqrt{3}w}\right]. 
\end{equation}

Let us focus on the classical solutions of the considered Bianchi I models. We will consider WKB approximation of the equation (\ref{BSF.01}) as it was 
performed in the section \ref{hamil_class} (recall that Hamiltonian constraint is equal to zero). We simply get the 
Hamilton-Jacobi equation of the form
\begin{equation}
\frac{1}{2}\left[  \left(  \frac{\partial S}{\partial x}\right)  ^{2}-\left(
\frac{\partial S}{\partial y}\right)  ^{2}-\left(  \frac{\partial S}{\partial
z}\right)  ^{2}-\left(  \frac{\partial S}{\partial\phi}\right)  ^{2}\right]
-e^{\sqrt{3}x}V\left(  \phi\right)  =0 \label{BSF.08}%
\end{equation}
where $S=S\left(  x,y,z,\phi\right)  ~$ describes a motion of a particle in the
$M^{4}$~space. Hamiltonian system is 
\begin{equation}
\dot{x}=\frac{\partial S}{\partial x}~,~\dot{y}=\frac{\partial S}{\partial
y}~,~\dot{z}=\frac{\partial S}{\partial z}~,~\dot{\phi}=\frac{\partial
S}{\partial\phi}. \label{BSF.09}%
\end{equation}
We easily find that for the potential $V\left(  \phi\right)  =0$ the equation (\ref{BSF.08}) possesses a solution
\begin{equation}
S_{0}\left(  x,y,z,\phi\right)  =c_{1}y+c_{2}z+c_{3}\phi+\varepsilon
\sqrt{c_{1}^{2}+c_{2}^{2}+c_{3}^{2}}x~,~\varepsilon=\pm1\text{.}
\label{BSF.10}%
\end{equation}
Applying it to the system of equations (\ref{BSF.09}) we obtain classical solutions 
\begin{align}
x\left(  t\right)  &=\varepsilon\sqrt{c_{1}^{2}+c_{2}^{2}+c_{3}^{2}}t+x_{0}\\
y\left(  t\right)  &=-c_{1}t+y_{0},\;
z\left(  t\right)  =-c_{2}t+z_{0},\;
\phi\left(  t\right) =-c_{3}t+\phi_{0}.
\end{align}%

The constant potential $V_0$ provides the solution of (\ref{BSF.08}) as 
\begin{align}
S_{V_{0}}\left(  x,y,z,\phi\right)  &=c_{1}y+c_{2}z+c_{3}\phi\nonumber\\ &+\varepsilon
\frac{2\sqrt{3}}{3}\left(  L\left(  x\right)  +\sqrt{c_{1-3}}\arctan
h\frac{L\left(  x\right)  }{\sqrt{c_{1-3}}}\right),
\end{align}
where we have defined the function $L\left(  x\right)  =\sqrt{c_{1}^{2}+c_{2}^{2}+c_{3}^{2}+2V_{0}%
e^{\sqrt{3}x}}~$ and the constant $c_{1-3}=c_{1}^{2}+c_{2}^{2}+c_{3}^{2}$. The Hamilton equations (\ref{BSF.09}) are found to be
\begin{equation}
\dot{x}=L\left(  x\right)  ~,~\dot{y}=-c_{1}~,~\dot{z}=-c_{2}~,~\phi=-c_{3}%
\end{equation}
carrying the exact solutions of the form  
\begin{align}
x\left(  t\right)  &=\frac{1}{3}\ln\left[  \frac{c_{1-3}}{2V_{0}}\left(
\tanh\left(  \frac{\sqrt{3}}{2}\sqrt{c_{1-3}}\left(  t+x_{0}\right)  \right)
-1\right)  \right]\\
y\left(  t\right)  &=c_{1}t+y_{0},\;
z\left(  t\right)  =c_{2}t+z_{0},\;
\phi\left(  t\right)  =c_{3}t+\phi_{0}.
\end{align}%
The classical solutions of (\ref{BSF.08}) with the exponential potential $V(\phi)=V_0 e^{\mu\phi}$ depend on the value of the constant $\mu$, as in the case 
of WDW solutions. Let us just consider the solution of the Hamilton-Jacobi equation (\ref{BSF.08}) for the value $\mu=-\sqrt{3}$. In order to it, we need to 
perform the coordinate transformation $\phi=\psi+x$ under which the Hamilton-Jacobi equation and the Hamiltonian system are now 
\begin{equation}
\frac{1}{2}\left[  \left(  \frac{\partial S}{\partial x}\right)  ^{2}-2\left(
\frac{\partial S}{\partial x}\right)  \left(  \frac{\partial S}{\partial\psi
}\right)  -\left(  \frac{\partial S}{\partial y}\right)  ^{2}-\left(
\frac{\partial S}{\partial z}\right)  ^{2}\right]  -V_{0}e^{-\sqrt{3}\psi}=0
\label{BSF.11A}%
\end{equation}
\begin{equation}
\dot{x}=\left(  \frac{\partial S}{\partial x}\right)  -\left(  \frac{\partial
S}{\partial\psi}\right)  ~,~\dot{y}=-\frac{\partial S}{\partial y}~,~\dot
{z}=-\frac{\partial S}{\partial z}~,~\dot{\psi}=-\frac{\partial S}{\partial
x}.
\end{equation}
One solves the equation (\ref{BSF.11A}) obtaining the Hamilton action 
\begin{equation}
S\left(  x,y,z,\psi\right)  =c_{1}x+c_{2}y+c_{3}z+\frac{\left(  c_{2}%
^{2}+c_{3}^{2}-c_{1}^{2}\right)  }{2c_{1}}\psi-\frac{\sqrt{3}V_{0}}{6c_{1}%
}e^{-\sqrt{3}\psi}%
\end{equation}
as well as the field equation
\begin{equation}
\dot{x}=\frac{\left(  c_{1}^{2}-c_{2}^{2}-c_{3}^{2}\right)  -V_{0}e^{\sqrt
{3}\psi}}{2c_{1}}~,~\dot{y}=-c_{2}~,~\dot{z}=-c_{3},~\dot{\psi}=-c_{1}.%
\end{equation}
The solutions of the above equations can be simply found  
\begin{align}
x\left(  t\right)  &=\frac{3}{2}c_{1}t-\frac{\left(  c_{2}^{2}+c_{3}%
^{2}\right)  }{2c_{1}}t+\frac{\sqrt{3}V_{0}}{6c_{1}^{2}}e^{-\sqrt{3}\psi_{0}%
}e^{\sqrt{3}c_{1}t}+x_{0},\\
y\left(  t\right)  &=-c_{2}t+y_{0},\;
z\left(  t\right)  =-c_{3}t+z_{0},\;
\psi\left(  t\right)  =-c_{1}t+\psi_{0},
\end{align}
where the quantities with the index $0$ are constants.
The details concerning the case $\left\vert \mu\right\vert \neq\sqrt{3}$ are given in \cite{tsamparlis2011geometric, wojnar2} and hence we will not
consider them here. The procedure is similar till obtaining the field equations (\ref{BSF.09}). In order to obtain analytical solutions of them, one needs to 
perform an extra transformation of the time variable in \cite{tsamparlis2011geometric}.

As an another brief example we will discuss Bianchi II models. Specifying, we will consider only the case of zero potential for which the scalar field behaves as 
stiff matter, that is, $p_{\phi}=\rho_{\phi}$. It arises to the WDW equation (\ref{B22})
\begin{equation}
\Psi_{,xx}-\Psi_{,yy}-\Psi_{,zz}-\Psi_{,\phi\phi}+e^{\frac{2\sqrt{3}}%
{3}\left(  2y+x\right)  }\Psi=0\mathrm{.} \label{BSF.20}%
\end{equation}
which can be solved by applying Lie invariants of the zeroth-order, similarly, as it was done for the Bianchi I models. One may get the solutions with respect to 
the various Lie algebras. Using for example \\
$\left\{K_{\left(  x\right)  }-\frac{1}{2}K_{\left(  y\right)  }+\mathbf{\nu}%
\Psi\partial_{\Psi};\;~K_{\left(  z\right)  }+\mu_{\left(  z\right)  }%
\Psi\partial_{\Psi};\;~K_{\left(  \phi\right)  }+\mu_{\left(  \phi\right)  }%
\Psi\partial_{\Psi}\right\}$
gives the invariant solution 
\begin{align}
\Psi_{1}\left(  x,y,z,\phi\right)  &=\exp\left(  \frac{2\mathbf{\nu}}{3}\left(
y+2x\right)  +\mu_{\left(  z\right)  }z+\mu_{\left(  \phi\right)  }%
\phi\right) \nonumber \\
&\times\left(  \Psi_{1}I_{\lambda}\left(  u\left(  x,y\right)  \right)
+\Psi_{2}K_{\lambda}\left(  u\left(  x,y\right)  \right)  \right)
\end{align}
for which we have defined the constant $\lambda=\frac{1}{3}\sqrt{12\mathbf{\nu}^{2}-9\left(  \mu_{\left(
z\right)  }^{2}+\mu_{\left(  \phi\right)  }^{2}\right)  }$ and the function $u\left(
x,y\right)  =\exp\left(  \frac{\sqrt{3}}{3}\left(  2y+x\right)  \right)$.
The functions $I_\lambda$ and $K_\lambda$ denote modified Bessel functions of the first and second kind, respectively. Choosing the algebras
 $\left\{  K_{\left(  z\right)
}\;;K_{\left(  x\right)  }-\frac{1}{2}K_{\left(  y\right)  };R_{\left(
xz\right)  }-\frac{1}{2}R_{\left(  yz\right)  }\right\}  $ or \\ $\left\{  K_{\left(  \phi\right)  }\;;K_{\left(
x\right)  }-\frac{1}{2}K_{\left(  y\right)  };~R_{\left(  x\phi\right)
}-\frac{1}{2}R_{\left(  y\phi\right)  }\right\}  $ gives also solutions in the terms of modified Bessel functions while the two Lie algebras
\begin{align*}
 \left\{  R_{\left(  z\phi\right)  },K_{\left(  x\right)
}-\frac{1}{2}K_{\left(  y\right)  },~R_{\left(  x\phi\right)  }-\frac{1}%
{2}R_{\left(  y\phi\right)  }\right\} \\
\left\{  R_{\left(
z\phi\right)  },K_{\left(  x\right)  }-\frac{1}{2}K_{\left(  y\right)
},~R_{\left(  xz\right)  }-\frac{1}{2}R_{\left(  yz\right)  }\right\} 
\end{align*}
 allows to solve the WDW equation and get the solution as
\begin{equation}
\Psi_{4}\left(  x,y,z,\phi\right)  =\Psi_{1}I_{0}\left(  u\left(  x,y\right)
\right)  +\Psi_{2}K_{0}\left(  u\left(  x,y\right)  \right)  \mathbf{.}%
\end{equation}
Applying the WKB approximation one may obtain classical solutions, similarly as it was done for Bianchi I.

\subsection{Conclusions}
We have discussed an another example of the usefulness of the Lie symmetries method in cosmological applications. As for hybrid gravity considered for FLRW spacetime, we 
were able to find an unknown potentials of a scalar field for some Bianchi models. Again we treated Lie symmetries as a criterion for selection models for which we could 
find exact solutions of Wheeler-DeWitt equations. Moreover, as WDW equations are invariant under the action of the three dimensional Lie algebra with
zero commutators, the Hamilton–Jacobi equations of the Hamiltonian system can be solved
by the method of separation of variables. It means that the field equations are Liouville integrable. Such analysis can be used to construct Wave Functions of the 
Universe as well as conservation laws (in the case when Lie symmetries as Noether ones). Existences  of symmetries gives rise to a straightforward interpretation 
of the Hartle criterion. It was shown \cite{CapHd} that symmetries generate oscillatory behaviors in a Wave Function of the Universe and then 
allow correlations among physical variables which gives rise to classically observable cosmological solutions. There also exists a possibility that one may use WDW 
equations to determine quantum potentials in the semi-classical approach of Bohmian mechanics \cite{bohma, bohmb}. The idea should be further investigated for cosmological purposes.

\section{Equilibrium and stability of relativistic stars}
In the previous parts we were focused on cosmological applications of some models of Extended Theories of Gravity. Here, we will consider astrophysical ones since 
there are also problematic issues concerning astrophysical objects like for instance neutron stars. Their structure and the relation between the mass and 
the radius are determined by equation of state (EoS) of dense matter. There are some propositions for its form, however it is still unknown.
That means that the relation between density and mass is not specified and hence a radius cannot be estimated. One gets its different values depending on 
a model taken into account. The problem is related to maximal masses of relativistic stars since GR predicts a maximal value for such objects.
The maximal mass of neutron stars is still an open question but recent observations 
estimate this limit as $2M_\odot$: for example the
pulsar PSR J1614-2230 has the limit $1.97$M$_\odot$ \cite{demorest}, another massive neutron star 
is Vela X-1 with the mass $\sim 1.8$M$_\odot$ \cite{rawls}. There are also indications of the existence of more massive neutron stars with 
masses $\sim 2.4$M$_\odot$, for instance B1957+20 \cite{van}. It should be also mentioned that a lot of EoS include 
hyperons which make the maximal mass limit for non-magnetic neutron stars 
significantly lower than expected $2M_\odot$ \cite{astash}.
There are a few ways to approach the problem of "hyperon puzzle", such as hyperon-vector coupling, chiral quark-meson coupling, existence of strong magnetic 
field inside the star and many others. For instance,
 it seems that the existence of neutron stars without strong magnetic field having masses larger than two solar mass is impossible in the framework 
of GR \cite{astash, astash2, astash3}. The topic is still controversial and under debate.

As neutron stars are very peculiar objects for testing 
theories of matter at high density regimes, data about their macroscopic properties like mass and radius can also be used for studying possible deviations from
GR. There exist suggestions \cite{astash3, eksi} that GR, being the only theory capable of describing strong gravitational field, is an extrapolation since the 
strength of gravity sourced by a star is many 
orders of magnitude larger that the one probed in the solar system weak field limit tests. From theoretical and experimental reasons one believes that GR 
should be modified when gravitational fields are strong and spacetime curvature is large \cite{berti}. Therefore, a promising route of investigation is to set a 
specific model of dense matter, i.e. equation of
state, and then to compute macroscopic properties of neutron stars in ETGs. Indeed, the predictions of alternative theories to GR concerning the structure of 
compact objects is currently an active research field \cite{oliveira2015neutron, palenzuela2015constraining, cisterna, cisterna2016slowly}. The results presented 
in the following parts can be also found in \cite{wojnar2016equilibrium, velten2016free}.

\subsection{Equilibrium and stability of relativistic stars in General Relativity}
Before we will discuss relativistic stars in theories different than General Relativity, let us present the problem in the Einstein theory 
\cite{weinberg, glende}. We will consider a spherical symmetric object whose geometry is given by the following metric
\begin{equation}\label{metryka_gwiazd}
 ds^2=-B(r)dt^2+A(r)dr^2+r^2d\theta^2+r^2\sin^2{\theta} d\phi^2.
\end{equation}
The matter of the star is assumed to be described by the perfect-fluid energy momentum tensor
\begin{equation}
T_{\mu\nu}=pg_{\mu\nu}+(p+\rho)u_\mu u_\nu,
\end{equation}
where $p$ and $\rho$ are pressure and the total energy density of the fluid. The four velocity $u^\mu$ of a co-moving (with the fluid) observer is normalized 
with the condition $u^\mu u_\mu = -1$. Additionally, in order to simplify calculations, we will make another assumptions, that is, the fluid is at rest so the only
non-vanishing component is $u_0=-(-g^{00})^{-\frac{1}{2}}=-\sqrt{B(r)}$. Moreover, since the metric is time-independent and one deals with spherical symmetry, we get that 
pressure $p$ and energy density $\rho$ are functions only of the radial coordinate $r$.

The Einstein's field equations ($\kappa = -8\pi G$)
\begin{equation}\label{mod}
R_{\mu\nu}-\frac{1}{2}Rg_{\mu\nu}=\kappa T_{\mu\nu}
\end{equation}
written for the considered system are
\begin{align}
-\frac{B''}{2A}+\frac{B'}{4A}\left(\frac{A'}{B}+\frac{B'}{B} \right)-\frac{B'}{rA}
=&\frac{\kappa}{2}(\rho+3p)B,\\
\frac{B''}{2B}-\frac{B'}{4B}\left(\frac{A'}{B}+\frac{B'}{B} \right)-\frac{A'}{rA}
=&\frac{\kappa}{2}(p-\rho)A,\\
 -1+\frac{r}{2A}\left(-\frac{A'}{B}+\frac{B'}{B}\right)+\frac{1}{A}
=&\frac{\kappa}{2}(p-\rho)r^2\label{nap},
\end{align}
where prime denotes $\frac{d}{dr}$. We have skipped the $\phi\phi$ equation as it is identical to $\theta\theta$ one. In order to find a form for $A(r)$, let us write
\begin{equation}
 \frac{R_{rr}}{2A}+\frac{R_{00}}{2B}+\frac{R_{\theta\theta}}{r^2}=-\frac{A'}{rA^2}-\frac{1}{r^2}+\frac{1}{Ar^2}=\kappa\rho
\end{equation}
which can be transformed into
\begin{equation}
 \left( \frac{r}{A} \right)'=1+\kappa\rho r^2.
\end{equation}
For $A(0)$ finite, the solution is
\begin{equation}\label{Adefin}
 A(r)=\left( 1-\frac{2G\mathcal{M}(r)}{r} \right)^{-1}
\end{equation}
where one defines 
\begin{equation}\label{masa_ns}
 \mathcal{M}(r)\equiv\int^r_0 4\pi \tilde{r}^2\rho(\tilde{r})d\tilde{r}.
\end{equation}

Using the hydrostatic equilibrium $\nabla^\mu T_{\mu\nu}=0$ which reads
\begin{equation}
 \frac{B'}{B}=-\frac{2p'}{p+\rho}
\end{equation}
and the equation (\ref{Adefin}) one finds that the equation (\ref{nap}) is 
\begin{equation}
 p'(r)=-\frac{G\mathcal{M}(r)\rho(r)}{r^2}\frac{\left(1+\frac{p(r)}{\rho(r)}\right)\left(1+\frac{4\pi r^3p(r)}{\mathcal{M}(r)}\right)}{\left(1-\frac{2G\mathcal{M}(r)}{r}\right)}.
\end{equation}
The stars that we are considering are assumed to be in convective equilibrium so the entropy per nucleon is nearly constant throughout the star. Moreover, they have such a 
chemical composition that it is constant. Pressure $p$ can be expressed as a function of the density $\rho$, th entropy per nucleon $s$, and the chemical composition, 
because of the equilibrium one sees that $p(r)$ can be regarded as a function of $\rho(r)$ alone. Due to that fact, one deals with a pair of first-order differential 
equation for $\rho(r)$ and $\mathcal{M}(r)$:
\begin{align}\label{uklad}
 \mathcal{M}'(r)&=4\pi r^2\rho(r),\\
 \frac{dp}{d\rho}\rho'(r)&=-\frac{G\mathcal{M}(r)\rho(r)}
 {r^2}\frac{\left(1+\frac{p(r)}{\rho(r)}\right)\left(1+\frac{4\pi r^3p(r)}{\mathcal{M}(r)}\right)}{1-\frac{2G\mathcal{M}(r)}{r}}.\nonumber
\end{align}
We are also equipped with an initial conditions provided by the equation (\ref{masa_ns})
\begin{equation}
 \mathcal{M}(0)=0.
\end{equation}
The above three equations, together with a given equation of state $p(\rho)$, determine the functions $\rho(r),\,\mathcal{M}(r),\,p(r)$ throughout the star, once we specify 
the value of $\rho(0)$. The system (\ref{uklad}) must be integrated out the center of the star until $p(\rho(r))=0$ at some point $r=R$. One interprets the value 
$R$ as the radius of the particular star with the central density $\rho(0)$.

The equations (\ref{uklad}) are called Tolman-Oppenheimer-Volkoff (TOV) equations \cite{OV, Tolman:1939dn, tolman1987relativity}.

Finalizing that part, let us just take a look at the stability problem of the considered system of the relativistic star (for details, see for example \cite{weinberg}). The 
equilibrium state of the star represented by the equations (\ref{uklad}) can be stable or unstable. We are interested in stable configurations. In our 
considerations we will need to recall the number of nucleons in the star which is defined as
 \begin{equation}
  N=\int \sqrt{g} J^{\,0}_Ndrd\theta d\phi=\int^R_0 4\pi r^2\sqrt{A(r)B(r)}J^{\,0}_N(r)dr,
 \end{equation}
where $J^{\,\mu}_N$ is the conserved nucleon number current. Using the relation between $J^{\,0}_N$ and the nucleon number density measured in a locally inertial 
reference frame at rest in the star $n=-u_\mu J^{\,\mu}_N=\sqrt{B}J^{\,0}_N$ as well as the form of the metric (\ref{metryka_gwiazd}) one gets
\begin{equation}
 N=\int^R_0 4\pi r^2\left(1-\frac{2G\mathcal{M}(r)}{r}\right)^{-\frac{1}{2}} n(r)dr.
\end{equation}
Similarly as for pressure, the number density $n(r)$ is a function of the density $\rho$, the chemical composition, and the entropy per nucleon $s$. Hence, the 
quantities $n(r)$ and $N$ are fixed for a star for given $\rho(0)$ together with chemical composition and constant $s$.

The stability criterion can be express by the following theory \cite{weinberg}
\begin{theorem}\label{theor_stab}
 A particular stellar configuration, with uniform entropy per nucleon and chemical composition, will satisfy the equations
 \begin{align}
 \mathcal{M}(r)&=\int^r_0 4\pi \tilde{r}^2\rho(\tilde{r})d\tilde{r},\nonumber\\
 p'(r)&=-\frac{G\mathcal{M}(r)\rho(r)}
 {r^2}\frac{\left(1+\frac{p(r)}{\rho(r)}\right)\left(1+\frac{4\pi r^3p(r)}{\mathcal{M}(r)}\right)}{1-\frac{2G\mathcal{M}(r)}{r}}\nonumber
\end{align}
for equilibrium, if and only if the quantity $\mathcal{M}$, defined by
\begin{equation}
 \mathcal{M}\equiv\int 4\pi r^2\rho(r)dr
\end{equation}
is stationary with respect to all variations of $\rho(r)$ that leave unchanged the quantity
\begin{equation}
  N=\int^R_0 4\pi r^2\left(1-\frac{2G\mathcal{M}(r)}{r}\right)^{-\frac{1}{2}} n(r)dr
\end{equation}
and that leave the entropy per nucleon and the chemical composition uniform and unchanged. The equilibrium is stable with respect to radial oscillations if and only 
if $\mathcal{M}$ is a minimum with respect to all such variations.
\end{theorem}
The proof of the theorem can be found in \cite{weinberg}. It is based on the Lagrange multiplier method. Since we will perform similar calculations 
as presented there for a model of Extended Theories of Gravity (see the subsection \ref{stab_section}), we will not repeat the proof of the theorem \ref{theor_stab}
here.

\subsection{Equilibrium of relativistic stars in Extended Theories of Gravity}
As already mentioned at the beginning of this section, the biggest challenge of modern astrophysics is the neutron stars' equation of state. Since General 
Relativity provides a limit on the neutron star's mass as not larger than $2M_{\odot}$, that condition and recent observations require stiff nuclear 
equation of state  \cite{nice20052}. The situation may differ in the case of Extended Theories of Gravity. Before considering any
model of ETGs, one may try to understand how different modifications of the TOV equations (\ref{uklad}) contribute to the  maximal mass value of a relativistic star. In
\cite{velten2016free} we considered parametrized TOV equations containing five parameters $\{\sigma,\alpha,\beta,\chi,\gamma\}$
\begin{align}
\mathcal{M}'(r)& = 4 \pi {r}^{2} (\rho + \sigma p),\\
p'(r) &= -\frac{G(1+\alpha) \mathcal{M}(r) \rho}{r^2}
\frac{\left(1+\frac{\beta p}{\rho}\right)\left(1+\frac{\chi 4\pi r^3 p}{\mathcal{M}(r)}\right)}{1 - \frac{\gamma 2G\mathcal{M}(r)}{r}}.
\end{align}
It was showed on the mass-radius diagrams how varying in parameters values shifts neutron stars' maximal masses and changes their sizes (radii).
The introduced parameters can be interpreted in the following way: $\alpha$ is viewed as a part of the effective gravitational constant, that 
is, $G_{\text{eff}}=G(1+\alpha)$. The larger $\alpha$, the smaller radius of the star while its maximal mass is also reduced. $\beta$ is a coupling to the 
inercial pressure while $\chi$ measures the active gravitational effects of pressure. Both extra contributions reduce the maximal mass; the latter has no effect on the 
radius. The parameter $\gamma$ is an intrinsic curvature contribution (it is zero in Newtonian physics and $1$ in GR). $\sigma$ changes the way of computations of the 
mass function - there can appear for example some gravitational effect of pressure. The detailed discussion may be found in \cite{velten2016free}.
 That exercise 
visualized that the problem of observed neutron stars' masses bigger than predicted ones can be also explained by geometric modifications of the Einstein's field 
equations. From now on, we will focus on a specific modification of Einstein's equations which will provide generalized TOV equations.

Many gravitational models of Extended Theories of Gravity (ETGs) can be recast in the form proposed in \cite{mim, mim2, mim3}
\begin{equation}\label{mod1}
 \sigma(\Psi^i)(G_{\mu\nu}-W_{\mu\nu})=\kappa T_{\mu\nu}.
\end{equation}
The tensor $G_{\mu\nu}=R_{\mu\nu}-\frac{1}{2}Rg_{\mu\nu}$ is the Einstein tensor, $\kappa=-8\pi G$, the factor $\sigma(\Psi^i)$ is a coupling to the gravity
while $\Psi^i$ represents for instance curvature invariants or other fields, like 
scalar ones. The symmetric tensor $W_{\mu\nu}$ stands for additional geometric terms which may appear in a specific ETG under consideration. Non-symmetric parts 
coming from a considered theory could be also included in the tensor $W_{\mu\nu}$ but then one should also add antisymmetric elements into energy-momentum tensor (for 
instance fermion fields). We will consider in that chapter only cases for which the tensor $W_{\mu\nu}$ is symmetric one. It is 
important to note that (\ref{mod1}) represents a parameterization of gravitational theories at the level of field equations. 
The energy-momentum tensor $T_{\mu\nu}$ is treated as the one of a perfect fluid, that is $T_{\mu\nu}=pg_{\mu\nu}+(p+\rho)u_\mu u_\nu$, where $p$ and 
$\rho$ are the pressure and the energy density of the fluid. The four velocity $u^\mu$ of the co-moving (with the fluid) observer is normalized 
with the condition $u^\mu u_\mu = -1$. One could make an assumption on the equation of state for $p$ and $\rho$, but we will not do it in order to keep our considerations
as general as it is possible.

It is worth noting that (\ref{mod1}) does not include all the possible alternatives to GR at the level of the equations of motion. However, 
most of the main proposals like, for instance, scalar tensor theories, $f(R)$ and hybrid gravity theories (see the chapter \ref{hg_chapter}), can be reshaped in this 
form as well as theories which have a time dependent effective gravitational coupling $\sigma \equiv \sigma(t)$ and $W_{\mu\nu}=0$. 

One may also add a coupling to the matter source (as it appears often in the so-called Einstein frame) but here we will not consider that case. 
From the structure of (\ref{mod1}) one sees that GR is immediately recovered for $\sigma(\Psi^i)=1$ in the geometric units and $W_{\mu\nu}=0$. The modified Einstein's field 
equations (\ref{mod1}) can be written for the later 
convenience as
\begin{equation}\label{mod}
G_{\mu\nu}=\kappa T^{eff}_{\mu\nu}=\frac{\kappa}{ \sigma} T_{\mu\nu}+W_{\mu\nu}.
\end{equation}
We would like to note that one cannot postulate that the energy-momentum tensor of the matter $T_{\mu\nu}$ is conserved. Rather, due to 
the Bianchi identity $\nabla_\mu G^{\mu\nu}=0$, the effective energy-momentum tensor $T^{eff}_{\mu\nu}$ is covariantly conserved i.e., $\nabla_\mu T_{eff}^{\mu\nu}=0$. In some special cases of 
ETG \cite{koivisto} one deals with the conservation of the matter energy-momentum tensor but in general it does not have to be true.


The simplest configuration for a star is the static and spherically symmetric geometry 
\begin{equation}
 ds^2=-B(r)dt^2+A(r)dr^2+r^2d\theta^2+r^2\sin^2{\theta} d\phi^2.
\end{equation}

From the normalization condition one has that $u_0=-\sqrt{B(r)}$. As the metric is time independent and spherically symmetric, the pressure $p$ and energy 
density $\rho$ are functions of the radial coordinate $r$ only. Hence we will assume that the coupling function $\sigma$ and 
the geometric contributions $W_{\mu\nu}$ are also independent of the coordinates $(t,\theta,\phi)$. The symbol
prime $(^{\prime})$ denotes the derivative with respect to the coordinate $r$.

We calculate in detail the components of (\ref{mod}). The components of the Ricci tensor read
\begin{align}
 R_{tt}&=-\frac{B''}{2A}+\frac{B'}{4A}\left(\frac{A'}{B}+\frac{B'}{B} \right)-\frac{B'}{rA}
=\frac{\kappa}{2\sigma}(\rho+3p)B+W_{tt}+\frac{BW}{2},\\
 R_{rr}&=\frac{B''}{2A}-\frac{B'}{4B}\left(\frac{A'}{B}+\frac{B'}{B} \right)-\frac{A'}{rA}
=\frac{\kappa}{2\sigma}(p-\rho)A+W_{rr}-\frac{AW}{2},\\
 R_{\theta\theta}&=-1+\frac{r}{2A}\left(-\frac{A'}{B}+\frac{B'}{B}\right)+\frac{1}{A}
=\frac{\kappa}{2\sigma}(p-\rho)r^2+W_{\theta\theta}-\frac{r^2W}{2},
\end{align}
where $W=-B^{-1}W_{tt}+A^{-1}W_{rr}+2r^{-2}W_{\theta\theta}$ is a trace of the tensor $W_{\mu\nu}$. Let us notice that the $R_{\phi\phi}$ equation is the same as 
the $R_{\theta\theta}$ one multiplied by the factor $\sin^{2}{\theta}$ and hence we concluded that $r^{-2}\sin^{-2}{\theta}W_{\phi\phi}=W_{\theta\theta}$.
Using the above equations to write
\begin{equation}
 \frac{R_{rr}}{2A}+\frac{R_{00}}{2B}+\frac{R_{\theta\theta}}{r^2}=-\frac{A'}{rA^2}-\frac{1}{r^2}+\frac{1}{Ar^2}=\frac{\kappa\rho}{\sigma}+r^2B^{-1}W_{tt}
\end{equation}
we obtain the following relation
\begin{equation}\label{forA}
 \left( \frac{r}{A} \right)'=1+\kappa  r^2\frac{\rho(r)}{\sigma(r)}+r^2B^{-1}(r)W_{tt}(r).
\end{equation}
Then we may solve equation (\ref{forA}) and write the solution as
\begin{equation}\label{mod_geo}
 A(r)=\left( 1-\frac{2Gm(r)}{r} \right)^{-1},
\end{equation}
where the mass function $m(r)$ is defined here as
\begin{equation}
m(r)=\int^r_0\left( 4\pi r^2\frac{\rho(\tilde{r})}{\sigma(\tilde{r})}-\frac{\tilde{r}^2W_{tt}(\tilde{r})}{2GB(\tilde{r})} \right)d\tilde{r}.
\end{equation}
It is clearly different from the usual definition given by GR (\ref{uklad}). Let us recall that the above quantity (similarly as in GR case \cite{weinberg}) 
is interpreted as total energy of 
a star together with the one coming from gravitational field.

For the further purposes we will also need the relations
\begin{align}
 \frac{A'}{A}=\frac{1-A}{r}-\frac{\kappa Ar}{\sigma}Q,\label{for1}\\
 \frac{B'}{B}=\frac{A-1}{r}-\frac{\kappa Ar}{\sigma}\Pi,\label{for2}
\end{align}
where we have defined  new quantities
\begin{eqnarray}\label{def}
Q(r):=\rho(r)+\frac{\sigma(r)W_{tt}(r)}{\kappa B(r)},\\
\label{def2} \Pi(r):=p(r)+\frac{\sigma(r)W_{rr}(r)}{\kappa A(r)}.
\end{eqnarray}

The hydrostatic equilibrium $\nabla_\mu T^{\mu\nu}_{eff}=0$ reads then
\begin{equation}\label{equil}
\kappa(\sigma^{-1}\nabla_\mu T^{\mu\nu}-\sigma^{-2}T^{\mu\nu}\nabla_\mu\sigma)+\nabla_\mu W^{\mu\nu}=0,
\end{equation}
or, more explicitly,
\begin{align}
&\kappa\sigma^{-1}\left(p'+(p+\rho)\frac{B'}{2B}\right)-\kappa p\frac{\sigma'}{\sigma^2}-\frac{A'}{A^2}W_{rr}+A^{-1}W'_{rr}\nonumber \\
&+\frac{2W_{rr}}{Ar}+
\frac{B'}{2B}\left(\frac{W_{rr}}{A}+\frac{W_{tt}}{B}\right)-\frac{2W_{\theta\theta}}{r^2}=0.
\end{align}
Let us notice that from (\ref{def2}) and with the help of (\ref{for1})
\begin{equation}\label{tov1}
 \left(\frac{\Pi}{\sigma}\right)'=\frac{p'}{\sigma}-\frac{p\sigma'}{\sigma^2}+\frac{W'_{rr}}{\kappa A}-\frac{W_{rr}(1-A)}{\kappa rA}+\frac{rW_{rr}Q}{\sigma}.
\end{equation}
This equation is the basic structure for deriving the generalized hydrostatic equilibrium for stars in ETG. Together with (\ref{for2}) and 
definition (\ref{def}), the equation (\ref{tov1}) can be written as
\begin{eqnarray}\label{tov}
  \left(\frac{\Pi}{\sigma}\right)'&=&-\frac{Gm}{r^2}\left(\frac{Q}{\sigma}+\frac{\Pi}{\sigma}\right)
  \left(1+\frac{4\pi r^3\frac{\Pi}{\sigma}}{m}\right)\left(1-\frac{2Gm}{r}\right)^{-1}\nonumber\\ 
	&+&\frac{2\sigma}{\kappa r}\left(\frac{W_{\theta\theta}}{r^2}-\frac{W_{rr}}{A}\right).
\end{eqnarray}

The above equation (\ref{tov}) and 
\begin{equation}\label{mr}
m(r)= \int^r_0 4\pi \tilde{r}^2\frac{Q(\tilde{r})}{\sigma(\tilde{r})} d\tilde{r}.
\end{equation}
have a similar functional form as the standard GR result. It is worth noting that such equations determine completely the stellar equilibrium since the 
assumption that pressure is expressed as a function of density only, i.e., the entropy per nucleon and the chemical composition as constant throughout 
the star. Such assumptions will also be used in the analysis of stability of these systems.

\subsection{Stability of relativistic stars in scalar - tensor theory of gravity}\label{stab_section}

In scalar-tensor theories the gravitational interaction is mediated not only by the metric field (as in GR), but also by scalar field $\phi$. Among 
many realizations of scalar-tensor theories, a simple prototype is the $k$-essence class in which the scalar field is said to be minimally coupled to
the geometric sector. 

The theory can be written according to the following action
\begin{equation}
 S=\frac{1}{2\kappa}\int d^4x\sqrt{-g}( R-\nabla_\mu\phi\nabla^\mu\phi-2V(\phi))+S_m[g_{\mu\nu},\psi].
\end{equation}

The field equations derived from it with respect to the metric $g_{\mu\nu}$ and the scalar field $\phi$ are
\begin{eqnarray}\label{fieldEQ}
 G_{\mu\nu}+\frac{1}{2}g_{\mu\nu}\nabla_\alpha\phi\nabla^\alpha\phi-\nabla_\mu\phi\nabla_\nu\phi+g_{\mu\nu}V(\phi)&=&\kappa T_{\mu\nu},\label{st1}\\
 V'(\phi)-\Box\phi&=&0,
\end{eqnarray}
respectively. From the Klein-Gordon we see that the scalar field $\phi$ depends on matter contribution ($\rho$) via the d'Alembertian operator. 
For the $k$-essence case we identify $\sigma_k=1$ and  
\begin{equation}
W_{\mu\nu}=-\frac{1}{2}g_{\mu\nu}\nabla_\alpha\phi\nabla^\alpha\phi+\nabla_\mu\phi\nabla_\nu\phi-g_{\mu\nu}V(\phi)
\end{equation}
from which we can write the following components
\begin{align}
 W_{tt}&=\frac{1}{2}B \nabla_\alpha\phi\nabla^\alpha\phi +BV(\phi)=B(C+2V),\\
 W_{rr}&=AC,\\
 W_{\theta\theta}&=-r^2 (C+2V).
\end{align}
In the above expressions we have defined $V\equiv V(\phi)$ and 
\begin{equation}
C\equiv C(Q,\phi,\phi')=\frac{1}{2}A^{-1}\phi'^2-V(\phi). 
\end{equation}
Let us remind that $A$ is a function of the $Q$ (\ref{mod_geo}).
Hence, the last term appearing in the generalized TOV equation (\ref{tov}) is $-\frac{4\sigma}{\kappa r}(C+V)=-2\sigma\frac{\phi'^2}{\kappa Ar}$. Moreover,
in the $k-$essence case, the functions $Q$ and $\Pi$ will have the form
\begin{align}
 Q&=\rho(r)+\kappa^{-1}(C+2V),\\
 \Pi&=p(r)+\kappa^{-1}C.
\end{align}
Notice that the second law of thermodynamics will differ in ETG's \cite{bamba}.
Let us calculate in detail the stability analysis. We then assume that the particle number $N^\alpha=n u^\alpha$ is supposed to be conserved
 \begin{equation}
  \nabla_\alpha(n u^\alpha)=u^\alpha \nabla_\alpha n + n\nabla_\alpha u^\alpha=0.
 \end{equation}
The crucial issue here is that we are dealing with effective energy-momentum tensor (from the Bianchi identities $\nabla_\mu G^{\mu\nu}=0$), therefore
\begin{eqnarray}\label{thermo}
u_\nu \nabla_\mu T^{\mu\nu}_{eff} &=&\sigma^{-1} \left(u^\mu\nabla_\mu p-n u^\mu\nabla_{\mu}\left(\frac{p+\rho}{n}\right)+\rho u^\mu\nabla_\mu\sigma\right) \nonumber\\
&+&u_\nu\nabla_\mu W^{\mu\nu},
\end{eqnarray}
and
\begin{eqnarray}\label{eq_therm}
-n u^\mu\left( p\nabla_\mu\left(\frac{1}{n}\right)+\nabla_\mu\left(\frac{\rho}{n}\right)+\frac{\rho}{n}\nabla_\mu\sigma \right) +\sigma u^\mu W^\nu_{\;\mu;\nu}=0.
\end{eqnarray}
As we are working with modified field equations of the specific form (\ref{st1}), the coupling term $\nabla_\mu\sigma$ in the above formula vanishes.
Furthermore, we will show that in the case of k-essence the tensor $\nabla^\nu W_{\mu\nu}=0$. The only non-vanishing terms that undergo
infinitesimal changes with respect to the infinitesimal changes of the energy density are 
\begin{equation}
 0=\delta\left( \frac{\rho}{n} \right)+p\delta\left(\frac{1}{n} \right),
\end{equation}
and consequently,
\begin{equation}\label{deltanr}
 \delta n(r) =\frac{n(r)}{p(r)+\rho(r)}\delta\rho(r).
\end{equation}
However, as we have already mentioned, the ETG that we are considering has a very special forms of the effective energy-momentum tensor (\ref{fieldEQ}). 
As $W_{\mu\nu}$ is symmetric and one also
deals with the K-G 
equations, we notice that
\begin{equation}
 \nabla^\mu W_{\mu\nu}=\nabla^\mu\phi(\nabla_\mu\nabla_\nu\phi-\nabla_\nu\nabla_\mu\phi)=0
\end{equation}
where we have used the K-G equation $\Box\phi=V'$. One may also compute it explicitly for the component $\mu=r$
\begin{equation}
 \nabla_\nu W^\nu_{\;r}=C'+(C+V)(\frac{A-1}{r}-\kappa Ar\Pi+\frac{4}{r}):=C'+D,
\end{equation} 
while the derivative $C'=\frac{dC(\phi,\phi')}{dr}$ after applying K-G equation, gives rise to
$$C'=-(C+V)(\frac{A-1}{r}-\kappa Ar\Pi+\frac{4}{r})=-D.$$

Therefore, component $\mu=r$ of equation (\ref{eq_therm}) resembles the GR form
\begin{equation}\label{new}
 n'(r)=n\frac{\rho'}{\rho+p}.
\end{equation}
Now on, we are going to use the Lagrange multipliers method following the procedure presented in \cite{weinberg}. The nucleon number $N$ remains unchanged but 
we should remember that it
also depends on the modified geometry (see the formula (\ref{mod_geo}) and below). It reads as 
$N=\int^R_0 4\pi r^2 [1-2Gm(r)/r]^{-1/2}n(r)dr$ . Then, we find
\begin{align}\label{rown_stab}
 0&=\delta m-\lambda \delta N=\int_0^\infty 4\pi r^2 \delta Q dr \nonumber\\
 &-\lambda \int_0^\infty 4\pi r^2\left(1-\frac{2Gm(r)}{r}\right)^{-\frac{1}{2}}\delta n(r)dr\nonumber\\
 &-
 \lambda G\int_0^\infty 4\pi r\left(1-\frac{2Gm(r)}{r}\right)^{-\frac{3}{2}}n(r)\delta m(r)dr,
\end{align}

Let us notice that  the integrands vanish outside the radius $R+\delta R$. 
It allows us to write the integration intervals as $[0,\infty]$ instead of $[0,R]$, where $R$ is a radius of a star. The variation does not change 
the entropy per nucleon as well as leaves the 
chemical composition uniform. Because of the form (\ref{deltanr}), one needs to understand the relation between $\delta\rho$ and $\delta Q$. Let us discuss it.

From the relation $\rho=Q-\kappa^{-1}(C+2V)$ we notice that $\rho$ is a function of $Q,\,\phi$ and $\phi'$. Hence, we obtain
 \begin{equation}
  \delta  \rho=\delta Q -\kappa^{-1}(\delta C + 2 V' \delta\phi).
 \end{equation}
Since $\phi=\phi(r)$ only, we may write $\phi'=\partial_\mu\phi=\nabla_\mu\phi$. The term $\delta C$ turns out to be
\begin{align}
 \delta C=&-\frac{1}{2}A^{-2}\phi'^2\delta A +A^{-1}\phi'\delta\phi'-V'\delta\phi=\\
 -&\frac{G}{r}\phi'^2\int^R_0 (4\pi r^2 \delta Q dr) +A^{-1}\phi'\delta\phi'-\Box\phi\delta\phi\nonumber\\
 =&-\frac{G}{r}\phi'^2\int^R_0 (4\pi r^2 \delta Q dr)+\nabla_\mu\phi\delta\nabla^\mu\phi-\Box\phi\delta\phi,
\end{align}
where we have used the K-G equation $\Box\phi=V'$. Then
 \begin{equation}
  \delta \rho=\delta Q -\kappa^{-1}\left(-\frac{G}{r}\phi'^2\int^R_0 (4\pi r^2 \delta Q dr)+\nabla_\mu\phi\delta\nabla^\mu\phi+\Box\phi\delta\phi \right).
 \end{equation}
 Let us notice that $\nabla_\mu\phi\delta\nabla^\mu\phi+\Box\phi\delta\phi=\nabla_\mu(\delta\phi\nabla^\mu\phi)$. Hence
  \begin{equation}
  \delta \rho=\delta Q -\kappa^{-1}\left(-\frac{G}{r}\phi'^2\int^R_0 (4\pi r^2 \delta Q dr)+\nabla_\mu(\delta\phi\nabla^\mu\phi)\right).
 \end{equation}
Now we are ready to write $\delta n$ with respect to $\delta Q$ and $\delta\phi$:
 \begin{align}\label{deltanr}
 \delta n(r)& =\frac{n(r)}{p(r)+\rho(r)}\\
 &\times\left(\delta Q -\kappa^{-1}\left(-\frac{G}{r}\phi'^2\int^R_0 (4\pi r^2 \delta Q dr)+\nabla_\mu(\delta\phi\nabla^\mu\phi)\right)\right)\nonumber.
\end{align}
We have used that $\delta m(r')=\int_0^\infty 4\pi r'^2 \delta Q dr'$. The equation (\ref{rown_stab}) now reads
 \begin{align}\label{eq_stab1}
 0&=\int_0^\infty 4\pi r^2 \delta Q dr
 -\lambda G\int_0^\infty 4\pi rA^{\frac{3}{2}}n(r)\int_0^\infty (4\pi \tilde{r}^2 \delta Q d\tilde{r})dr\nonumber\\
 -&\lambda \int_0^\infty 4\pi r^2A^{\frac{1}{2}} \frac{n(r)}{p(r)+\rho(r)}\nonumber\\
 &\times\left(\delta Q -\kappa^{-1}\left(-\frac{G}{r}\phi'^2\int^\infty (4\pi \tilde{r}^2 \delta Q d\tilde{r})+\nabla_\mu(\delta\phi\nabla^\mu\phi)\right)\right)dr
\end{align}
Before going further, let us discuss the term
 \begin{equation}
  \int_0^\infty 4\pi r^2A^{\frac{1}{2}}\frac{n(r)}{p(r)+\rho(r)}\kappa^{-1}\nabla_\mu(\delta\phi\nabla^\mu\phi)dr.
 \end{equation}
We may write it as
\begin{align*}
    &\kappa^{-1}\int_0^\infty\nabla^\mu \left( 4\pi r^2A^{\frac{1}{2}}\frac{n(r)}{p(r)+\rho(r)} \delta\phi\nabla_\mu\phi \right)dr\\
    -&\kappa^{-1}\int_0^\infty \delta\phi\nabla^\mu\phi \nabla_\mu\left( 4\pi r^2A^{\frac{1}{2}}\frac{n(r)}{p(r)+\rho(r)} \right)dr\\
    =& \kappa^{-1}\int_0^\infty\partial^\mu \left( 4\pi r^2A^{\frac{1}{2}}\frac{n(r)}{p(r)+\rho(r)} \delta\phi\partial_\mu\phi \right)dr \\
  +&  \kappa^{-1}\int_0^\infty\left[4\pi r^2A^{\frac{1}{2}} \Gamma^\mu_{\mu\nu} \frac{n(r)}{p(r)+\rho(r)} 
  -\partial_\nu\left( 4\pi r^2A^{\frac{1}{2}}\frac{n(r)}{p(r)+\rho(r)}\right) \right] \delta\phi\partial^\nu\phi dr.
\end{align*}
The term $\int_0^\infty\partial^\mu \left( 4\pi r^2A^{\frac{1}{2}}\frac{n(r)}{p(r)+\rho(r)} \delta\phi\partial_\mu\phi \right)dr$ is a constant. By choosing 
a suitable boundary condition for the scalar field $\phi$, it may vanish. We will neglect it in te further analysis. Interchanging the $r$ and 
$r'$ integrals in the equation (\ref{eq_stab1}) we will get the following one
 \begin{align}
  0=\delta m-\lambda\delta N&=\int_0^\infty4\pi r^2\left[1-\frac{\lambda n(r)}{p(r)+\rho(r)}A^{\frac{1}{2}}-\lambda G \int_r^\infty4\pi r'n(r')A^{\frac{3}{2}}dr'\right.\nonumber\\
  &-\left.\lambda G\kappa^{-1}\int^\infty_r4\pi \tilde{r} A^{\frac{1}{2}}\frac{n}{p+\rho}\phi'^2 \right]\delta Q(r)dr\nonumber\\
  &-
 \lambda\kappa^{-1}\int_0^\infty\partial^\nu\phi \left[4\pi r^2A^{\frac{1}{2}} \Gamma^\mu_{\mu\nu} \frac{n(r)}{p(r)+\rho(r)} \right.\nonumber\\
  &-\left. \partial_\nu\left( 4\pi r^2A^{\frac{1}{2}}\frac{n(r)}{p(r)+\rho(r)}\right) \right] \delta\phi dr.
 \end{align}
In order to have the vanishing right hand side of the above equation, both terms containing the variations $\delta Q$ and $\delta\phi$ must vanish 
independently. The term with $\delta Q$ will vanish if
\begin{align}\label{eq1}
  \frac{1}{\lambda}=\frac{n(r)}{p(r)+\rho(r)}A^{\frac{1}{2}}+
 G\int_r^\infty 4\pi r' n(r')A^{\frac{3}{2}}dr'\\\nonumber
 +G\kappa^{-1}\int^\infty_r4\pi \tilde{r} A^{\frac{1}{2}}\frac{n}{p+\rho}\phi'^2 dr
\end{align}
while the second one with $\delta\phi$ vanishes when
\begin{equation}\label{eq2}
 4\pi r^2A^{\frac{1}{2}} \Gamma^\mu_{\mu r} \frac{n(r)}{p(r)+\rho(r)} 
  -\partial_r\left( 4\pi r^2A^{\frac{1}{2}}\frac{n(r)}{p(r)+\rho(r)}\right)=0.
\end{equation}
We will start with (\ref{eq1}). Deriving it with respect to $r$ and using $ n'(r)=n\frac{\rho'}{\rho+p}$ one has:
\begin{align*}
 0=&-4\pi GrA-\frac{p'}{(p+\rho)^2}+\frac{G}{p+\rho}A(4\pi r Q -\frac{m}{r^2})\\
 &-4\pi r G\kappa^{-1}\frac{\phi'^2}{p+\rho}.
\end{align*}
Applying the following relations to the above expression:
\begin{align}
 \frac{A-1}{r}=A\frac{2Gm}{r^2},\\
 p+\rho=\Pi_k+Q_k - 2\kappa^{-1}(C+V),\\
 2(C+V)= A^{-1}\phi'^2,\\
   \Pi'_k=p'+\kappa^{-1}C'=p'-\kappa^{-1}(C+V)\left(\frac{A-1}{r}-\kappa Ar\Pi+\frac{4}{r}\right)\label{Capital}
\end{align}
we find that
\begin{align}\label{stabilityTOVkessence}
 \Pi'=-\frac{AGm}{r^2}(\Pi+Q)(1+4\pi r^3\frac{\Pi}{m})-4\frac{C+V}{\kappa r},
\end{align}
which is a form of generalized TOV equation derived in the previous section for the k-essence model.

 Let us come back to the equation (\ref{eq2}). Writing the derivative with respect to $r$ explicitly and 
 computing the gamma term, that is, $\Gamma^\mu_{\mu r}=\frac{2}{r}-\frac{1}{2}\left(\kappa Ar(\Pi+Q)\right)$ we will again obtain, after applying
 $ n'(r)=n\frac{\rho'}{\rho+p}$ and (\ref{Capital})
 \begin{align*}\label{stabilityTOVkessence}
 \Pi'=-\frac{AGm}{r^2}(\Pi+Q)(1+4\pi r^3\frac{\Pi}{m})-4\frac{C+V}{\kappa r}
\end{align*}
 which finally proves that the relativistic star's system provided by the k-essence model is a stable configuration.

\subsection{Remarks}
Let us here conclude our investigation. We have shown that Extended Theories of Gravity based on the phenomenological field
equations (\ref{mod1}) provide the stellar equilibrium equations for static, spherically symmetric
geometries given by the equations (\ref{tov1}) and (\ref{mr}). They are the analogous version of the TOV equations for any ETG. 
Such equations can now be further applied to specific gravitational theories. The differences between our equations and the ones provided by GR are in the 
definition of the mass $m(r)$ as one deals with the coupling $\sigma$ and the additional term $W_{tt}$ and in the definition of pressure. Due to that fact,
one needs to introduce effective quantities in order to obtain TOV-like form (\ref{uklad}). For the particular case shown in (\ref{tov}) the TOV
structure is preserved only if one finds a suitable theory in 
which $W_{\theta \theta}= W_{rr} r^2 / A$ and regarded that we identify $Q$ and $\Pi$, as the effective density and effective pressure, respectively.

Concerning the stability of such systems, we argue that this analysis should be implemented case by case only, i.e., it is difficult to achieve general
results without specifying the functions $W_{\mu\nu}$ and $\sigma(\Psi^i)$. As an example showing the applicability of our results, we worked on the specific 
class of $k-$essence theories. For this case, we generalized the stability theorem found for instance in \cite{weinberg} taking into account 
the new functions $Q$ ans $\Pi$. We found that the specific k-essence case leads to stable configurations.

The considered example shows that the equilibrium (\ref{tov}) is recovered from the Lagrange multiplier method with the reformulated stability criterion.
Contrary to the standard case, even assuming uniform entropy per nucleon and chemical composition, the interpretation of the mass function $m$
should be identified with effective energy density $Q$. The same analysis should be also applied to the definition of the nucleon number $N$.

The investigation of the stability of stellar systems in ETG and other modifications of gravity that cannot be written in the form (\ref{mod1}) should be
further examined. A very interesting case is the scalar-tensor gravity with non-minimally coupled scalar field. Due to that fact the equation 
(\ref{thermo}) will have a much complex form than the GR and k-essence cases. The work is in progress.



\appendix

\part*{Appendix} 


\chapter{Lie symmetry method}\label{app_lie}

Laws of physics are often written in a form of differential equations. Solving them allows us to determine a behavior of a physical system if we know initial 
conditions in the case of ordinary differential equations or boundary conditions for partial ones. One immediately comes to the conclusion that the knowledge of 
methods whose applications result in a solution of a differential equation is particularly important for physicists. Some equations that we face are well-known 
differential equations with given solutions or we are just lucky to find a way to write them in a form of a known ones, already classified. The problem arises 
when we deal with differential equations of an unfamiliar type. Fortunately, there exist tools which may help. The ones that we are using are Lie symmetries methods.

A subclass of Lie symmetries are very well-known Noether symmetries which have reached rightful place in physics. The application of Noether theorem has been 
proven to have crucial importance for research in quantum and particle physics as well as in cosmology 
\cite{capozziello1996nother, capozziello2008f, vakili2008noether, TsamC02} (see below). Unfortunately, Noether symmetries might be 
applied only in special cases: for systems which are modeled with a Lagrangian.
Lie symmetries method unlike the Noether symmetries approach can be used in the case of differential equations which do not arise from Lagrangian of a physical 
system, that is, they are not obtained from variational principle. Moreover, it may happen that the system does not admit any Noether symmetries but it admits Lie 
ones and we are still able to solve or simplify the differential equations. In the following chapter we are 
going to summarize Lie symmetries methods which have been used in the thesis.


\section{One-parameter point transformations and Lie symmetries group}
Let us consider two points $P$ and $Q$ living in a neighborhood $U$ in a smooth manifold $M$, $\text{dim}M=n$, with coordinates $(x,y)$ and
$(\tilde{x},\tilde{y})$, respectively. The following transformation of the coordinates of the point $P$ into the coordinates of $Q$ on $U$ 
\begin{equation}
 \tilde{x}=\tilde{x}(x,y),\;\;\;\tilde{y}=\tilde{y}(x,y)
\end{equation}
is called a point transformation \cite{hydon, thesis, StephaniB}. The functions $\tilde{x}(x,y)$ and $\tilde{y}(x,y)$ are 
independent. They map points $(x,y)$ into points $(\tilde{x},\tilde{y})$. 

One is particularly interested in one-parameter point transformations which depend on one (or more) arbitrary parameter $\epsilon\in\mathbb{R}$
\begin{equation}\label{one_param}
 \tilde{x}=\tilde{x}(x,y;\epsilon),\;\;\;\tilde{y}=\tilde{y}(x,y;\epsilon).
\end{equation}
Moreover, we want them to be invertible and that repeated applications produce a transformation of the same family 
for some $\hat{\epsilon}=\hat{\epsilon}(\tilde{\epsilon},\epsilon)$:
\begin{equation}
 \hat{x}=\hat{x}(\tilde{x},\tilde{y};\tilde{\epsilon})=\hat{x}(x,y;\hat{\epsilon}).
\end{equation}
The identity of the transformation is given by, for example, $\epsilon=0$:
\begin{equation}
 \tilde{x}=\tilde{x}(x,y;0),\;\;\;\tilde{y}=\tilde{y}(x,y;0).
\end{equation}
The transformations (\ref{one_param}) with the above properties form a one - parameter group of point transformations. 

We call the transformation 
\begin{equation}\label{sym}
\mathtt{T}: x\mapsto\tilde{x}(x)
\end{equation}
a symmetry, if it satisfies the following conditions \cite{hydon}:
\begin{itemize}
 \item The transformation preserves the structure.
 \item The transformation is a diffeomorphism. 
 \item The transformation maps the object to itself (the symmetry condition).
\end{itemize}
Now on, let us consider an infinite set of symmetries $\mathtt{T}_\epsilon$ (one-parameter group of point transformations)
\begin{equation}\label{liegr}
 \mathtt{T}_\epsilon: x^s\mapsto\tilde{x}^s(x^1,...,x^n;\epsilon),\;\;s=1,...,n.
\end{equation}
We will call the set of symmetries $\mathtt{T}_\epsilon$ a one-parameter local Lie group if the following conditions are satisfied \cite{hydon}
\begin{itemize}
 \item $\mathtt{T}_0$ is the trivial symmetry, so that $\tilde{x}^s=x^s$ when $\epsilon=0$.
 \item $\mathtt{T}_\epsilon$ is a symmetry for every $\epsilon$ in some neighborhood of zero.
 \item $\mathtt{T}_\epsilon\mathtt{T}_\delta=\mathtt{T}_{\epsilon+\delta}$ for every $\epsilon,\,\delta$ sufficiently close to zero.
 \item Each $\tilde{x}^s$ may be represented as a Taylor series in $\epsilon$ in some neighborhood of $\epsilon=0$:
 \begin{equation}
  \tilde{x}^s(x^1,...,x^n;\epsilon)=x^s+\epsilon\xi^s(x^1,...,x^n)+\mathcal{O}(\epsilon^2),\;\;s=1,...,n.
 \end{equation}
\end{itemize}
One may visualize the one-parameter group on an $x-y$ plane. Let us consider an arbitrary point $A=(x_0,y_0)$ on a plane with $\epsilon=0$. 
Varying the parameter $\epsilon$, the images $(\tilde{x}_0,\tilde{y}_0)$ of the point $A$ will move along some curve \cite{StephaniB}. Let us take another
initial points and repeat the procedure: one gets a family of curves. Each curve represents points which can be transformed into each other under 
the action of the group. That curve is called the orbit of the group. The family of the curves is characterized by a field of their tangent vectors $\mathbf{X}$.
In order to see it, let us consider infinitesimal transformations: taking an arbitrary point $(x,y)$ and representing the transformations (\ref{one_param}) as a 
Taylor series
\begin{align}
 \tilde{x}(x,y;\epsilon)&=x+\epsilon\xi(x,y)+...=x+\epsilon\mathbf{X}x+...,\\
 \tilde{y}(x,y;\epsilon)&=y+\epsilon\eta(x,y)+...=x+\epsilon\mathbf{X}y+...
\end{align}
One defines functions $\xi$ and $\eta$ 
\begin{equation}\label{inf_fun}
 \xi(x,y)=\left.\frac{\partial\tilde{x}}{\partial\epsilon}\right|_{\epsilon=0},\;\;
\eta(x,y)=\left.\frac{\partial\tilde{y}}{\partial\epsilon}\right|_{\epsilon=0},
 \end{equation}
with the operator (tangent vector) $\mathbf{X}$ as
 \begin{equation}\label{gener}
  \mathbf{X}=\xi(x,y)\frac{\partial}{\partial x}+\eta(x,y)\frac{\partial}{\partial y}.
 \end{equation}
The operator $\mathbf{X}$ is called the infinitesimal generator of the transformation.

As a simple example of a one-parameter group let us consider the rotations
\begin{equation}
 \tilde{x}=x\cos{\epsilon}-y\sin{\epsilon},\;\;\tilde{y}=x\sin{\epsilon}+y\cos{\epsilon},
\end{equation}
for which, from the definitions (\ref{inf_fun}) one has $\xi(x,y)=-y,\;\eta(x,y)=x$ so the infinitesimal generator of the rotation transformations is 
\begin{equation}
 \mathbf{X}=-y\frac{\partial}{\partial x}+x\frac{\partial}{\partial y}.
\end{equation}
Before starting the discussion on Lie symmetries of differential equations, we need to introduce a prolongation of the infinitesimal generator (\ref{gener}): 
\begin{defin}
 The prolongation up to the $n$th derivative of the infinitesimal generator (\ref{gener}) of a point transformation is a vector
 \begin{equation}\label{prolong}
  \mathbf{X}^{(n)}=\xi(x,y)\frac{\partial}{\partial x}+\eta(x,y)\frac{\partial}{\partial y}+\eta^{(1)}\frac{\partial}{\partial y'}+...+
  \eta^{(n)}\frac{\partial}{\partial y^{(n)}},
 \end{equation}
where the functions $\eta^{(n)}(x,y,y',...,y^{(n)})$ are defined as
\begin{equation}\label{ety}
 \eta^{(n)}=\frac{d\eta^{(n-1)}}{dx}-y^{(n)}\frac{d\xi}{dx}.
\end{equation}

\end{defin}
One should notice that the functions $\eta^{(n)}$ are not the $n$th derivative of $\eta$ but they are polynomials in the derivatives $y',...,y^{(n)}$. Since 
the expressions of (\ref{ety}) are complicated for higher $n$, let us just write the first two steps:
\begin{align}
 \eta^{(1)}=&\,\eta_{,x}+(\eta_{,y}-\xi_{,x})y'-\xi_{,y}y'^{2},\\
 \eta^{(2)}=&\,\eta_{,xx}+(2\eta_{,xy}-\xi_{,xx})y'+(\eta_{,yy}-2\xi_{,xy})y'^2\nonumber \\
 -&\xi_{,yy}y'^3+(\eta_{,y}-2\xi_{,x}-3\xi_{,y}y')y'',
\end{align}
where the coma, for example in $\eta_{,x}$, denotes the partial derivative with respect to $x$.

\section{Ordinary differential equations and Lie point symmetries}
Let us start with the following theorem \cite{StephaniB}:
\begin{theorem}
 We will say that an ordinary differential equation (ODE) 
\begin{equation}\label{ode}
 H(x,y,y',...,y^{(n)}):=y^{(n)}-\omega(x,y,y',...,y^{(n-1)})=0,
\end{equation}
where $y=y(x),\,y'=\frac{dy}{dx},...,\,y^{(n)}=\frac{d^ny}{dx^n}$, admits a group of symmetries with generator $\mathbf{X}$ if and only if
\begin{equation}\label{war_sym}
 \mathbf{X}^{(n)}H\equiv0,\;\; \text{mod}H=0
\end{equation}
is held, where $\mathbf{X}^{(n)}$ is the $n$th prolongation of $\mathbf{X}$.
\end{theorem}
A point transformation (\ref{one_param}) is a symmetry transformation (a symmetry) of the $n$th order ODE (\ref{ode}) if it maps solutions into solutions. It
means that the image $\tilde{y}(\tilde{x})$ of any solution $y(x)$ is again a solution: (\ref{ode}) does not change under a symmetry transformation, so
\begin{equation}\label{ode2}
H(\tilde{x},\tilde{y},\tilde{y}',...,\tilde{y}^{(n)})=0.
\end{equation}
It is important to notice that the existence of a symmetry is independent of the choice of variables that we use for expressing the ODE and its solutions. It might 
happen that we are dealing with a complicated looking differential equation with several symmetries found by the procedure explained below. That may mean that our 
differential equation is a simple one but given in unsuitable variables. Using symmetries, we can transform the equation into an easier form.

The $n$th order ODE (\ref{ode2}) is valid for all values of the parameter $\epsilon$ hence the differentiation of it with respect to $\epsilon$ gives
\begin{align}\label{warun}
 0=&\left.\frac{\partial H(\tilde{x},\tilde{y},\tilde{y}',...,\tilde{y}^{(n)})}{\partial \epsilon}\right|_{\epsilon=0}\nonumber\\
 =&\left.\left( \frac{\partial H}{\partial\tilde{x}}\frac{\partial\tilde{x}}{\partial\epsilon} +
 \frac{\partial H}{\partial\tilde{y}}\frac{\partial\tilde{y}}{\partial\epsilon} +...+
 \frac{\partial H}{\partial\tilde{y}^{(n)}}\frac{\partial\tilde{y}^{(n)}}{\partial\epsilon}
 \right)\right|_{\epsilon=0}
\end{align}
and using the definitions of (\ref{ety}) with $(\partial H/\partial\tilde{x})|_{\epsilon=0}=(\partial H/\partial x)$, the 
condition (\ref{warun}) is
\begin{equation}
 \xi\frac{\partial H}{\partial x}+ \eta \frac{\partial H}{\partial y}+\eta' \frac{\partial H}{\partial y'}+...+\eta^{(n)}\frac{\partial H}{\partial y^{(n)}}=0,
\end{equation}
or simply
\begin{equation}\label{warun2}
 \mathbf{X}^{(n)}H=0.
\end{equation}
The ODE $H=y^{(n)}-\omega(x,y,y',...,y^{(n-1)})=0$ is invariant under the infinitesimal transformation, it means, if $H=0$ holds and it admits 
a group of symmetries with generators $\mathbf{X}$, then 
$\mathbf{X}^{(n)}H=0$ also holds. The converse is also true \cite{StephaniB}.

Applying the definition (\ref{prolong}) into the symmetry condition (\ref{war_sym}) we may write it as 
\begin{align}\label{war_sym1}
 \eta^{(n)}=&\mathbf{X}^{(n)}\omega\nonumber\\
 =&\left(\xi\frac{\partial}{\partial x}+\eta\frac{\partial}{\partial y}+\eta^{(1)}\frac{\partial}{\partial y'}+...+
  \eta^{(n-1)}\frac{\partial}{\partial y^{(n-1)}}\right)\omega
\end{align}
with $\eta^{(i)}$ given by (\ref{ety}). The $n$-derivative $y^{(n)}$ appearing in $\eta^{(n)}$ must be substituted by $\omega$ \cite{StephaniB, hydon}. This equation reduces to a system of
partial differential equations (PDE's)
after equating to zeroes terms which are multiplied by powers of $y^{(n-1)},\;y^{(n-2)},...,\;\text{and so on}$ because the functions $\xi(x,y)$ and $\eta(x,y)$
are independent of the derivatives of $y$. The system of PDE's determining $\xi(x,y)$ and $\eta(x,y)$ can usually be solved.

As an example we will consider the simplest second-order ODE $y''=0$. The linearized symmetry condition \cite{hydon} is
\begin{equation*}
 \eta^{(2)}=0\;\;\text{when}\;\;y''=0
\end{equation*}
that is,
\begin{equation}\label{exam1}
 \eta_{,xx}+(2\eta_{,xy}-\xi_{,xx})y'+(\eta_{,yy}-2\xi_{,xy})y'^2-\xi_{,yy}y'^3=0.
\end{equation}
The condition (\ref{exam1}) splits into the system of determining equations:
\begin{equation}\label{exam2}
 \eta_{,xx}=0,\;\;2\eta_{,xy}-\xi_{,xx}=0,\;\;\eta_{,yy}-2\xi_{,xy}=0,\;\;\xi_{,yy}=0,
\end{equation}
with the general solution of the last one:
\begin{equation*}
 \xi(x,y)=A(x)y+B(x)
\end{equation*}
 with the arbitrary functions $A$ and $B$.
The third equation of (\ref{exam2}) gives ($C$ and $D$ are also arbitrary functions)
\begin{equation*}
 \eta(x,y)=A'(x)y^2+C(x)y+D(x)
\end{equation*}
and using these results to the remaining equations in (\ref{exam1}) one obtains
\begin{equation*}
 A'''(x)y^2+C''(x)y+D''(x)=0,\;\;3A''(x)y+2C'(x)-B''(x)=0.
\end{equation*}
Since the unknown functions in the above equations are independent of $y$, one equates powers of $y$ obtaining a system of ODEs:
\begin{equation*}
 A''(x)=0,\;\;C''(x)=0,\;\;D''(x)=0,\;\;B''(x)=2C'(x)
\end{equation*}
which is easily solved. Hence, for every one-parameter Lie group of symmetries of the equation $y''(x)=0$ the function $\xi$ and $\eta$ are
\begin{align*}
 \xi(x,y)&=c_1+c_3x+c_5y+c_7x^2+c_8xy,\\
 \eta(x,y)&=c_2+c_4y+c_6x+c_7xy+c_8y^2,
\end{align*}
where $c_i,\;i\in\{1,...,8\}$ are constants. The most general infinitesimal generator is of the form
\begin{equation*}
 \mathbf{X}=\sum^{i=1}_{8}c_i\mathbf{X}_i,
\end{equation*}
with the vectors
\begin{align*}
 \mathbf{X}_i=\partial_x,\;\;\mathbf{X}_2=\partial_y\;\;\mathbf{X}_3=x\partial_x,\;\;\mathbf{X}_4=y\partial_y,\;\;\mathbf{X}_5=y\partial_x,\\
 \mathbf{X}_6=x\partial_y,\;\;\mathbf{X}_7=x^2\partial_x+xy\partial_y,\;\;\mathbf{X}_8=xy\partial_x+y^2\partial_y.
\end{align*}

Now on, when we are familiar with finding Lie point symmetries we may use them for simplifications of problems which come down to solving ODEs. In the next two 
subsections we will present two methods of reducing an order of ODE's (for more details and examples see \cite{hydon}).

\subsection{Reducing order of ODE's by canonical coordinates}
We will say that the generator $\mathbf{X}=\xi(x,y)\frac{\partial}{\partial x}+\eta(x,y)\frac{\partial}{\partial y}$ can be written in its normal form 
$\mathbf{X}=\partial_s$ if there exists a system of coordinates $\{r(x,y),s(x,y)\}$ such that
\begin{equation*}
 \mathbf{X}r=0,\;\;\;\mathbf{X}s=1,
\end{equation*}
that is,
\begin{align}
 \xi(x,y)r_{,x}+\eta(x,y)r_{,y}=&0,\nonumber\\
 \xi(x,y)s_{,x}+\eta(x,y)s_{,y}=&1,\nonumber\\
 r_{,x}s_{,y}-r_{,y}s_{,x}\neq0.\nonumber
\end{align}
The last equation is the non-degeneracy condition: it ensures that the change of coordinates is invertible in some neighborhood of $(x,y)$. The coordinates 
$\{r(x,y),s(x,y)\}$ are called canonical coordinates. Using canonical coordinates allows to reduce an order of ODE as it is presented below.

Let the vector $\mathbf{X}$ be an infinitesimal generator of a one-parameter Lie group of symmetries of the ODE 
\begin{equation}\label{od1}
y^{(n)}=\omega(x,y,y',...,y^{(n-1)}),\;\;\;n\geq0 
\end{equation}
and let $\{r(x,y),s(x,y)\}$ are canonical coordinates so that $\mathbf{X}=\partial_s$. One may write the ODE for some function $\Omega$ in terms of canonical coordinates:
\begin{equation*}
 s^{(n)}=\Omega(r,s,\dot{s},...,s^{(n-1)}),\;\;\dot{s}=\frac{ds}{dr},\;\;s^{(k)}=\frac{d^ks}{dr^k}.
\end{equation*}
But the considered ODE is invariant under the Lie group of translations in $s$ so from the symmetry condition
\begin{equation*}
 \Omega_{,s}=0\;\;\;\text{so}\;\;\;s^{(n)}=\Omega(r,\dot{s},s^{(n-1)}).
\end{equation*}
Let us introduce $v=\dot{s}$; then the above equation is an ODE of order $n-1$:
$$
v^{(n-1)}=\Omega(r,v,...,v^{(n-2)}),\;\;\;v^{(k)}=\frac{d^{k+1}s}{dr^{k+1}}.
$$

\subsection{Reducing order of ODE's by Lie invariants}
If a non-constant function $I(x,y,y',...,y^{(k)})$ satisfies 
\begin{equation}\label{inv}
 \mathbf{X}^{(k)}I=0,
\end{equation}
where $\mathbf{X}^{(k)}$ is the prolongation of the infinitesimal generator $\mathbf{X}$ of a one-parameter Lie group of symmetries of the ODE (\ref{od1}), then we say 
that $I$ is a $k$th order differential invariant of the group generated by $\mathbf{X}$. \cite{hydon}. Since in canonical coordinates $\mathbf{X}=\partial_s$,
the differential invariant is of the form
\begin{equation}
 I=F(r,\dot{s},...,s^{(k)})=F(r,v,...,v^{(k-1)})
\end{equation}
for some function $F$. The zeroth order differential invariant is the canonical coordinate $r(x,y)$. Moreover, all first-order invariants are functions of $r(x,y)$ and 
$v(x,y,y')$, and higher order invariant are functions of $r,\;v$ and derivatives of $v$ with respect to $r$. One may show \cite{hydon} that the condition
(\ref{inv})
$$
\xi I_{,x}+\eta I_{,y}+...+\eta^{(k)}I_{,y^{(k)}}=0
$$
is equivalent to
\begin{equation}\label{inv2}
 \frac{dx}{d\xi}=\frac{dy}{\eta}=...=\frac{dy^{(k)}}{\eta^{(k)}}.
\end{equation}
One says that $I$ is a first integral of (\ref{inv2}). Is is worth to note that the canonical coordinate $r$ is a first integral of 
\begin{equation}
  \frac{dx}{d\xi}=\frac{dy}{\eta}
\end{equation}
and $v$ is a first integral of 
\begin{equation}
  \frac{dx}{d\xi}=\frac{dy}{\eta}=\frac{dy'}{\eta^{(1)}}.
\end{equation}
From the zeroth order invariant $r$ and first order invariant $v$ one may define the following differential invariants
$$
\frac{dv}{dr},...,\;\frac{d^{n-1}v}{dr^{n-1}},\;\;\;\;\;\text{where}\;\;\frac{dv}{dr}=\frac{v_{,x}+v_{,y}y'+v_{,y'}y''}{u_{,x}+u_{,y}y'}
$$
which are functions of different derivatives of $y$ appearing in (\ref{od1}). It allows us to rewrite (\ref{od1}) in terms of invariants giving us a 
result which is $(n-1)$th order ODE 
\begin{equation}
 \frac{d^{n-1}v}{dr^{n-1}}=\Omega\left( r,v,\frac{dv}{dr},...,\frac{d^{n-2}v}{dr^{n-2}} \right).
\end{equation}

\subsection{Noether symmetries}\label{ap_noether}
There exists a special case of Lie point symmetries which are very important in physics. They are called Noether symmetries whose first differential invariants  
(first integrals) have physical meaning; for example, when the symmetry is time translation (or rotation), one deals with conservation of 
 energy (or angular momentum).
 
 Let 
\begin{equation}\label{action_ap}
 S=\int^{t_2}_{t_1} L(q^k,\dot{q}^k,t)dt
\end{equation}
be an action of a physical system whose dynamics is described by the function $L(q^k,\dot{q}^k,t)$ called Lagrangian. The dot denotes the derivative with respect to the 
 time variable $t$. The equations of motion (Euler - Lagrange 
equations) derived from variationl principle are \cite{ingarden}
\begin{equation}\label{el_lag}
 \frac{d}{dt}\frac{\partial L}{\partial \dot{q}^i}-\frac{\partial L}{\partial q^i}=0.
\end{equation}
Now on, let us defined a Noether symmetry \cite{StephaniB}:
\begin{defin}
 A Noether symmetry is a Lie point transformation that leaves the action $S$ invariant up to an additive constant $\hat{V}(\epsilon)$ with $\epsilon$ being 
 the group parameter.
\end{defin}
Let $\tilde{S}=S+\hat{V}(\epsilon)$ be an action obtained by mapping the action $S$ by a point transformation 
and $\hat{V}(\epsilon)=\int^{t_2}_{t_1}\frac{d\hat{V}(q^k,t,\epsilon)}{dt}dt$. One sees that $S$ and $\tilde{S}$ leads to the same equations of motion (\ref{el_lag})
hence Noether symmetries leave the differential equations invariant. Expanding $\tilde{S}$ in Taylor series with respect to $\epsilon$ with the infinitesimal 
prolonged generator given by the
vector field $\mathbf{X}^{(1)}=\xi\frac{\partial}{\partial t}+\eta^a\frac{\partial}{\partial q^a}+\dot{\eta}^a\frac{\partial}{\partial \dot{q}^a}$ we have:
\begin{align}
 \tilde{S}&=\int\tilde{L}d\tilde{t}=\int L(\tilde{q}^k,\dot{\tilde{q}}^k,\tilde{t})d\tilde{t}\\
 &=\int[L(q^k,\dot{q}^k,t)+\epsilon\mathbf{X}L+\mathcal{O}(\epsilon^2)]
 \left( dt+\epsilon\frac{d\xi}{dt}dt+\mathcal{O}(\epsilon^2) \right)\\
 &=S+\epsilon\int\frac{dV(q^k,t)}{dt}dt+...
\end{align}
Collecting terms linear in $\epsilon$ one gets the condition for $\tilde{S}=S+\hat{V}(\epsilon)$ being true; let us write it as a theorem:
\begin{theorem}\label{theo_noether}
 The infinitesimal generator $\mathbf{X}$ is a Noether symmetry if there exists a function $V=V(q^k,t)$ such that the following condition is satisfied:
 \begin{equation}\label{noe_cond}
 \mathbf{X}^{(1)}L+\frac{d\xi}{dt}L=\frac{dV}{dt},
\end{equation}
where $\mathbf{X}^{(1)}$ is the first prolongation of the generator $\mathbf{X}$.
\end{theorem}
Noether symmetries are called variational symmetries if $V=0$. For every Noether symmetry there exist a first integral, it means
\begin{defin}
 If $\mathbf{X}=\xi\frac{\partial}{\partial t}+\eta^a\frac{\partial}{\partial q^a}$ is the generator of a Noether symmetry then
 \begin{equation}
  I_{N}=\xi E_H -\eta^k L_{,\dot{q}^k}+V(q^i,t)
 \end{equation}
is a first integral that satisfies $\mathbf{X}^{(1)}I_{N}=0$. The quantity $E_H=\dot{q}^k L_{,\dot{q}^k}-L$ is called Hamiltonian of the dynamical system. 
\end{defin}
One should notice that the condition (\ref{noe_cond}) will not give us all possible Lie symmetries; it may happen that a system does not admit any Noether symmetries and 
due to that fact no conserved quantities may be used to reduce the order of the differential equations. But there can still exist Lie symmetries from which one may 
construct Lie invariants.

\subsection{Linear ODE's}
Lie point symmetries are a very useful tool for solving differential equations: one finds symmetries from the condition (\ref{war_sym1}) and applying 
Lie invariants method we are able to reduce order of an ODE which can help to solve a differential equation. 
Although, they are not helpful in the case of linear ODE's of order $n\geq2$. It happens that one or more determining equations has the same for as
the ODE that we wanted to solve. One usually needs to know the general 
solution of the ODE in order to find Lie point symmetries \cite{hydon}:
\begin{theorem}\label{theo_linear}
 Every homogeneous linear ODE of order $n\geq3$ has infinitesimal generators of the form
 \begin{equation}
  \mathbf{X}_1=y\partial_y,\;\;\mathbf{X}_2=y_1\partial_y,...,\;\;\mathbf{X}_{n+1}=y_n\partial_y,
 \end{equation}
where $\{y_1,...,y_n\}$ is a set of functionally independent solutions of the ODE. If the ODE can be mapped into the ODE $y^{(n)}=0$ by a point 
transformation, then it admits three extra infinitesimal generators:
\begin{equation}
 X_{n+2}=\partial_x,\;\;\;X_{n+3}=x\partial_x,\;\;\;X_{n+4}=x^2\partial_x+(n-1)xy\partial_y.
\end{equation}

\end{theorem}

Let us consider the fourth order ODE:
\begin{equation}\label{pert}
  \lambda^{(4)}+\frac{5}{\tau}\lambda'''+\left( \frac{2}{\tau^2}+v \right)\lambda''+\left( \frac{v}{\tau}-\frac{2}{\tau^3} \right)\lambda'=0.
\end{equation}
It comes from the system of equations which gives arise when one deals with the linear approximation of Einstein's field equations for 
perturbations of the spatial metric tensor in the early stages of expansion of the Universe \cite{landau}
$$
h^i_j=\lambda(\tau)P^i_j+\mu(\tau)Q^i_j.
$$
In the above $\tau$ denotes conformal time, $v$ is a constant, and the tensors $P^i_j$ and $Q^i_j$ are defined by a scalar function used to express
the perturbation of the density of the matter filling an isotropic universe \cite{landau}. 

We will show that in the case of (\ref{pert}) the Lie symmetry method does not work as it has been already mentioned. As a first step 
let us simplify the equation by $y(\tau)=\lambda'$
\begin{equation}\label{pert2}
  y'''+\frac{5}{\tau}y''+\left( \frac{2}{\tau^2}+v \right)y'+\left( \frac{v}{\tau}-\frac{2}{\tau^3} \right)y=0.
\end{equation}
We will look for a vector field $X=\xi(\tau,y)\partial_\tau+\eta(\tau,y)\partial_y$ which is a symmetry vector of (\ref{pert2}). Before using the symmetry 
condition (\ref{war_sym1}) for
$$
\omega=\frac{5}{\tau}y''-\left( \frac{2}{\tau^2}-v \right)y'-\left( \frac{v}{\tau}+\frac{2}{\tau^3} \right)y
$$ 
one needs to find $\eta^{(3)}$ from (\ref{ety}):
\begin{align*}
 \eta^{(3)}&=-\left[\left\{ \left(y'\xi_{,yyy}-\eta_{,yyy}+3 \xi_{,xyy}\right)y'-3 \eta_{,xyy}+3 \xi_{,xxy}\right\}y'\right.\\
 &\left.-3 \eta_{,xxy}+\xi_{,xxx}\right]y' 
 +y''' \left(-4 y'\xi_{,y}+\eta_{,y}-3 \xi_{,x}\right)-3 y''^2 \xi_{,y}\\
 &+3 y''
   \left(y' \left(-2 y' \xi_{,yy}+\eta_{,yy}-3 \xi_{,xy}\right)+\eta_{,xy}-\xi_{,xx}\right)+\eta_{,xxx}.
\end{align*}
The symmetry conditions (\ref{war_sym1}) for (\ref{pert2}) splits into a system of partial differential equations for the functions $\alpha(\tau),\,\beta(\tau),\,\gamma(\tau)$:
\begin{align}
\beta'''+\frac{5}{\tau}\beta''+\left( \frac{2}{\tau^2}+v \right)\beta'+\left( \frac{v}{\tau}-\frac{2}{\tau^3} \right)\beta=0,\label{de1}\\
\gamma'''+\frac{5}{\tau}\gamma''+\left( \frac{2}{\tau^2}+v \right)\gamma'+\left( \frac{v}{\tau}-\frac{2}{\tau^3} \right)\gamma=0,\label{de2}\\
-5\alpha+\frac{5}{\tau}\alpha'+3\tau^2(\beta'-\alpha'')=0,\\
-4\frac{\alpha}{\tau^3}+\frac{2}{\tau^2}(2+v\tau^2)\alpha'+\frac{1}{\tau}(10\beta'-5\alpha''-\tau\alpha''')=0,\\
(6-v\tau^2)\alpha-\tau(v\tau^2-2)\beta+3\tau(v\tau^2-2)\alpha'=0
\end{align}
for 
$$
 \xi(\tau,y)=\alpha(\tau)\;\; \text{and}\;\;\eta(\tau,y)=\beta(\tau)y+\gamma(\tau).
$$
We notice that the equations (\ref{de1}) and (\ref{de2}) have the form of the ODE (\ref{pert2}) that we wanted to solve. The same happens when we 
use the method in order to solve (\ref{pert}).

One may try to reduce the order of (\ref{pert}) with the help of the theorem \ref{theo_linear} and canonical coordinates. We know that the equation 
admits the symmetry vector $X=\lambda\partial_\lambda$ from which we get that the canonical coordinates are
\begin{equation}
 s=\mathrm{ln}\lambda,\;\;\;r=\eta.
\end{equation}
If one denotes $u=\frac{ds}{dr}$, then $\lambda'=\lambda u$ and the ODE (\ref{pert}) is now the third order ODE of the form
\begin{align*}
  u'''+\left(4u+\frac{5}{\eta}\right)u''&+3u'^2\left( 6u^2+v+\frac{15}{\eta}+\frac{2}{\eta^2} \right)\\
  &+
\left( u^3+\frac{5}{\eta}u^2+(v+\frac{2}{\eta^2})u+\frac{v}{\eta}-\frac{2}{\eta^3} \right)u=0.
\end{align*}

\section{Lie algebra}
It may happen the an ODE has many symmetries and some of them belong to $m$-parameter Lie group
\begin{equation}
 \tilde{x}=\tilde{x}(x,y,\delta),\;\;\;\tilde{y}=\tilde{y}(x,y,\delta),\;\;\;\;\delta=\epsilon^\beta\partial_\beta,\;\;\beta=1,...,m
\end{equation}
with an infinitesimal generator defined as
\begin{equation}
 \mathbf{X}_\beta=\xi_\beta(x,y)\partial_x+\eta_\beta(x,y)\partial_y.
\end{equation}
Symmetries belonging to an $m$-parameter Lie group can be regarded as a composition of symmetries from $m$ one-parameter Lie groups.

Let $\mathcal{L}$ denotes a set of all infinitesimal generators of one (or more)-parameter Lie group of point symmetries of an ODE of order $n\geq2$. Since the 
linearized symmetry condition is linear in $\xi$ and $\eta$ one has that $\mathcal{L}$ is a vector space 
\begin{equation*}
 \mathbf{X}_1,\mathbf{X}_2\in\mathcal{L}\;\;\;\rightarrow\;\;\;c_1\mathbf{X}_1+c_2\mathbf{X}_2\in\mathcal{L},\;\;\forall c_1,c_2 \in \mathbb{R}
\end{equation*}
The dimension $m$ of the vector space is a number of arbitrary constants appearing in the general solution of the linearized symmetry condition. 
Every $\mathbf{X}\in\mathcal{L}$ may be written in the form
$$
\sum^m_{i=1}c_i \mathbf{X}_i,\;\;\;c_i\in R,
$$
where $\{\mathbf{X}_1,...,\mathbf{X}_m\}$ is a basis for $\mathcal{L}$. Similarly,
the set of point symmetries generated by all $\mathbf{X}\in\mathcal{L}$ forms an $m$-parameter
Lie group.

Let $\mathbf{X}_1,...,\mathbf{X}_4\in\mathcal{L}$. A first-order operator
\begin{equation}
 [\mathbf{X}_1,\mathbf{X}_2]=\mathbf{X}_1\mathbf{X}_2-\mathbf{X}_2\mathbf{X}_1
\end{equation}
is called a commutator of $\mathbf{X}_1$ with $\mathbf{X}_2$. It is antisymmetric, bilinear and satisfies the Jacobi identity:
\begin{align}
 [\mathbf{X}_1,\mathbf{X}_2]&=-[\mathbf{X}_2,\mathbf{X}_1],\\
 [c_1\mathbf{X}_1+c_2\mathbf{X}_2,\mathbf{X}_3]&=c_1[\mathbf{X}_1,\mathbf{X}_3]+c_2[\mathbf{X}_2,\mathbf{X}_3],\nonumber\\
 [\mathbf{X}_1,c_2\mathbf{X}_2+c_3\mathbf{X}_3]&=c_2[\mathbf{X}_1,\mathbf{X}_2]+c_3[\mathbf{X}_1,\mathbf{X}_3],\\
0=[\mathbf{X}_1,[\mathbf{X}_2,\mathbf{X}_3]]&+[\mathbf{X}_2,[\mathbf{X}_3,\mathbf{X}_1]]+[\mathbf{X}_3,[\mathbf{X}_1,\mathbf{X}_2]].
\end{align}
Moreover, $\mathcal{L}$ is closed under the commutator $\mathbf{X}_i,\mathbf{X}_j\in\mathcal{L}\rightarrow[\mathbf{X}_i,\mathbf{X}_j]\in\mathcal{L}$ and
the commutator of any two generators $\mathbf{X}_1,\mathbf{X}_2\in\mathcal{L}$ in the basis is a linear combinations of the basis generators
\begin{equation}
 [\mathbf{X}_i,\mathbf{X}_j]=c^k_{ij}\mathbf{X}_k.
\end{equation}
The constants $c^k_{ij}$ are called structure constants. From antisymmetry and the Jacobi identity one gets that
\begin{align*}
c^q_{ij}=&-c^q_{ji},\\
c^q_{ij}c^l_{kq}+c^q_{jk}c^l_{iq}+&c^q_{ki}c^l_{jq}=0,\;\;\forall i,j,k,l.
\end{align*}
If $ [\mathbf{X}_i,\mathbf{X}_j]=0,\;(c^k_{ij}=0)$, we say that the generators $\mathbf{X}_i$ and $\mathbf{X}_j$ commute.

A finite dimensional linear space $\mathcal{L}$ with a commutator as a product on $\mathcal{L}$ satisfying above conditions forms a Lie algebra.

One may show \cite{hydon} that an $n$th order ODE can be reduced with $m\leq n$ Lie point symmetries, which are generated by $\mathcal{L}$,
to an ODE of order $n-m$ (or to an algebraic equation if $n=m$).
The considered differential equation is written in terms of the differential invariants of each generator, it means the ODE can be written in terms of functions that are 
invariant under all of its symmetry generators.

One should mention \cite{hydon} that if $\mathbf{X}_1$ and $\mathbf{X}_2$ generate Lie point symmetries, then so does $[\mathbf{X}_1,\mathbf{X}_2]$. It can be used in
order to find more Lie symmetries of the differential equation.

\section{Lie point symmetries of PDE's}

Finding Lie point symmetries of partial differential equations (PDE's) is a very similar procedure as the case of ODE's. Due to that fact, we will shortly give 
basic notions with necessary formulas for the simplest case, it means we are going to consider PDE's with one dependent variable $u$ and two independent ones, $t$ 
and $x$.

A point transformation \cite{hydon,StephaniB} is a diffeomorphism
$$
\mathtt{T}:\,\big(x,t,u(x,t)\big)\mapsto\Big(\tilde{x}\big(x,t,u(x,t)\big),\tilde{t}\big(x,t,u(x,t)\big),\tilde{u}\big(x,t,u(x,t)\big)\Big)
$$
which maps the surface $u=u(x,t)$ into 
\begin{align*}
 \tilde{x}&=\tilde{x}\big(x,t,u(x,t)\big),\\
 \tilde{t}&=\tilde{t}\big(x,t,u(x,t)\big), \\
 \tilde{u}&=\tilde{u}\big(x,t,u(x,t)\big).
\end{align*}
If $H(x,t,u,u_x,u_t,...,u_\sigma)=u_\sigma-\omega(x,t,u,u_x,u_t,...)=0$ is an $n$th order PDE, where $u_\sigma$ is one of the $n$th order derivatives of $u$ and $\omega$ is 
independent of $u_\sigma$, then the point transformation $T$ is its point symmetry if 
\begin{equation}\label{pde_sym1}
H(\tilde{x},\tilde{t},\tilde{u},\tilde{u}_{\tilde{x}},\tilde{u}_{\tilde{t}},...)=0\;\text{when}\;H(x,t,u,u_x,u_t,...)=0\;\text{holds}. 
\end{equation}
As for ODE's, we are looking for one-parameter Lie groups of point symmetries \cite{hydon}, it means, one searches for point symmetries that have the form
\begin{align}
 \tilde{x}&=x+\epsilon\xi(x,t,u)+\mathcal{O}(\epsilon^2),\\
 \tilde{t}&=t+\epsilon\tau(x,t,u)+\mathcal{O}(\epsilon^2),\\
\tilde{u}&=u+\epsilon\eta(x,t,u)+\mathcal{O}(\epsilon^2),
 \end{align}
with the infinitesimal generator
\begin{equation}
 \mathbf{X}=\xi\partial_x+\tau\partial_t+\eta\partial_u.
\end{equation}
The first two prolongations of the above generator are
\begin{align}
 \mathbf{X}^{(1)}&=\mathbf{X}+\eta^{(x)}\partial_{u_{,x}}+\eta^{(t)}\partial_{u_{,t}},\\
 \mathbf{X}^{(2)}&=\mathbf{X}^{(1)}+\eta^{(xx)}\partial_{u_{,xx}}+\eta^{(xt)}\partial_{u_{,xt}}+\eta^{(tt)}\partial_{u_{,tt}}
\end{align}
with long expressions for $\eta^{(ij...)}$
\begin{align}
 \eta^{(x)}&=\eta_{,x}+(\eta_{,u}-\xi_{,x})u_{,x}-\tau_{,x}u_{,t}-\xi_{,u} u^2_{,x}-\tau_{,u}u_{,x}u_{,t},\\
 \eta^{(t)}&=\eta_{,t}-\xi_{,t}u_{,x}+(\eta_{,u}-\tau_{,t})u_{,t} - \xi_{,u}u_{,x}u_{,t}-\tau_{,u}u^2_{,t},\\
 \eta^{(xx)}&=\eta_{,xx}+(2\eta_{,xu}-\xi_{,xx})u_{,x}-\tau_{,xx}u_{,t}+(\eta_{,uu}-2\xi_{,xu})u^2_{,x}\nonumber\\
 &-2\tau_{,xu}u_{,x}u_{,t}-\xi_{,uu}u^3_{,x}-\tau_{,uu}u^2_{,x}u_{,t}+(\eta_{,u}-2\xi_{,x})u_{,xx}\nonumber\\
 &-2\tau_{,x}u_{,xt}-3\xi_{,u}u_{,x}u_{,xx}-\tau_{,u}u_{,t}u_{,xx}-2\tau_{,u}u_{,x}u_{,xt},\\
 \eta^{(xt)}&=\eta_{,xt}+(\eta_{,tu}-\xi_{,xt})u_{,x} + (\eta_{,xu}-\tau_{,xt})u_{,t}-\eta_{,tu}u^2_{,x}\nonumber\\
 &+(\eta_{,uu}-\xi_{,xu}-\tau_{,tu})u_{,x}u_{,t}-\tau_{,xu}u^2_{,t}-\xi_{,uu}u^2_{,x}u_{,t}-\tau_{,uu}u_{,x}u^2_{,t}\nonumber\\
 &-\xi_{,t}u_{,xx}-\xi_{,u}u_{,t}u_{,xx}+ (\eta_{,u}-\xi_{,x}-\tau_{,t})u_{,xt} - 2\xi_{,u}u_{,x}u_{,xt}\nonumber\\
 &-2\tau_{,u}u_{,t}u_{,xt}-\tau_{,x}u_{,tt}-\tau_{,u}u_{,x}u_{,tt},\\
  \eta^{(tt)}&=\eta_{,tt}-\xi_{,tt}u_{,x}+ (2\eta_{,tu}-\tau_{,tt})u_{,t}-2\xi_{,tu}u_{,x}u_{,t}\nonumber\\
  &+ (\eta_{,uu}-2\tau_{,tu})u^2_{,t} - \xi_{,uu}u_{,x}u^2_{,t}-\tau_{,uu}u^3_{,t}-2\xi_{,t}u_{,xt}\nonumber\\
  &-2\xi_{,u}u_{,t}u_{,xt} + (\eta_{,u}-2\tau_{,t}) u_{,tt} -\xi_{,u}u_{,x}u_{,tt} - 3\tau_{,u}u_{,t}u_{,tt}.
\end{align}
For higher terms and more general case there exist recurrence formulas \cite{StephaniB} but one should use some computer algebra.

The linearized symmetry conditions is obtained when we differentiate the symmetry condition (\ref{pde_sym1}) with respect to $\epsilon$ at $\epsilon=0$ so one has
\begin{equation}
 \mathbf{X}^{(n)}H(x,t,u,u_x,u_t,...,u_\sigma)\equiv0,\;\;\;\text{mod}H(x,t,u,u_x,u_t,...,u_\sigma)=0.
\end{equation}
The above condition gives, after eliminating $u_\sigma$, a linear system of determining equations for $\xi$, $\tau$, and $\eta$.

Assuming that we have already found a Lie point symmetry of a PDE $H(x,t,u,u_{,x},...)=0$ one may use Lie invariant method as it was done for ODE's. The difference
is that one reduces a number of variables instead of reducing order of differential equation.

\section{Handful of useful theorems}\label{useful_theo}
We would like to give a few extra notions and theorems which are very helpful in the case when one applies Lie symmetries method to considered problems. 
It has been shown \cite{TsamAnd} that 
when one deals with differential equations derived from a Lagrangian, one may relate Lie symmetries with conformal algebra of a metric of a Riemannian space which is 
a phase space of the physical system.

Let us introduce a conformal Killing vector (CKV) of a metric $G_{ij}$:
\begin{defin}
 A vector field $u^i$ is a conformal Killing vector if it satisfies
 \begin{equation}
  \mathcal{L}_u G_{ij}=2\psi G_{ij},
 \end{equation}
where $\mathcal{L}_u$ is a Lie derivative with respect to the vector field $u^i$ and $\psi$ is called a conformal factor.

We will say that $u^i$ is
\begin{itemize}
 \item  a Killing vector if $\psi=0$,
 \item a homothetic vector if $\psi_{,i}=0$,
 \item a special conformal Killing vector if $\psi_{;ij}=0$,
 \item  a proper conformal Killing vector if $\psi_{;ij}\neq0$.
\end{itemize}
\end{defin}
Two metrics are called conformally related, it means one has the relation 
$$\bar{G}_{ij}=N^2{G}_{ij},$$
$N^2$ being a conformal factor.
 Moreover, if $u^i$ is a CKV of the metric $\bar{G}_{ij}$ then it is also a CKV of the conformally related metric $G_{ij}$ with the conformal factor $\psi$ defined as
 $$
 \psi=\bar{\psi}N^2 - N N_{,i} u^i.
 $$

 Let us again consider the action (\ref{action_ap}) with the Lagrangian of the form 
 \begin{equation}\label{lagr_ap}
  L(q^i,\dot{q}^i)=\frac{1}{2}G_{ij}\dot{q}^i\dot{q}^j-V(q^k)
 \end{equation}
which is a Lagrangian of a particle which moves under the action of the potential $V(q^k)$ in a Riemannian space with the metric $G_{ij}$. The dot represents a derivative 
with respect to time $t$ which is a parameter of a curve along which the particle moves. We will perform two transformations: the first one is a coordinate 
transformation of the form
\begin{equation}
 d\tau=N^2(q^i)dt
\end{equation}
with the $\tau$-derivative denoted from now on as a prime $'$. The next one is a conformal transformation of the metric $G_{ij}=N^{-2}\bar{G}_{ij}$. Defining a new 
potential $\bar{V}(q^k)=N^{-2}(q^k)V(x^k)$ one gets that the new Lagrangian
\begin{equation}
  \tilde{L}(q^i,q'^i)=\frac{1}{2}\bar{G}_{ij}q'^iq'^j-\bar{V}(q^k)
\end{equation}
has the same form as the original one (\ref{lagr_ap}). We will say that such Lagrangians are conformally related. There exists a theorem \cite{TsamC01} which is 
very useful in the case of cosmology:
\begin{theorem}\label{theo_confor}
 The Euler-Lagrange equations for two conformal Lagrangians transform covariantly under the conformal transformation relating the Lagrangians if and only if the 
 Hamiltonian vanishes.
\end{theorem}
The equations of motion coming from the Lagrangians $L$ in the variables $(t,q^i)$ and $\tilde{L}$ in the variables $(\tau,q^i)$ are of the same form. In the 
other words, it results from the above theorem that physical systems with vanishing energy have conformally invariant equations of motion. 
In the case of FRLW cosmology, the Hamiltonian of the considered system is the $(0,0)$ Einstein equation being also a constraint. The more detailed discussion on 
the FRLW cosmology and also scalar field cosmology is given in \cite{TsamC01}.

 There are two another important results concerning conformally related metrics and Lagrangians. If one considers Noether symmetries of the Lagrangian (\ref{lagr_ap}),
 they can be obtained \cite{TsamAnd} from the homothetic algebra of the metric $G_{ij}$. The same result can be applied to the Lagrangian $\tilde{L}$ and the metric 
 $\bar{G}_{ij}$. One should also mention that the conformal algebras of the metrics $G_{ij}$ and $\bar{G}_{ij}$ are spanned by the same conformal Killing vectors 
 \cite{yano} with subalgebras of homothetic and Killing vectors 
 different for each metric.
 
Let us now consider a Klein - Gordon equation
\begin{equation}\label{KG_ap}
 \nabla w = V(q^i) w,
\end{equation}
where $\nabla$ is the Laplace operator of the Riemannian space with metric $G_{ij}$ having the following form
\begin{equation}
 \nabla w = \frac{1}{\sqrt{|G|}}\frac{\partial}{\partial q^i}\left( \sqrt{|G|}G^{ij}\frac{\partial}{\partial q^j}  \right)w,
\end{equation}
$G$ being the determinant and $G^{ij}$ the inverse of the metric $G_{ij}$. It was shown \cite{pal_heat, AnIJGMP} that there exists a relation between Lie point symmetries 
of Klein - Gordon equation and conformal algebra:
\begin{theorem}\label{KG_theo}
 The Lie point symmetries of Klein - Gordon equation (\ref{KG_ap}) are generated from the conformal Killing vectors of the metric $G_{ij}$ which defines the Laplace 
 operator in the following way:
 \begin{itemize}
  \item for the dimension of the Riemannian space $n>2$ the generic Lie symmetry vector is
  \begin{equation}
   \mathbf{X}=u^i(q^k)\partial_i+\frac{2-n}{n}\psi(q^k)w +a_ow+b(q^k)\partial_w
  \end{equation}
where $u^i$ is a CKV with conformal factor $\psi(q^k)$, the function $b(q^k)$ is a solution of the K-G equation (\ref{KG_ap}). Moreover, the following condition is 
satisfied for the potential $V(q^k)$:
\begin{equation}\label{pot_theo}
 u^k V_{,k}+2\psi V - \frac{2-n}{n}\nabla\psi=0;
\end{equation}
 \item for $n=2$ the generic Lie symmetry vector is
  \begin{equation}
   \mathbf{X}=u^i(q^k)\partial_i +a_ow+b(q^k)\partial_w
  \end{equation}
where $u^i$ is a CKV with conformal factor $\psi(q^k)$, the function $b(q^k)$ is a solution of the K-G equation (\ref{KG_ap}). Moreover, the following condition is 
satisfied for the potential $V(q^k)$:
\begin{equation}
 u^k V_{,k}+2\psi V=0;
\end{equation}
 \end{itemize}

\end{theorem}
The above theorem allows us to construct Lie point symmetries of Klein - Gordon equations from known conformal algebras and to find an unknown potential function 
appearing in (\ref{KG_ap}) without solving determining equations. It is extremely useful in Extended Theories of Gravity when one deals with not determined actions 
of a theory: if one is able to transform the action into scalar - tensor action, it is possible to find the Lagrangian due to the fact that we may express unknown 
parts by the potential (for example $f(R)$ gravity in metric or Palatini formalism, Hybrid Gravity etc.).

Let us consider a PDE of the form $H(x,t,u,u_x,u_t,...,u_\sigma)$. Applying a Lie symmetry to the considered PDE gives us a new differential equation $\tilde{H}$ which is 
different than $H$. It may happen that it also admits Lie symmetries that are not symmetries of the original equation. It has been proved \cite{govinder,and_bianchi}
that
\begin{theorem}
 If two Lie point symmetries $\mathbf{X}_1,\mathbf{X}_2$ of a PDE 
 $$H(x,t,u,u_x,u_t,...,u_\sigma)=0$$ 
 commute as 
 \begin{equation}
  [\mathbf{X}_1,\mathbf{X}_2]=c\mathbf{X}_1,\;\;c\;\text{is a constant},
 \end{equation}
then the reduction variables defined by $\mathbf{X}_1$ will reduce $\mathbf{X}_2$ to a point symmetry of the PDE (called $H_1$, say) obtained from $H$ via these 
variables. However, the variables defined by $\mathbf{X}_2$ reduce $\mathbf{X}_1$ to an expression that has no relevance for the PDE (called $H_2$, say) obtained from 
$H$ via these reduction variables.
\end{theorem}



\chapter{$\Lambda$CDM model as a $2$D dynamical system of the Newtonian type}\label{apB}
Dynamical systems theory is another approach that we use for an examination of cosmological models. In contrast to the Lie symmetries method described in the 
Chapter \ref{app_lie}, we are interested in the evolutionary behavior of the Universe, that is, we would like to get to know special points of the evolution such 
as cosmological singularities or steady states instead of looking for exact solutions of given systems \cite{wainwright2005dynamical}.
The another reason for applying dynamical systems approach are difficulties in finding solutions of differential equations that describe cosmological 
models. It often happens in modified cosmological models that exact solutions are impossible to obtain because complicated equations appear.
Such analysis provides an interesting approach to theoretical cosmology: one may examine the Universe's evolutionary paths in a phase plane. Many authors showed that one may
treat the whole Universe as a fictious point particle moving in 
a one-dimensional potential well (see for instance 
\cite{borowiec2016inflationary, hrycyna2015cosmological, szydlowski2009scalar, copeland2006dynamics, sami2012cosmological, coley1999dynamical, carloni2005cosmological}).
Due to that fact, the considered evolution of the system reduces to a simple $2$-dimensional dynamical system of the Newtonian 
type. The evolution for all admissible initial conditions is represented by trajectories in a phase space \cite{bor_kam}, that is, the 
considered potential function gives us full information about the dynamics \cite{zoo_marek}.

We would like to briefly give basic notions on phase portraits and critical points.
Moreover, we depict the method for the standard model of cosmology, that is, $\Lambda$CDM model, as a simple example.

\section{Phase portraits of linear systems in $\mathbb{R}^2$}
Let us briefly discuss simple examples of phase portraits of linear systems in $2$-dimensional vector space $\mathbb{R}^2$. We will consider the linear 
system of the form
\begin{equation}\label{linsys1}
 \dot{\mathbf{x}}=A\mathbf{x}
\end{equation}
where $\mathbf{x}\in\mathbb{R}^2$ is a vector field while $A$ is a $2\times2$ matrix. One may reduce the system (\ref{linsys1}) to an uncoupled 
linear system by the diagonalization procedure
\begin{equation}\label{linsys}
 \dot{\mathbf{x}}=B\mathbf{x},\;\;\;\;B=P^{-1}AP
\end{equation}
where $P^{-1}AP=\text{diag}[\lambda_1,\lambda_2]$, $P$ is a matrix consisting of generalized eigenvectors of $A$
and $\lambda_i,\;i=\{1,2\}$, stand for eigenvalue of the matrix $A$. One may generalize the problem to 
the $n$-dimensional system (see for example \cite{perko}).
Let us assume that the form of the matrix $B$ is one of the following possibilities
\begin{equation}
 B=\begin{bmatrix}
  \lambda & 0 \\
  0       & \mu 
 \end{bmatrix},\;\;
 B=\begin{bmatrix}
  \lambda & 1 \\
  0       & \lambda 
 \end{bmatrix},\;\;
 B=\begin{bmatrix}
  a & -b \\
  b & a 
 \end{bmatrix}.
\end{equation}
Then, one may use the fundamental theorem of linear systems:
\begin{theorem}\label{lineasys}
Let $B$ be an $n\times n$ matrix. Then for a given $\mathbf{x}_0\in\mathbb{R}^n$, the initial value problem
\begin{align}
  \dot{\mathbf{x}}=B\mathbf{x},\\
  \mathbf{x}(0)=\mathbf{x}_0\label{initial}
\end{align}
has a unique solution given by
\begin{equation}
  \dot{\mathbf{x}}=e^{Bt}\mathbf{x_0},
\end{equation}
\end{theorem}
where 
\begin{equation*}
e^{Bt}=\displaystyle\sum_{k=0}^{\infty}\frac{B^kt^k}{k!}
\end{equation*}
is a $n\times n$ matrix which can be computed in terms of
the eigenvalues and eigenvectors of $B$. It can be 
 showed \cite{perko, hirsch} that the solution of the (\ref{linsys}) with the initial 
condition (\ref{initial}) is
\begin{align*}
  \mathbf{x}(t)=\begin{bmatrix}
  e^{\lambda t} & 0 \\
  0       & e^{\mu t} 
 \end{bmatrix}\mathbf{x}_0,\;\;\;\;
 \mathbf{x}(t)=e^{\lambda t}\begin{bmatrix}
  1 & t \\
  0       & 1 
 \end{bmatrix}\mathbf{x}_0,\\
 \mathbf{x}(t)=e^{at}\begin{bmatrix}
  \cos{bt} & -\sin{bt} \\
  \sin{bt} & \cos{bt} 
 \end{bmatrix}\mathbf{x}_0,
\end{align*}
respectively. 

Due to that solutions one may draw different types of phase portraits with respect to the eigenvalues of the matrices. We will draw phase portraits of the 
linear system (\ref{linsys}); phase portraits of the linear system (\ref{linsys1}) is obtained from the drawn ones for (\ref{linsys}) under the linear 
transformation of coordinates $\mathbf{x}=P\mathbf{y}$.

\begin{figure}
	\centering
	\includegraphics[scale=0.85]{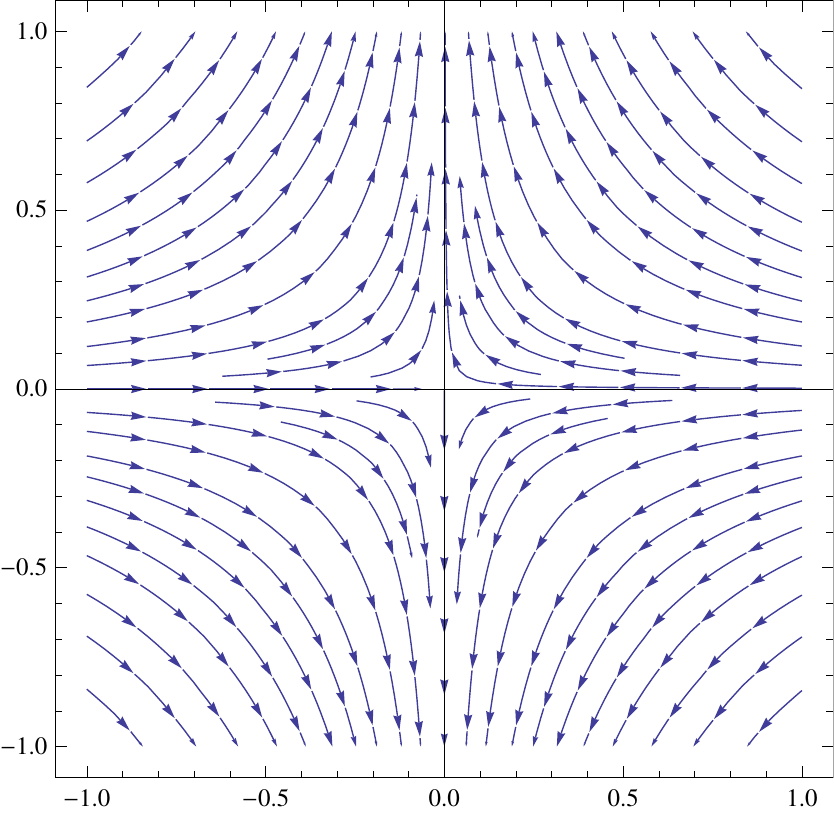}
	\caption{Phase portrait of the system $\dot{x}=-x,\;\dot{y}=y$. The equilibrium point (critical point of the system, it means
 $\dot{x}=\dot{y}=0$) at the origin is called a \textbf{saddle point}. As $\lambda<0$, solutions along the line $y=0$ decay to $0$ (stable line) while 
 $\mu>0$ corresponds to the growing solutions along $x=0$ (unstable line).}
	\label{saddle1}
\end{figure}

The first case
\begin{equation}
 B=\begin{bmatrix}
  \lambda & 0 \\
  0       & \mu 
 \end{bmatrix}
 \end{equation}
 includes \textbf{saddle point} at the origin (see the figure \ref{saddle1}) for $\lambda<0<\mu$. The arrows are reversed when $\mu<0<\lambda$. 
 Considering the matrix $A$, if it has two real eigenvalues of opposite sign, $\lambda<0<\mu$, then the phase portrait of th system (\ref{linsys1})
 is linearly equivalent to the phase portrait \ref{saddle1}: it is obtained by a linear transformation of coordinates.
 Separatrices of the system are
 the four non-zero trajectories (solution curves) approaching the equilibrium point at the origin as $t\rightarrow\pm\infty$.
 
  \begin{figure}
	\centering
	\includegraphics[scale=0.85]{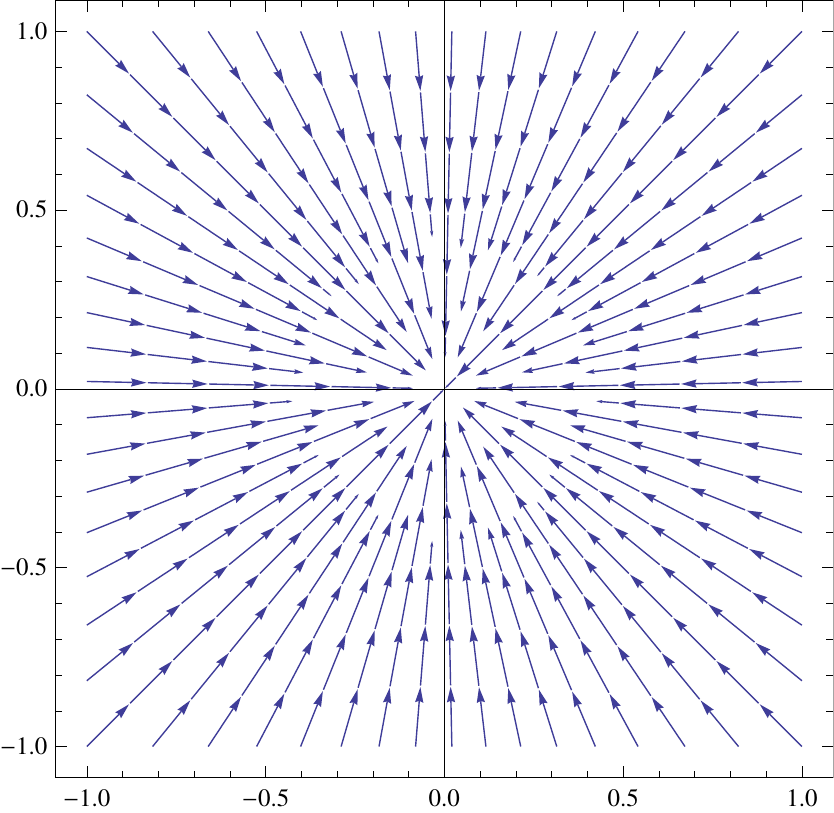}
	\caption{Phase portrait of the system $\dot{x}=-x,\;\dot{y}=-y$, that is, $\lambda=\mu<0$. The equilibrium point is a \textbf{stable node} at 
	the origin and in 
	that case it is called a \textbf{proper node}. If $\lambda=\mu>0$, the arrows are reversed and the origin is an \textbf{unstable node}.}
	\label{proper_node}
\end{figure}

  \begin{figure}
	\centering
	\includegraphics[scale=0.85]{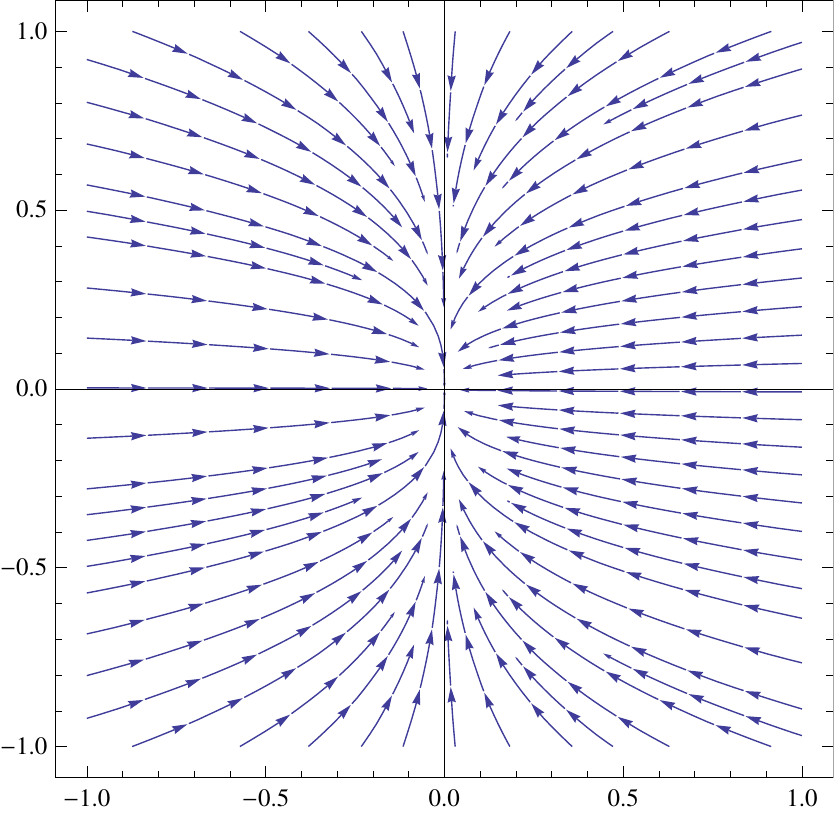}
	\caption{Phase portrait of the system $\dot{x}=-3x,\;\dot{y}=-y$, that is, $\lambda<\mu<0$. The equilibrium point is a \textbf{stable node} at
	the origin and in 
	that case it is called a \textbf{improper node}. If $\lambda>\mu>0$, the arrows are reversed and the origin is an \textbf{unstable node}. 
	}
	\label{improper_node1}
\end{figure}

  \begin{figure}
	\centering
	\includegraphics[scale=0.85]{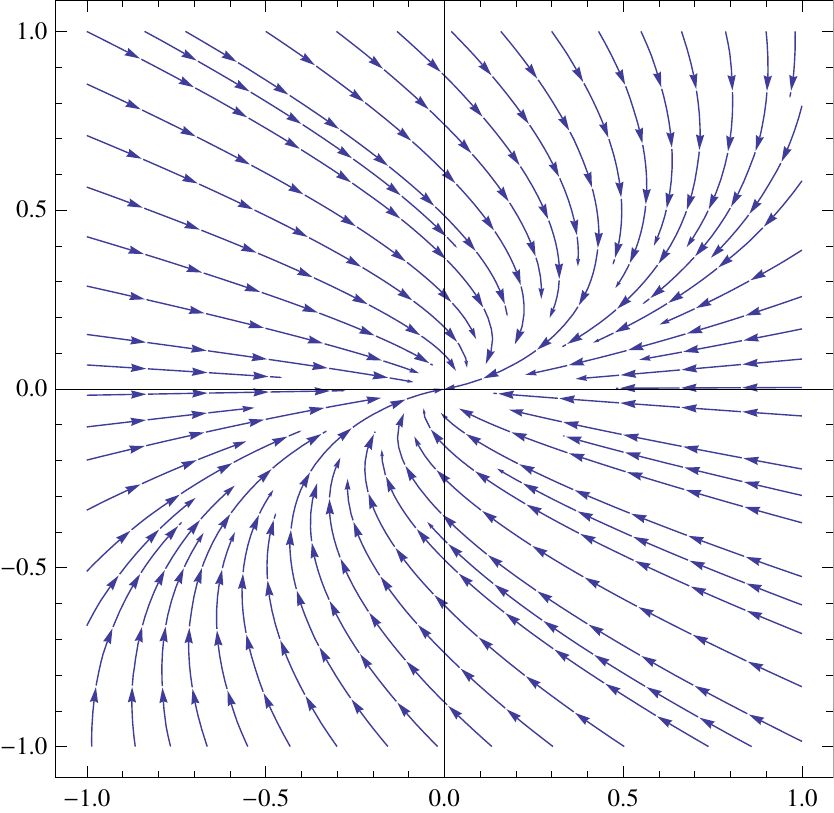}
	\caption{Phase portrait of the system $\dot{x}=-x+y,\;\dot{y}=-y$, that is, $\lambda<0$. The equilibrium point is a \textbf{stable node} at
	the origin and in 
	that case it is called a \textbf{improper node}. If $\lambda>0$, the arrows are reversed and the origin is an \textbf{unstable node}.}
	\label{improper_node2}
\end{figure}
 
 If one deals with matrices of the forms
 \begin{equation}
 B=\begin{bmatrix}
  \lambda & 0 \\
  0       & \mu 
 \end{bmatrix},\;\;\;\;\text{or}\;\;\;
 B=\begin{bmatrix}
  \lambda & 1 \\
  0       & \lambda 
 \end{bmatrix} 
 \end{equation}
with $\lambda\leq\mu<0$ for the first one and for $\lambda<0$ in the case of the second matrix, the phase portraits are shown in the figures 
\ref{proper_node}, \ref{improper_node1}, \ref{improper_node2}. The origin is a \textbf{stable node} in each of these cases; the case $\lambda=\mu$ 
is called a proper node (Fig. \ref{proper_node}) while in the two other cases (Fig. \ref{improper_node1} and Fig. \ref{improper_node2})
one deals with improper nodes. Moreover, one has the arrows in the pictures \ref{proper_node}, \ref{improper_node1}, \ref{improper_node2} reversed
if $\lambda\geq\mu>0$ or if $\lambda>0$ and the 
origins are \textbf{unstable nodes}. Let us notice that the stability of a node is given by the sign of the eigenvalues, that is, the node is stable if 
$\lambda\leq\mu<0$ and unstable if $\lambda\geq\mu>0$.

\begin{figure}
	\centering
	\includegraphics[scale=0.85]{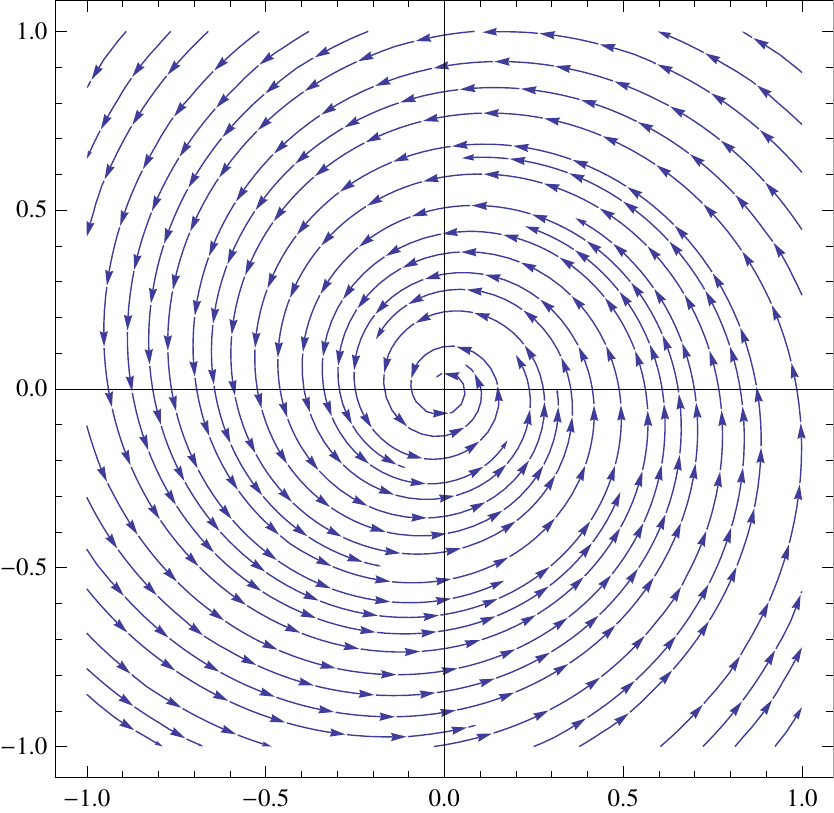}
	\caption{Phase portrait of the system $\dot{x}=-\frac{1}{2}x-3y,\;\dot{y}=3x-\frac{1}{2}y$, that is, $\lambda_{\pm}=a\pm ib$ with $a<0$ and $b>0$.
	The equilibrium point is a \textbf{stable focus} at
	the origin. If $a>0$, the arrows are reversed and the origin is an \textbf{unstable focus}.}
	\label{stable_focus}
\end{figure}

Another case includes a pair of complex conjugate eigenvalues of the matrix $A$ with nonzero real part. The diagonal matrix $B$ of the matric $A$ is then
\begin{equation}
  B=\begin{bmatrix}
  a & -b \\
  b & a 
 \end{bmatrix}\;\;\;\;\text{with}\;\;\;\;a<0.
\end{equation}
The phase portrait of the system (\ref{linsys1}) for $b>0$ (counterclockwise direction) is drawn in the figure \ref{stable_focus}. The clockwise direction 
happens when $b<0$. The origin is a \textbf{stable focus}; it will be \textbf{unstable}, that is, 
the trajectories will spiral away from the 
origin (with increasing $t$) if $a>0$. 

The last case refers to
\begin{equation}
   B=\begin{bmatrix}
  0 & -b \\
  b & 0 
 \end{bmatrix}.
\end{equation}
We say that the system (\ref{linsys}) has a center in the origin (see the figure \ref{center} for $b>0$). It happens when the matrix $A$ has a pair of pure imaginary 
complex conjugate eigenvalues $\lambda_{\pm}=\pm ib$.
The trajectories lie on circles $|\mathbf{x}(t)|=\text{constant}$. If one or both of the eigenvalues of $A$ is zero ($\mathrm{det}\,A=0$), the origin is 
called a degenerate equilibrium point of (\ref{linsys1}).
  \begin{figure}
	\centering
	\includegraphics[scale=0.85]{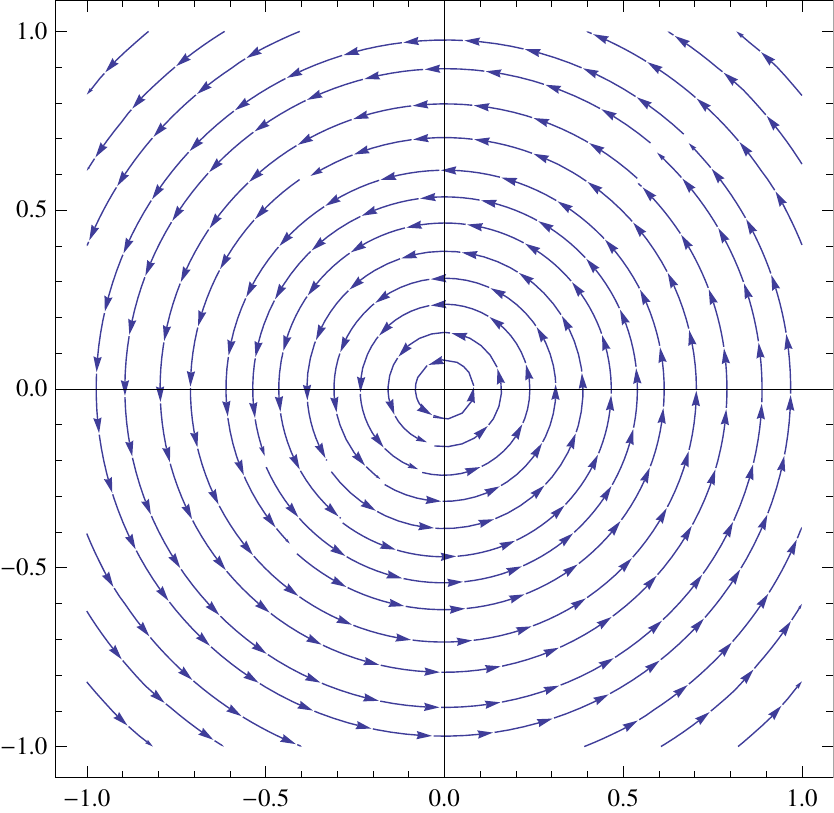}
	\caption{Phase portrait of the system $\dot{x}=-y,\;\dot{y}=x$, that is, $\lambda_{\pm}=\pm ib$ with $b>0$ (counterclockwise direction).
	The system is said to have a \textbf{center} at
	the origin. If $b<0$, the arrows are reversed (clockwise direction).}
	\label{center}
\end{figure}

Concluding, the linear system (\ref{linsys1}) has one of the following possibilities: a saddle, a node, a focus or a center at the origin if the matrix $A$ is similar to one of the matrices B
considered above, that is, its phase portrait is linearly equivalent to one of the phase portraits of the linear system (\ref{linsys}). One may easily 
determine a kind of an equilibrium point if $\mathrm{det}\,A\neq0$; for that purpose let us recall a theorem \cite{perko}

\begin{theorem}
 Let $\delta=\mathrm{det}\,A$ is a determinant of the matrix $A$ while $\tau=\mathrm{Tr}\,A$ is its trace. Considering the linear system (\ref{linsys1}) one says that
 \begin{itemize}
  \item The system (\ref{linsys1}) has a saddle point at the origin if $\delta<0$.
  \item If $\delta>0$ and $\tau^2-4\delta\geq0$ then (\ref{linsys1}) has a node at the origin: it is stable if $\tau<0$ and unstable if $\tau>0$.
  \item If $\delta>0$,  $\tau^2-4\delta<0$, and $\tau\neq0$ then (\ref{linsys1}) has a focus at the origin which is stable 
  if $\tau<0$ and unstable if $\tau>0$.
  \item If $\delta>0$ and $\tau=0$ then (\ref{linsys1}) has a center at the origin.
 \end{itemize} 
\end{theorem}
Moreover, we will say that a stable node or focus of (\ref{linsys1}) is a \textbf{sink} of the linear system while an unstable node or focus of (\ref{linsys1}) 
is called a \textbf{source} of the linear system.

In the further considerations we will be interested in Newtonian equations of motion
\begin{equation}
 \ddot{x}=-\frac{\partial V}{\partial x}
\end{equation}
with the first integral
\begin{equation}
 \frac{\dot{x}^2}{2}+V(x)=E.
\end{equation}
Such a system describes a unit-mass particle moving in a $1$-dimensional potential energy $V(x)$ with energy level $E$ on a half line $x:\,x\leq0$. It can
be reduced to the $2$-dimensional Newtonian type dynamical system
\begin{align}
 \dot{x}&=y,\nonumber\\
 \dot{y}&=-\frac{\partial V}{\partial x}, \\
 E&=\frac{y^2}{2}+V(x)\label{sys_energy}.
\end{align}
The system's matrix $A$ after the linearization procedure becomes
\begin{equation}
   A=\begin{bmatrix}
  0 & 1 \\
  -\frac{\partial^2V}{\partial x^2} & 0 
 \end{bmatrix}.
\end{equation}
while the characteristic equation simply gives the eigenvalues
\begin{equation}
 \lambda_\pm=\pm\frac{\partial^2V}{\partial x^2}.
\end{equation}
In the following subsection we will show how the dynamical system approach may be used to examine cosmological models. We will focus on the $\Lambda$CDM model 
which was briefly deiscussed in the Introduction \ref{introduction}.

\section{LCDM model as a dynamical system}

Let us recall the main ingredients of the $\Lambda$CDM model describing our Universe pretty well from radiation dominated epoch till 
nowadays accelerating expansion. On the large scale it is isotropic and homogeneous - that feature is described by the spatially flat $(k=0)$ FRLW
metric (\ref{frlw}) - and filled with pressureless substance, that is, dust. Additionally, one introduces another one possessing negative 
pressure, so-called cosmological constant $\Lambda$, in order to explain the late time acceleration. The dynamics of the model consists of the following
equations ($H=\dot{a}/a$)
\begin{align}\label{lcdmsystem}
 \ddot{a}=-\frac{1}{6}(\rho+3p)a=-\frac{\partial V}{\partial a},\\
 \dot{\rho}=-3H(\rho+p),\\
 H^2=\frac{1}{3}\rho-\frac{k}{a^2},
\end{align}
where the first equation (Raychaudhuri equation) comes from the Einstein's field equations $(j,j),\,j=1,2,3$ for the perfect fluid energy momentum
tensor (\ref{perfect}), the second one is a result of Bianchi identity (conservation of the energy-momentum tensor) while the last one is the 
first integral (Friedmann equation) of the two first equations. Together with the equation of state $p_i=\omega_i\rho_i,\,i=m,\Lambda$ one finds that the potential is of 
the form $V=-\frac{1}{6}a^2\rho$.
The energy density $\rho$ consists of two fluids: dust
($\omega_m=0$ concluding baryonic matter and cold dark matter) and cosmological constant ($\omega_\Lambda=-1$)
\begin{equation}
 \rho=\rho_m+\rho_\Lambda=\rho_{m,0}a^{-3}+\Lambda
\end{equation}
and therefore the potential is 
\begin{equation}\label{pot_lcdm}
 V=-\frac{1}{6}(\rho_{m,0}a^{-1}+\Lambda a^2).
\end{equation}
Let us remind that quantities labeled by the index $'0'$ correspond to the present epoch values.
 We have neglected the radiation density parameter as it is very small today ($\Omega_{red}\sim10^{-5}$).
  \begin{figure}
	\centering
	\includegraphics[scale=1.00]{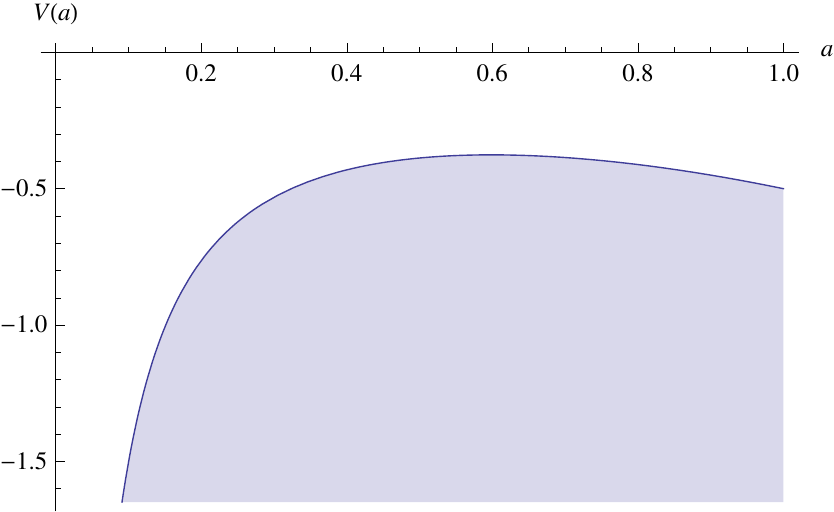}
	\caption{The potential $V(a)=-\frac{1}{2}(\Omega_{m,0} a^{-1}+\Omega_{\Lambda,0}a^2)$ of the $\Lambda$CDM model. In
	this example we have used the approximated values $\Omega_{m,0}=0.3,\;\Omega_{\Lambda,0}=0.7$. The shaded domain $E-V<0$, $E$ being total energy of 
	the system, is forbidden for a classical particle.}
	\label{center}\label{LCDM}
\end{figure}
  \begin{figure}
	\centering
	\includegraphics[scale=1.00]{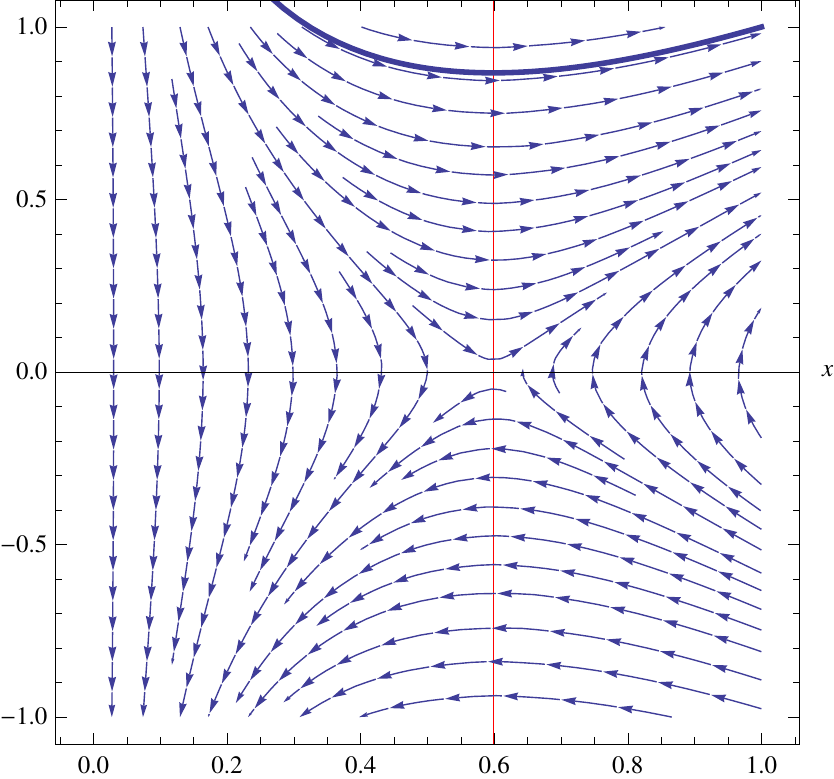}. 
	\caption{The phase portrait of the $\Lambda$CDM model obtained from (\ref{sys1}) and (\ref{sys2}). The thick line represents the flat model, it means
	the system with the total energy $E\sim\Omega_{k,0}=0$. The vertical line divides the trajectories into decelerating (left) and accelerating 
	(right) parts.}
	\label{center}\label{LCDMfaza}
\end{figure}
Having a look at the equations (\ref{lcdmsystem}) suggests that \cite{szydlowski2012generic} one may interpret 
cosmological evolutionary paths of $\Lambda$CDM model as a motion of a fictitious particle of unit mass. The motion takes place in configuration space 
$\{a:a\geq0\}$ in a one-dimensional potential parametrized by the scale factor $a$. The Universe accelerates when the potential is a decreasing function of 
the scale factor (Fig. \ref{LCDM}) while it decelerates when the potential grows. The extremum of the potential function corresponds to the zero 
acceleration case (static universe).

In order to obtain a $2$-dimensional dynamical system describing the considered cosmological model (\ref{lcdmsystem}) we need to replace
all dimensional quantities by dimensionless ones. We introduce density parameters as the values $\Omega_{i,0}=\frac{\rho_i}{3H^2_0}$. The quantity $H_0$ is a
present-day value of the Hubble's function equaled to $67.27\frac{\text{km}}{\text{sMpc}}$ \cite{planck2015planck}. Furthermore, we define a dimensionless scale 
factor $x=\frac{a}{a_0}$ measuring the value of $a$ in the units of the present value $a_0$ and the 
reparametrized cosmological time $t\rightarrow\tau:\;dt H_0=d\tau$. Then we may write:
\begin{align}
 \frac{dx}{d\tau}&=y,\label{sys1}\\
 \frac{dy}{d\tau}&=-\frac{\partial V}{\partial x},\label{sys2}\\
\frac{y^2}{2}+V(x)&=E\nonumber
\end{align}
where now
\begin{align*}
V(x)=-\frac{1}{2}(\Omega_{m,0}x^{-3}+\Omega_{\Lambda,0})x^2,\;\;\;\;E=\frac{1}{2}\Omega_{k,0}.
\end{align*}
One should mention that the density parameters $\Omega_{i,0}$ are not independent. They satisfy the constraint coming from $\frac{H^2}{H_0}=1$ for $a=1$ (as we assume 
that value for a present day). Hence, one has only one parameter to estimate, for example $\Omega_{\Lambda,0}=1-\Omega_{m,0}$.

The phase portrait of the above system is drawn in the picture \ref{LCDMfaza}.
 The critical point of the system is a saddle point $[a_{static}\sim0.6;0]$ which 
 represents the Einstein static universe - the extremum of the potential. A vertical line passing through the saddle point divides each trajectory 
 into two parts: decelerating phase of the Universe (on the 
 left from the critical point $a_{static}$) and accelerating one (on the right from the saddle point). It also shows that the model includes a singularity at 
 the origin $a_{\text{sing}}(t=0)=0$.
 
 The above example shows how the dynamical system theory may be used for cosmological purposes. $\Lambda$CDM system is rather simple one while modified theories 
 of gravity usually posses Friedmann evolutionary equations of the form much more complicated than presented here. They are
first order ordinary non-linear differential equations on the scale factor $a(t)$. Therefore,
 considering a geometrical structure of a phase space may help us to understand the evolution of a universe described by a model under consideration.
In order to perform
such   analysis,   one   needs   to   represent dynamics   of     cosmological   models   in   terms   of
dynamical   system   theory.  Phase   diagrams   will   allow   to   find   critical   points   which   correspond   to
extremes of the effective potentials whose diagram will provide information about the velocity of
cosmic expansion and classically forbidden regions.



\cleardoublepage

\label{app:bibliography} 

\manualmark
\markboth{\spacedlowsmallcaps{\bibname}}{\spacedlowsmallcaps{\bibname}} 
\refstepcounter{dummy}

\addtocontents{toc}{\protect\vspace{\beforebibskip}} 
\addcontentsline{toc}{chapter}{\tocEntry{\bibname}}

\bibliographystyle{unsrtnat}

\bibliography{Bibliography} 

\begin{thebibliography}{201}
\providecommand{\natexlab}[1]{#1}
\providecommand{\url}[1]{\texttt{#1}}
\expandafter\ifx\csname urlstyle\endcsname\relax
  \providecommand{\doi}[1]{doi: #1}\else
  \providecommand{\doi}{doi: \begingroup \urlstyle{rm}\Url}\fi

\bibitem[Einstein(1915)]{einstein1915field}
Albert Einstein.
\newblock The field equations of gravitation.
\newblock \emph{Sitzungsber. Preuss. Akad. Wiss. Berlin (Math. Phys.) 844-847},
  1915.

\bibitem[Einstein(1916)]{einstein1916found}
Albert Einstein.
\newblock The foundation of the general theory of relativity.
\newblock \emph{Annalen Phys. 49, 769-822, [Annalen Phys. 14, 517(2005)]},
  1916.

\bibitem[Dyson et~al.(1920)Dyson, Eddington, and
  Davidson]{dyson1920determination}
Frank~W Dyson, Arthur~S Eddington, and Charles Davidson.
\newblock A determination of the deflection of light by the sun's gravitational
  field, from observations made at the total eclipse of may 29, 1919.
\newblock \emph{Philosophical Transactions of the Royal Society of London A:
  Mathematical, Physical and Engineering Sciences}, 220\penalty0
  (571-581):\penalty0 291--333, 1920.

\bibitem[Pound and Rebka~Jr(1959)]{pound1959gravitational}
Robert~V Pound and GA~Rebka~Jr.
\newblock Gravitational red-shift in nuclear resonance.
\newblock \emph{Physical Review Letters}, 3\penalty0 (9):\penalty0 439, 1959.

\bibitem[Pound and Rebka~Jr(1960)]{pound1960apparent}
Robert~V Pound and GA~Rebka~Jr.
\newblock Apparent weight of photons.
\newblock \emph{Physical Review Letters}, 4\penalty0 (7):\penalty0 337, 1960.

\bibitem[Hubble(1929)]{hubble1929relation}
Edwin Hubble.
\newblock A relation between distance and radial velocity among extra-galactic
  nebulae.
\newblock \emph{Proceedings of the National Academy of Sciences}, 15\penalty0
  (3):\penalty0 168--173, 1929.

\bibitem[Schwarzschild(1916)]{schwarzschild1916gravitationsfeld}
Karl Schwarzschild.
\newblock {\"U}ber das gravitationsfeld eines massenpunktes nach der
  einsteinschen theorie.
\newblock \emph{Sitzungsberichte der K{\"o}niglich Preu{\ss}ischen Akademie der
  Wissenschaften (Berlin), 1916, Seite 189-196}, 1:\penalty0 189--196, 1916.

\bibitem[Schwarzschild(1999)]{schwarzschild1999gravitational}
Karl Schwarzschild.
\newblock On the gravitational field of a mass point according to einstein's
  theory.
\newblock \emph{arXiv preprint physics/9905030}, 1999.

\bibitem[Abbott et~al.(2016)Abbott, Abbott, Abbott, Abernathy, Acernese,
  Ackley, Adams, Adams, Addesso, Adhikari, et~al.]{abbott2016observation}
BP~Abbott, Richard Abbott, TD~Abbott, MR~Abernathy, Fausto Acernese, Kendall
  Ackley, Carl Adams, Thomas Adams, Paolo Addesso, RX~Adhikari, et~al.
\newblock Observation of gravitational waves from a binary black hole merger.
\newblock \emph{Physical review letters}, 116\penalty0 (6):\penalty0 061102,
  2016.

\bibitem[Copeland et~al.(2006)Copeland, Sami, and
  Tsujikawa]{copeland2006dynamics}
Edmund~J Copeland, Mohammad Sami, and Shinji Tsujikawa.
\newblock Dynamics of dark energy.
\newblock \emph{International Journal of Modern Physics D}, 15\penalty0
  (11):\penalty0 1753--1935, 2006.

\bibitem[Huterer and Turner(1999)]{huterer1999prospects}
Dragan Huterer and Michael~S Turner.
\newblock Prospects for probing the dark energy via supernova distance
  measurements.
\newblock \emph{Physical Review D}, 60\penalty0 (8):\penalty0 081301, 1999.

\bibitem[Ade et~al.(2014)Ade, Aghanim, Alves, Armitage-Caplan, Arnaud, Ashdown,
  Atrio-Barandela, Aumont, Aussel, Baccigalupi, et~al.]{ade2014planck}
Peter~AR Ade, N~Aghanim, MIR Alves, C~Armitage-Caplan, M~Arnaud, M~Ashdown,
  F~Atrio-Barandela, J~Aumont, H~Aussel, C~Baccigalupi, et~al.
\newblock Planck 2013 results. i. overview of products and scientific results.
\newblock \emph{Astronomy \& Astrophysics}, 571:\penalty0 A1, 2014.

\bibitem[Capozziello and De~Laurentis(2011)]{report}
Salvatore Capozziello and Mariafelicia De~Laurentis.
\newblock Extended theories of gravity.
\newblock \emph{Physics Reports}, 509\penalty0 (4):\penalty0 167--321, 2011.

\bibitem[Capozziello and Faraoni(2010)]{capozziello2010beyond}
Salvatore Capozziello and Valerio Faraoni.
\newblock \emph{Beyond Einstein gravity: A Survey of gravitational theories for
  cosmology and astrophysics}, volume 170.
\newblock Springer Science and Business Media, 2010.

\bibitem[Starobinsky(1980)]{starobinsky1980new}
Alexei~A Starobinsky.
\newblock A new type of isotropic cosmological models without singularity.
\newblock \emph{Physics Letters B}, 91\penalty0 (1):\penalty0 99--102, 1980.

\bibitem[Guth(1981)]{guth1981inflationary}
Alan~H Guth.
\newblock Inflationary universe: A possible solution to the horizon and
  flatness problems.
\newblock \emph{Physical Review D}, 23\penalty0 (2):\penalty0 347, 1981.

\bibitem[Bull et~al.(2015)Bull, Akrami, Adamek, Baker, Bellini, Jim{\'e}nez,
  Bentivegna, Camera, Clesse, Davis, et~al.]{bull2015beyond}
Philip Bull, Yashar Akrami, Julian Adamek, Tessa Baker, Emilio Bellini,
  Jose~Beltr{\'a}n Jim{\'e}nez, Eloisa Bentivegna, Stefano Camera,
  S{\'e}bastien Clesse, Jonathan~H Davis, et~al.
\newblock Beyond lcdm: Problems, solutions, and the road ahead.
\newblock \emph{arXiv preprint arXiv:1512.05356}, 2015.

\bibitem[Clarkson and Maartens(2010)]{clarkson2010inhomogeneity}
Chris Clarkson and Roy Maartens.
\newblock Inhomogeneity and the foundations of concordance cosmology.
\newblock \emph{Classical and Quantum Gravity}, 27\penalty0 (12):\penalty0
  124008, 2010.

\bibitem[Ellis et~al.(1978)Ellis, Maartens, and Nel]{ellis1978expansion}
GFR Ellis, R~Maartens, and SD~Nel.
\newblock The expansion of the universe.
\newblock \emph{Monthly Notices of the Royal Astronomical Society},
  184\penalty0 (3):\penalty0 439--465, 1978.

\bibitem[Will(1981)]{will1981theory}
CM~Will.
\newblock Theory and experiment in gravitational physics, 1981.

\bibitem[De~Felice and Tsujikawa(2010)]{de2010f}
Antonio De~Felice and Shinji Tsujikawa.
\newblock f(r) theories.
\newblock \emph{Living Rev. Rel}, 13\penalty0 (3):\penalty0 1002--4928, 2010.

\bibitem[Capozziello et~al.(2006)Capozziello, Cardone, and
  Troisi]{capozziello2006dark}
Salvatore Capozziello, VF~Cardone, and A~Troisi.
\newblock Dark energy and dark matter as curvature effects?
\newblock \emph{Journal of Cosmology and Astroparticle Physics}, 2006\penalty0
  (08):\penalty0 001, 2006.

\bibitem[Capozziello et~al.(2007)Capozziello, Cardone, and
  Troisi]{capozziello2007low}
Salvatore Capozziello, Vincenzo~F Cardone, and Antonio Troisi.
\newblock Low surface brightness galaxy rotation curves in the low energy limit
  of rn gravity: no need for dark matter?
\newblock \emph{Monthly Notices of the Royal Astronomical Society},
  375\penalty0 (4):\penalty0 1423--1440, 2007.

\bibitem[Sotiriou and Faraoni(2010)]{sotiriou2010f}
Thomas~P Sotiriou and Valerio Faraoni.
\newblock f(r) theories of gravity.
\newblock \emph{Reviews of Modern Physics}, 82\penalty0 (1):\penalty0 451,
  2010.

\bibitem[Borowiec et~al.(2012)Borowiec, Kamionka, Kurek, and
  Szyd{\l}owski]{bor_kam}
Andrzej Borowiec, Micha{\l} Kamionka, Aleksandra Kurek, and Marek
  Szyd{\l}owski.
\newblock Cosmic acceleration from modified gravity with palatini formalism.
\newblock \emph{Journal of Cosmology and Astroparticle Physics}, 2012\penalty0
  (02):\penalty0 027, 2012.

\bibitem[Flanagan(2004)]{flanagan2004palatini}
Eanna~E Flanagan.
\newblock Palatini form of 1/r gravity.
\newblock \emph{Physical review letters}, 92\penalty0 (7):\penalty0 071101,
  2004.

\bibitem[Iglesias et~al.(2007)Iglesias, Kaloper, Padilla, and
  Park]{iglesias2007not}
Alberto Iglesias, Nemanja Kaloper, Antonio Padilla, and Minjoon Park.
\newblock How (not) to use the palatini formulation of scalar-tensor gravity.
\newblock \emph{Physical Review D}, 76\penalty0 (10):\penalty0 104001, 2007.

\bibitem[Olmo(2008)]{olmo2008hydrogen}
Gonzalo~J Olmo.
\newblock Hydrogen atom in palatini theories of gravity.
\newblock \emph{Physical Review D}, 77\penalty0 (8):\penalty0 084021, 2008.

\bibitem[Barausse et~al.(2008)Barausse, Sotiriou, and
  Miller]{barausse2008curvature}
Enrico Barausse, Thomas~P Sotiriou, and John~C Miller.
\newblock Curvature singularities, tidal forces and the viability of palatini f
  (r) gravity.
\newblock \emph{Classical and Quantum Gravity}, 25\penalty0 (10):\penalty0
  105008, 2008.

\bibitem[Olmo(2005)]{olmo2005gravity}
Gonzalo~J Olmo.
\newblock The gravity lagrangian according to solar system experiments.
\newblock \emph{Physical review letters}, 95\penalty0 (26):\penalty0 261102,
  2005.

\bibitem[Sotiriou(2006{\natexlab{a}})]{sotiriou2006nearly}
Thomas~P Sotiriou.
\newblock The nearly newtonian regime in non-linear theories of gravity.
\newblock \emph{General Relativity and Gravitation}, 38\penalty0 (9):\penalty0
  1407--1417, 2006{\natexlab{a}}.

\bibitem[Ferraris et~al.(1994)Ferraris, Francaviglia, and
  Volovich]{ferraris1994universality}
Marco Ferraris, Mauro Francaviglia, and Igor Volovich.
\newblock The universality of vacuum einstein equations with cosmological
  constant.
\newblock \emph{Classical and Quantum Gravity}, 11\penalty0 (6):\penalty0 1505,
  1994.

\bibitem[Sotiriou(2006{\natexlab{b}})]{sotiriou2006f}
Thomas~P Sotiriou.
\newblock f(r) gravity and scalar - tensor theory.
\newblock \emph{Classical and Quantum Gravity}, 23\penalty0 (17):\penalty0
  5117, 2006{\natexlab{b}}.

\bibitem[Olmo and Sanchis-Alepuz(2011)]{olmo2011hamiltonian}
Gonzalo~J Olmo and Helios Sanchis-Alepuz.
\newblock Hamiltonian formulation of palatini f(r) theories {\`a} la
  brans-dicke theory.
\newblock \emph{Physical Review D}, 83\penalty0 (10):\penalty0 104036, 2011.

\bibitem[Olmo et~al.(2009)Olmo, Sanchis-Alepuz, and
  Tripathi]{olmo2009dynamical}
Gonzalo~J Olmo, Helios Sanchis-Alepuz, and Swapnil Tripathi.
\newblock Dynamical aspects of generalized palatini theories of gravity.
\newblock \emph{Physical Review D}, 80\penalty0 (2):\penalty0 024013, 2009.

\bibitem[Olmo and Singh(2009)]{olmo2009covariant}
Gonzalo~J Olmo and Parampreet Singh.
\newblock Covariant effective action for loop quantum cosmology {\`a} palatini.
\newblock \emph{Journal of Cosmology and Astroparticle Physics}, 2009\penalty0
  (01):\penalty0 030, 2009.

\bibitem[Stelle(1977)]{stelle1977renormalization}
KS~Stelle.
\newblock Renormalization of higher-derivative quantum gravity.
\newblock \emph{Physical Review D}, 16\penalty0 (4):\penalty0 953, 1977.

\bibitem[Barraco et~al.(1999)Barraco, Dominguez, and
  Guibert]{barraco1999conservation}
DE~Barraco, E~Dominguez, and R~Guibert.
\newblock Conservation laws, symmetry properties, and the equivalence principle
  in a class of alternative theories of gravity.
\newblock \emph{Physical Review D}, 60\penalty0 (4):\penalty0 044012, 1999.

\bibitem[Koivisto(2006{\natexlab{a}})]{koivisto2006note}
Tomi Koivisto.
\newblock A note on covariant conservation of energy--momentum in modified
  gravities.
\newblock \emph{Classical and Quantum Gravity}, 23\penalty0 (12):\penalty0
  4289, 2006{\natexlab{a}}.

\bibitem[Capozziello et~al.(2009{\natexlab{a}})Capozziello, De~Laurentis,
  Francaviglia, and Mercadante]{capozziello2009dark}
Salvatore Capozziello, Mariafelicia De~Laurentis, M~Francaviglia, and
  S~Mercadante.
\newblock From dark energy and dark matter to dark metric.
\newblock \emph{Foundations of Physics}, 39\penalty0 (10):\penalty0 1161--1176,
  2009{\natexlab{a}}.

\bibitem[Capozziello et~al.(2015{\natexlab{a}})Capozziello, De~Laurentis,
  Fatibene, Ferraris, and Garruto]{capozziello2015extended}
S~Capozziello, MF~De~Laurentis, L~Fatibene, M~Ferraris, and S~Garruto.
\newblock Extended cosmologies.
\newblock \emph{arXiv preprint arXiv:1509.08008}, 2015{\natexlab{a}}.

\bibitem[Fatibene and Francaviglia(2013)]{fatibene2013mathematical}
Lorenzo Fatibene and Mauro Francaviglia.
\newblock Mathematical equivalence vs. physical equivalence between extended
  theories of gravitations.
\newblock \emph{arXiv preprint arXiv:1302.2938}, 2013.

\bibitem[Capozziello et~al.(2015{\natexlab{b}})Capozziello, Fatibene, and
  Garruto]{capozziello2015equivalence}
S~Capozziello, L~Fatibene, and S~Garruto.
\newblock Equivalence among frames in extended gravity.
\newblock \emph{arXiv preprint arXiv:1512.08535}, 2015{\natexlab{b}}.

\bibitem[Fatibene et~al.(2012)Fatibene, Francaviglia, and
  Magnano]{fatibene2012characterization}
L~Fatibene, M~Francaviglia, and G~Magnano.
\newblock On a characterization of geodesic trajectories and gravitational
  motions.
\newblock \emph{International Journal of Geometric Methods in Modern Physics},
  9\penalty0 (05):\penalty0 1220007, 2012.

\bibitem[Einstein et~al.(1989)Einstein, Beck, and Havas]{einstein1989swiss}
Albert Einstein, Anna Beck, and Peter Havas.
\newblock \emph{The Swiss Years: Writing 1900-1909:...}
\newblock Princeton University Press, 1989.

\bibitem[Misner et~al.(1973)Misner, Thorne, and Wheeler]{misner1973gravitation}
Charles~W Misner, Kip~S Thorne, and John~Archibald Wheeler.
\newblock \emph{Gravitation}.
\newblock Macmillan, 1973.

\bibitem[Allemandi et~al.(2005)Allemandi, Borowiec, Francaviglia, and
  Odintsov]{alle_bor1}
Gianluca Allemandi, Andrzej Borowiec, Mauro Francaviglia, and Sergei~D
  Odintsov.
\newblock Dark energy dominance and cosmic acceleration in first-order
  formalism.
\newblock \emph{Physical Review D}, 72\penalty0 (6):\penalty0 063505, 2005.

\bibitem[Allemandi et~al.(2004{\natexlab{a}})Allemandi, Borowiec, and
  Francaviglia]{alle_bor2}
Gianluca Allemandi, Andrzej Borowiec, and Mauro Francaviglia.
\newblock Accelerated cosmological models in first-order nonlinear gravity.
\newblock \emph{Physical Review D}, 70\penalty0 (4):\penalty0 043524,
  2004{\natexlab{a}}.

\bibitem[Allemandi et~al.(2004{\natexlab{b}})Allemandi, Borowiec, and
  Francaviglia]{alle_bor3}
Gianluca Allemandi, Andrzej Borowiec, and Mauro Francaviglia.
\newblock Accelerated cosmological models in ricci squared gravity.
\newblock \emph{Physical Review D}, 70\penalty0 (10):\penalty0 103503,
  2004{\natexlab{b}}.

\bibitem[Allemandi et~al.(2006)Allemandi, Capone, Capozziello, and
  Francaviglia]{allemandi2006conformal}
Gianluca Allemandi, Monica Capone, Salvatore Capozziello, and Mauro
  Francaviglia.
\newblock Conformal aspects of the palatini approach in extended theories of
  gravity.
\newblock \emph{General Relativity and Gravitation}, 38\penalty0 (1):\penalty0
  33--60, 2006.

\bibitem[Faraoni and Gunzig(1999)]{faraoni1999einstein}
Valerio Faraoni and Edgard Gunzig.
\newblock Einstein frame or jordan frame?
\newblock \emph{International journal of theoretical physics}, 38\penalty0
  (1):\penalty0 217--225, 1999.

\bibitem[Jackiw and Polychronakos(1999)]{jackiw1999fluid}
R~Jackiw and AP~Polychronakos.
\newblock Fluid dynamical profiles and constants of motion from d-branes.
\newblock \emph{Communications in mathematical physics}, 207\penalty0
  (1):\penalty0 107--129, 1999.

\bibitem[Ogawa(2000)]{ogawa2000remark}
Naohisa Ogawa.
\newblock Remark on the classical solution of the chaplygin gas as d-branes.
\newblock \emph{Physical Review D}, 62\penalty0 (8):\penalty0 085023, 2000.

\bibitem[Chaplygin(1904)]{chap}
Sergey Chaplygin.
\newblock On gas jets.
\newblock \emph{Sci. Mem. Moscow Univ. Math. Phys.}, 21\penalty0 (1), 1904.

\bibitem[Kamenshchik et~al.(2001)Kamenshchik, Moschella, and
  Pasquier]{kamenshchik2001alternative}
Alexander Kamenshchik, Ugo Moschella, and Vincent Pasquier.
\newblock An alternative to quintessence.
\newblock \emph{Physics Letters B}, 511\penalty0 (2):\penalty0 265--268, 2001.

\bibitem[Bento et~al.(2002)Bento, Bertolami, and Sen]{bento}
MC~Bento, O~Bertolami, and AA~Sen.
\newblock Generalized chaplygin gas, accelerated expansion, and
  dark-energy-matter unification.
\newblock \emph{Physical Review D}, 66\penalty0 (4):\penalty0 043507, 2002.

\bibitem[Lu(2009)]{lu2009cosmology}
Jianbo Lu.
\newblock Cosmology with a variable generalized chaplygin gas.
\newblock \emph{Physics Letters B}, 680\penalty0 (5):\penalty0 404--410, 2009.

\bibitem[Bili{\'c} et~al.(2002)Bili{\'c}, Tupper, and
  Viollier]{bilic2002unification}
Neven Bili{\'c}, Gary~B Tupper, and Raoul~D Viollier.
\newblock Unification of dark matter and dark energy: the inhomogeneous
  chaplygin gas.
\newblock \emph{Physics Letters B}, 535\penalty0 (1):\penalty0 17--21, 2002.

\bibitem[Popov(2010)]{popov2010dark}
VA~Popov.
\newblock Dark energy and dark matter unification via superfluid chaplygin gas.
\newblock \emph{Physics Letters B}, 686\penalty0 (4):\penalty0 211--215, 2010.

\bibitem[Naji et~al.(2014)Naji, Pourhassan, and Amani]{naji2014effect}
J~Naji, B~Pourhassan, and Ali~R Amani.
\newblock Effect of shear and bulk viscosities on interacting modified
  chaplygin gas cosmology.
\newblock \emph{International Journal of Modern Physics D}, 23\penalty0
  (02):\penalty0 1450020, 2014.

\bibitem[Kremer and Alves(2004)]{kremer2004palatini}
Gilberto~M Kremer and Daniele~SM Alves.
\newblock Palatini approach to 1/r gravity and its implications to the late
  universe.
\newblock \emph{Physical Review D}, 70\penalty0 (2):\penalty0 023503, 2004.

\bibitem[Gorini et~al.(2003)Gorini, Kamenshchik, and Moschella]{gorini2003can}
Vittorio Gorini, Alexander Kamenshchik, and Ugo Moschella.
\newblock Can the chaplygin gas be a plausible model for dark energy?
\newblock \emph{Physical Review D}, 67\penalty0 (6):\penalty0 063509, 2003.

\bibitem[Avelino et~al.(2014)Avelino, Bolejko, and Lewis]{avelino2014nonlinear}
PP~Avelino, K~Bolejko, and GF~Lewis.
\newblock Nonlinear chaplygin gas cosmologies.
\newblock \emph{Physical Review D}, 89\penalty0 (10):\penalty0 103004, 2014.

\bibitem[Kahya and Pourhassan(2015)]{kahya2015universe}
EO~Kahya and B~Pourhassan.
\newblock The universe dominated by the extended chaplygin gas.
\newblock \emph{Modern Physics Letters A}, 30\penalty0 (13):\penalty0 1550070,
  2015.

\bibitem[Fabris et~al.(2011)Fabris, Velten, Ogouyandjou, and
  Tossa]{fabris2011ruling}
JC~Fabris, HES Velten, C~Ogouyandjou, and J~Tossa.
\newblock Ruling out the modified chaplygin gas cosmologies.
\newblock \emph{Physics Letters B}, 694\penalty0 (4):\penalty0 289--293, 2011.

\bibitem[Hoppe()]{hoppe9311059supermembranes}
Jens Hoppe.
\newblock Supermembranes in four-dimensions.
\newblock \emph{arXiv preprint hep-th/9311059}.

\bibitem[Jackiw and Polychronakos(2000)]{jackiw2000supersymmetric}
R~Jackiw and AP~Polychronakos.
\newblock Supersymmetric fluid mechanics.
\newblock \emph{Physical Review D}, 62\penalty0 (8):\penalty0 085019, 2000.

\bibitem[Benaoum()]{Benaoum}
H.~B. Benaoum.
\newblock Accelerated universe from modi ed chaplygin gas and tachyonic.
\newblock \emph{arXiv preprint hep-th/0205140}.

\bibitem[Chimento(2004)]{chimento2004extended}
Luis~P Chimento.
\newblock Extended tachyon field, chaplygin gas, and solvable k-essence
  cosmologies.
\newblock \emph{Physical Review D}, 69\penalty0 (12):\penalty0 123517, 2004.

\bibitem[Kamenshchik et~al.(2000)Kamenshchik, Moschella, and
  Pasquier]{kamenshchik2000chaplygin}
Alexander Kamenshchik, Ugo Moschella, and Vincent Pasquier.
\newblock Chaplygin-like gas and branes in black hole bulks.
\newblock \emph{Physics Letters B}, 487\penalty0 (1):\penalty0 7--13, 2000.

\bibitem[Collaboration et~al.(2015)]{planck2015planck}
Planck Collaboration et~al.
\newblock Planck 2015 results. xiii. cosmological parameters.
\newblock \emph{arXiv preprint arXiv:1502.01589}, 2015.

\bibitem[Borowiec et~al.(2016)Borowiec, Stachowski, Szyd{\l}owski, and
  Wojnar]{borowiec2016inflationary}
Andrzej Borowiec, Aleksander Stachowski, Marek Szyd{\l}owski, and Aneta Wojnar.
\newblock Inflationary cosmology with chaplygin gas in palatini formalism.
\newblock \emph{Journal of Cosmology and Astroparticle Physics}, 2016\penalty0
  (01):\penalty0 040, 2016.

\bibitem[Szydlowski et~al.(2015)Szydlowski, Stachowski, Borowiec, and
  Wojnar]{szyd_stach}
Marek Szydlowski, Aleksander Stachowski, Andrzej Borowiec, and Aneta Wojnar.
\newblock Do sewn singularities falsify the palatini cosmology?
\newblock \emph{arXiv preprint arXiv:1604.02632}, 2015.

\bibitem[Suzuki et~al.(2012)Suzuki, Rubin, Lidman, Aldering, Amanullah,
  Barbary, Barrientos, Botyanszki, Brodwin, Connolly, et~al.]{suzuki2012hubble}
N~Suzuki, D~Rubin, C~Lidman, G~Aldering, R~Amanullah, K~Barbary, LF~Barrientos,
  J~Botyanszki, M~Brodwin, N~Connolly, et~al.
\newblock The hubble space telescope cluster supernova survey. v. improving the
  dark-energy constraints above z> 1 and building an early-type-hosted
  supernova samplebased.
\newblock \emph{The Astrophysical Journal}, 746\penalty0 (1):\penalty0 85,
  2012.

\bibitem[Metropolis et~al.(1953)Metropolis, Rosenbluth, Rosenbluth, Teller, and
  Teller]{metropolis1953equation}
Nicholas Metropolis, Arianna~W Rosenbluth, Marshall~N Rosenbluth, Augusta~H
  Teller, and Edward Teller.
\newblock Equation of state calculations by fast computing machines.
\newblock \emph{The journal of chemical physics}, 21\penalty0 (6):\penalty0
  1087--1092, 1953.

\bibitem[Hastings(1970)]{hastings1970monte}
W~Keith Hastings.
\newblock Monte carlo sampling methods using markov chains and their
  applications.
\newblock \emph{Biometrika}, 57\penalty0 (1):\penalty0 97--109, 1970.

\bibitem[Schwarz et~al.(1978)]{schwarz1978estimating}
Gideon Schwarz et~al.
\newblock Estimating the dimension of a model.
\newblock \emph{The annals of statistics}, 6\penalty0 (2):\penalty0 461--464,
  1978.

\bibitem[Kass and Raftery(1995)]{kass1995bayes}
Robert~E Kass and Adrian~E Raftery.
\newblock Bayes factors.
\newblock \emph{Journal of the american statistical association}, 90\penalty0
  (430):\penalty0 773--795, 1995.

\bibitem[Nojiri et~al.(2005)Nojiri, Odintsov, and Tsujikawa]{Nojiri:2005sx}
Shin’ichi Nojiri, Sergei~D Odintsov, and Shinji Tsujikawa.
\newblock Properties of singularities in the (phantom) dark energy universe.
\newblock \emph{Physical Review D}, 71\penalty0 (6):\penalty0 063004, 2005.

\bibitem[Singh and Vidotto(2011)]{Singh:2010qa}
Parampreet Singh and Francesca Vidotto.
\newblock Exotic singularities and spatially curved loop quantum cosmology.
\newblock \emph{Physical Review D}, 83\penalty0 (6):\penalty0 064027, 2011.

\bibitem[Dabrowski(2014)]{dabrowski2014singularities}
Mariusz~P Dabrowski.
\newblock Are singularities the limits of cosmology?
\newblock \emph{arXiv preprint arXiv:1407.4851}, 2014.

\bibitem[D{\k{a}}browski and Denkiewicz(2009)]{dkabrowski2009barotropic}
Mariusz~P D{\k{a}}browski and Tomasz Denkiewicz.
\newblock Barotropic index w-singularities in cosmology.
\newblock \emph{Physical Review D}, 79\penalty0 (6):\penalty0 063521, 2009.

\bibitem[Frampton et~al.(2011)Frampton, Ludwick, and
  Scherrer]{frampton2011little}
Paul~H Frampton, Kevin~J Ludwick, and Robert~J Scherrer.
\newblock The little rip.
\newblock \emph{Physical Review D}, 84\penalty0 (6):\penalty0 063003, 2011.

\bibitem[Frampton et~al.(2012)Frampton, Ludwick, and
  Scherrer]{frampton2012pseudo}
Paul~H Frampton, Kevin~J Ludwick, and Robert~J Scherrer.
\newblock Pseudo-rip: Cosmological models intermediate between the cosmological
  constant and the little rip.
\newblock \emph{Physical Review D}, 85\penalty0 (8):\penalty0 083001, 2012.

\bibitem[Szyd{\l}owski(2007)]{zoo_marek}
Marek Szyd{\l}owski.
\newblock Cosmological zoo — accelerating models with dark energy.
\newblock \emph{Journal of Cosmology and Astroparticle Physics}, 2007\penalty0
  (09):\penalty0 007, 2007.

\bibitem[Nojiri et~al.(2015)Nojiri, Odintsov, and
  Oikonomou]{nojiri2015singular}
S~Nojiri, SD~Odintsov, and VK~Oikonomou.
\newblock Singular inflation from generalized equation of state fluids.
\newblock \emph{Physics Letters B}, 747:\penalty0 310--320, 2015.

\bibitem[Odintsov and Oikonomou(2015)]{odintsov2015singular}
SD~Odintsov and VK~Oikonomou.
\newblock Singular inflationary universe from f(r) gravity.
\newblock \emph{Physical Review D}, 92\penalty0 (12):\penalty0 124024, 2015.

\bibitem[Bautin and I.~A.~Leontovich(1976)]{Bautin:1976mt}
N.~N. Bautin and eds. I.~A.~Leontovich.
\newblock \emph{Methods and Techniques for Qualitative Analysis of Dynamical
  Systems on the Plane}, volume~1.
\newblock Nauka, Moscow [In Russian], 1976.

\bibitem[Herrera et~al.(2013)Herrera, Olivares, and
  Videla]{herrera2013intermediate}
Ram{\'o}n Herrera, Marco Olivares, and Nelson Videla.
\newblock Intermediate inflation on the brane and warped dgp models.
\newblock \emph{The European Physical Journal C}, 73\penalty0 (6):\penalty0
  1--9, 2013.

\bibitem[Barrow and Graham(2015)]{barrow2015singular}
John~D Barrow and Alexander~AH Graham.
\newblock Singular inflation.
\newblock \emph{Physical Review D}, 91\penalty0 (8):\penalty0 083513, 2015.

\bibitem[Kiefer(2010)]{kiefer2010avoidance}
Claus Kiefer.
\newblock On the avoidance of classical singularities in quantum cosmology.
\newblock In \emph{Journal of Physics: Conference Series}, volume 222, page
  012049. IOP Publishing, 2010.

\bibitem[Harko et~al.(2012)Harko, Koivisto, Lobo, and Olmo]{Harko:2011nh}
Tiberiu Harko, Tomi~S Koivisto, Francisco~SN Lobo, and Gonzalo~J Olmo.
\newblock Metric-palatini gravity unifying local constraints and late-time
  cosmic acceleration.
\newblock \emph{Physical Review D}, 85\penalty0 (8):\penalty0 084016, 2012.

\bibitem[Capozziello et~al.(2015{\natexlab{c}})Capozziello, Harko, Koivisto,
  Lobo, and Olmo]{capozziello2015hybrid}
Salvatore Capozziello, Tiberiu Harko, Tomi~S Koivisto, Francisco~SN Lobo, and
  Gonzalo~J Olmo.
\newblock Hybrid metric-palatini gravity.
\newblock \emph{Universe}, 1\penalty0 (2):\penalty0 199--238,
  2015{\natexlab{c}}.

\bibitem[Capozziello et~al.(2014{\natexlab{a}})Capozziello, Harko, Lobo, Olmo,
  and Vignolo]{capozziello2014cauchy}
Salvatore Capozziello, Tiberiu Harko, Francisco~SN Lobo, Gonzalo~J Olmo, and
  Stefano Vignolo.
\newblock The cauchy problem in hybrid metric-palatini f(x)-gravity.
\newblock \emph{International Journal of Geometric Methods in Modern Physics},
  11\penalty0 (05):\penalty0 1450042, 2014{\natexlab{a}}.

\bibitem[Capozziello et~al.(2012{\natexlab{a}})Capozziello, Harko, Koivisto,
  Lobo, and Olmo]{capozziello2012wormholes}
Salvatore Capozziello, Tiberiu Harko, Tomi~S Koivisto, Francisco~SN Lobo, and
  Gonzalo~J Olmo.
\newblock Wormholes supported by hybrid metric-palatini gravity.
\newblock \emph{Physical Review D}, 86\penalty0 (12):\penalty0 127504,
  2012{\natexlab{a}}.

\bibitem[Capozziello et~al.(2013{\natexlab{a}})Capozziello, Harko, Koivisto,
  Lobo, and Olmo]{Capoz}
Salvatore Capozziello, Tiberiu Harko, Tomi~S Koivisto, Francisco~SN Lobo, and
  Gonzalo~J Olmo.
\newblock Cosmology of hybrid metric-palatini f (x)-gravity.
\newblock \emph{Journal of Cosmology and Astroparticle Physics}, 2013\penalty0
  (04):\penalty0 011, 2013{\natexlab{a}}.

\bibitem[B{\"o}hmer et~al.(2013)B{\"o}hmer, Lobo, and
  Tamanini]{bohmer2013einstein}
Christian~G B{\"o}hmer, Francisco~SN Lobo, and Nicola Tamanini.
\newblock Einstein static universe in hybrid metric-palatini gravity.
\newblock \emph{Physical Review D}, 88\penalty0 (10):\penalty0 104019, 2013.

\bibitem[Lima(2014)]{lima2014dynamics}
Nelson~A Lima.
\newblock Dynamics of linear perturbations in the hybrid metric-palatini
  gravity.
\newblock \emph{Physical Review D}, 89\penalty0 (8):\penalty0 083527, 2014.

\bibitem[Lima et~al.(2016)Lima, Smer-Barreto, and
  Lombriser]{lima2016constraints}
Nelson~A Lima, Vanessa Smer-Barreto, and Lucas Lombriser.
\newblock Constraints on decaying early modified gravity from cosmological
  observations.
\newblock \emph{arXiv preprint arXiv:1603.05239}, 2016.

\bibitem[Carloni et~al.(2015)Carloni, Koivisto, and Lobo]{carloni2015dynamical}
Sante Carloni, Tomi Koivisto, and Francisco~SN Lobo.
\newblock Dynamical system analysis of hybrid metric-palatini cosmologies.
\newblock \emph{Physical Review D}, 92\penalty0 (6):\penalty0 064035, 2015.

\bibitem[Azizi and Borhani(2015)]{azizi2015thermodynamics}
Tahereh Azizi and Najibe Borhani.
\newblock Thermodynamics in hybrid metric-palatini gravity.
\newblock \emph{Astrophysics and Space Science}, 357\penalty0 (2):\penalty0
  1--9, 2015.

\bibitem[Capozziello et~al.(2013{\natexlab{b}})Capozziello, Harko, Koivisto,
  Lobo, and Olmo]{capozziello2013virial}
Salvatore Capozziello, Tiberiu Harko, Tomi~S Koivisto, Francisco~SN Lobo, and
  Gonzalo~J Olmo.
\newblock The virial theorem and the dark matter problem in hybrid
  metric-palatini gravity.
\newblock \emph{Journal of Cosmology and Astroparticle Physics}, 2013\penalty0
  (07):\penalty0 024, 2013{\natexlab{b}}.

\bibitem[Capozziello et~al.(2013{\natexlab{c}})Capozziello, Harko, Koivisto,
  Lobo, and Olmo]{capozziello2013galactic}
Salvatore Capozziello, Tiberiu Harko, Tomi~S Koivisto, Francisco~SN Lobo, and
  Gonzalo~J Olmo.
\newblock Galactic rotation curves in hybrid metric-palatini gravity.
\newblock \emph{Astroparticle Physics}, 50:\penalty0 65--75,
  2013{\natexlab{c}}.

\bibitem[Borka et~al.(2015)Borka, Capozziello, Jovanovi{\'c}, and
  Jovanovi{\'c}]{borka2015probing}
D~Borka, S~Capozziello, P~Jovanovi{\'c}, and V~Borka Jovanovi{\'c}.
\newblock Probing hybrid modified gravity by stellar motion around galactic
  centre.
\newblock \emph{arXiv preprint arXiv:1504.07832}, 2015.

\bibitem[Fu et~al.(2016)Fu, Zhao, and Liu]{fu2016hybrid}
Qi-Ming Fu, Li~Zhao, and Yu-Xiao Liu.
\newblock Hybrid metric-palatini brane system.
\newblock \emph{arXiv preprint arXiv:1601.06546}, 2016.

\bibitem[Tamanini and Boehmer(2013)]{tamanini2013generalized}
Nicola Tamanini and Christian~G Boehmer.
\newblock Generalized hybrid metric-palatini gravity.
\newblock \emph{Physical Review D}, 87\penalty0 (8):\penalty0 084031, 2013.

\bibitem[Koivisto and Tamanini(2013)]{koivisto2013ghosts}
Tomi~S Koivisto and Nicola Tamanini.
\newblock Ghosts in pure and hybrid formalisms of gravity theories: a unified
  analysis.
\newblock \emph{Physical Review D}, 87\penalty0 (10):\penalty0 104030, 2013.

\bibitem[Borowiec et~al.(2015)Borowiec, Capozziello, De~Laurentis, Lobo,
  Paliathanasis, Paolella, and Wojnar]{wojnar}
Andrzej Borowiec, Salvatore Capozziello, Mariafelicia De~Laurentis,
  Francisco~SN Lobo, Andronikos Paliathanasis, Mariacristina Paolella, and
  Aneta Wojnar.
\newblock Invariant solutions and noether symmetries in hybrid gravity.
\newblock \emph{Physical Review D}, 91\penalty0 (2):\penalty0 023517, 2015.

\bibitem[Paliathanasis et~al.(2016)Paliathanasis, Karpathopoulos, Wojnar, and
  Capozziello]{wojnar2}
A~Paliathanasis, L~Karpathopoulos, A~Wojnar, and S~Capozziello.
\newblock Wheeler--dewitt equation and lie symmetries in bianchi scalar-field
  cosmology.
\newblock \emph{The European Physical Journal C}, 76\penalty0 (4):\penalty0
  1--14, 2016.

\bibitem[Dabrowski et~al.(2009)Dabrowski, Garecki, and Blaschke]{dabrowski}
Mariusz~P Dabrowski, Janusz Garecki, and David~B Blaschke.
\newblock Conformal transformations and conformal invariance in gravitation.
\newblock \emph{Annalen der Physik}, 18\penalty0 (1):\penalty0 13--32, 2009.

\bibitem[Olmo(2011)]{olmo}
Gonzalo~J Olmo.
\newblock Palatini approach to modified gravity: f(r) theories and beyond.
\newblock \emph{International Journal of Modern Physics D}, 20\penalty0
  (04):\penalty0 413--462, 2011.

\bibitem[Capozziello and De~Felice(2008)]{capozziello2008f}
Salvatore Capozziello and Antonio De~Felice.
\newblock f(r) cosmology from noether’s symmetry.
\newblock \emph{Journal of Cosmology and Astroparticle Physics}, 2008\penalty0
  (08):\penalty0 016, 2008.

\bibitem[Capozziello et~al.(2009{\natexlab{b}})Capozziello, Piedipalumbo,
  Rubano, and Scudellaro]{capozziello2009noether}
Salvatore Capozziello, Ester Piedipalumbo, Claudio Rubano, and Paolo
  Scudellaro.
\newblock Noether symmetry approach in phantom quintessence cosmology.
\newblock \emph{Physical Review D}, 80\penalty0 (10):\penalty0 104030,
  2009{\natexlab{b}}.

\bibitem[Kucukakca and Camci(2012)]{kucukakca2012noether}
Y~Kucukakca and U~Camci.
\newblock Noether gauge symmetry for f(r) gravity in palatini formalism.
\newblock \emph{Astrophysics and Space Science}, 338\penalty0 (1):\penalty0
  211--216, 2012.

\bibitem[Roshan and Shojai(2008)]{roshan2008palatini}
Mahmood Roshan and Fatimah Shojai.
\newblock Palatini f(r) cosmology and noether symmetry.
\newblock \emph{Physics Letters B}, 668\penalty0 (3):\penalty0 238--240, 2008.

\bibitem[Zhang et~al.(2010)Zhang, Gong, and Zhu]{zhang2010noether}
Yi~Zhang, Yun-gui Gong, and Zong-Hong Zhu.
\newblock Noether symmetry approach in multiple scalar fields scenario.
\newblock \emph{Physics Letters B}, 688\penalty0 (1):\penalty0 13--20, 2010.

\bibitem[Vakili(2008{\natexlab{a}})]{vakili2008noether}
Babak Vakili.
\newblock Noether symmetry in f(r) cosmology.
\newblock \emph{Physics Letters B}, 664\penalty0 (1):\penalty0 16--20,
  2008{\natexlab{a}}.

\bibitem[Vakili(2008{\natexlab{b}})]{vakili2008noether2}
Babak Vakili.
\newblock Noether symmetric f(r) quantum cosmology and its classical
  correlations.
\newblock \emph{Physics Letters B}, 669\penalty0 (3):\penalty0 206--211,
  2008{\natexlab{b}}.

\bibitem[Capozziello and Lambiase(2000)]{capozziello2000selection}
S~Capozziello and Gen Lambiase.
\newblock Selection rules in minisuperspace quantum cosmology.
\newblock \emph{General Relativity and Gravitation}, 32\penalty0 (4):\penalty0
  673--696, 2000.

\bibitem[Arnowitt et~al.(1959)Arnowitt, Deser, and
  Misner]{arnowitt1959dynamical}
Richard Arnowitt, Stanley Deser, and Charles~W Misner.
\newblock Dynamical structure and definition of energy in general relativity.
\newblock \emph{Physical Review}, 116\penalty0 (5):\penalty0 1322, 1959.

\bibitem[Kiefer(2007)]{Kiefer2007ria}
Claus Kiefer.
\newblock \emph{{Quantum Gravity}}.
\newblock Oxford University Press, New York, 2007.
\newblock ISBN 9780199212521.

\bibitem[Gourgoulhon(2012)]{gourgoulhon}
Eric Gourgoulhon.
\newblock \emph{3+1 formalism in general relativity: bases of numerical
  relativity}, volume 846.
\newblock Springer Science \& Business Media, 2012.

\bibitem[Capozziello et~al.(2012{\natexlab{b}})Capozziello, De~Laurentis, and
  Odintsov]{capozziello2012hamiltonian}
Salvatore Capozziello, Mariafelicia De~Laurentis, and Sergei~D Odintsov.
\newblock Hamiltonian dynamics and noether symmetries in extended gravity
  cosmology.
\newblock \emph{The European Physical Journal C}, 72\penalty0 (7):\penalty0
  1--21, 2012{\natexlab{b}}.

\bibitem[Hartle(1986)]{hartle1986ingravitation}
JB~Hartle.
\newblock Gravitation in astrophysics (eds) b carter and jb hartle, 1986.

\bibitem[Tsamparlis and Paliathanasis(2011{\natexlab{a}})]{TsamAnd}
Michael Tsamparlis and Andronikos Paliathanasis.
\newblock Two-dimensional dynamical systems which admit lie and noether
  symmetries.
\newblock \emph{Journal of Physics A: Mathematical and Theoretical},
  44\penalty0 (17):\penalty0 175202, 2011{\natexlab{a}}.

\bibitem[Paliathanasis et~al.(2014)Paliathanasis, Tsamparlis, Basilakos, and
  Capozziello]{TsamC02}
Andronikos Paliathanasis, Michael Tsamparlis, Spyros Basilakos, and Salvatore
  Capozziello.
\newblock Scalar-tensor gravity cosmology: Noether symmetries and analytical
  solutions.
\newblock \emph{Physical Review D}, 89\penalty0 (6):\penalty0 063532, 2014.

\bibitem[Tsamparlis et~al.(2013)Tsamparlis, Paliathanasis, Basilakos, and
  Capozziello]{TsamC01}
Michael Tsamparlis, Andronikos Paliathanasis, Spyros Basilakos, and Salvatore
  Capozziello.
\newblock Conformally related metrics and lagrangians and their physical
  interpretation in cosmology.
\newblock \emph{General Relativity and Gravitation}, 45\penalty0 (10):\penalty0
  2003--2022, 2013.

\bibitem[Paliathanasis and Tsamparlis(2014{\natexlab{a}})]{AnIJGMP}
Andronikos Paliathanasis and Michael Tsamparlis.
\newblock The geometric origin of lie point symmetries of the schr{\"o}dinger
  and the klein--gordon equations.
\newblock \emph{International Journal of Geometric Methods in Modern Physics},
  11\penalty0 (04):\penalty0 1450037, 2014{\natexlab{a}}.

\bibitem[Dirac(1996)]{dirac1996general}
Paul Adrien~Maurice Dirac.
\newblock \emph{General theory of relativity}.
\newblock Princeton University Press, 1996.

\bibitem[Matschull(1996)]{matschull1996dirac}
Hans-Juergen Matschull.
\newblock Dirac's canonical quantization programme.
\newblock \emph{arXiv preprint quant-ph/9606031}, 1996.

\bibitem[DeWitt and Graham(2015)]{halliwell}
Bryce~Seligman DeWitt and Neill Graham.
\newblock \emph{The many worlds interpretation of quantum mechanics}.
\newblock Princeton University Press, 2015.

\bibitem[Stephani and MacCallum(1989)]{StephaniB}
Hans Stephani and Malcolm MacCallum.
\newblock \emph{Differential equations: their solution using symmetries}.
\newblock Cambridge University Press, 1989.

\bibitem[Hydon(2000)]{hydon}
Peter~E. Hydon.
\newblock \emph{Symmetry Methods for Differential Equations: A Beginner's
  Guide}.
\newblock Cambridge University Press, New York, USA, 1st edition, 2000.

\bibitem[Leach(1980)]{LeachOSc}
PGL Leach.
\newblock The complete symmetry group of the one-dimensional time-dependent
  harmonic oscillator.
\newblock \emph{Journal of Mathematical Physics}, 21\penalty0 (2):\penalty0
  300--304, 1980.

\bibitem[Abraham-Shrauner et~al.(2006)Abraham-Shrauner, Govinder, and
  Arrigo]{Abraham}
Barbara Abraham-Shrauner, Keshlan~S Govinder, and Daniel~J Arrigo.
\newblock Type-ii hidden symmetries of the linear 2d and 3d wave equations.
\newblock \emph{Journal of Physics A: Mathematical and General}, 39\penalty0
  (20):\penalty0 5739, 2006.

\bibitem[Paliathanasis and Tsamparlis(2014{\natexlab{b}})]{TypeII}
Andronikos Paliathanasis and Michael Tsamparlis.
\newblock The reduction of the laplace equation in certain riemannian spaces
  and the resulting type ii hidden symmetries.
\newblock \emph{Journal of Geometry and Physics}, 76:\penalty0 107--123,
  2014{\natexlab{b}}.

\bibitem[Hartle and Hawking(1983)]{hartle1983wave}
James~B Hartle and Stephen~W Hawking.
\newblock Wave function of the universe.
\newblock \emph{Physical Review D}, 28\penalty0 (12):\penalty0 2960, 1983.

\bibitem[Vilenkin(1982)]{vilenkin1982creation}
Alexander Vilenkin.
\newblock Creation of universes from nothing.
\newblock \emph{Physics Letters B}, 117\penalty0 (1):\penalty0 25--28, 1982.

\bibitem[Vilenkin(1984)]{vilenkin1984quantum}
Alexander Vilenkin.
\newblock Quantum creation of universes.
\newblock \emph{Physical Review D}, 30\penalty0 (2):\penalty0 509, 1984.

\bibitem[Capozziello et~al.(2013{\natexlab{d}})Capozziello, Harko, Lobo, and
  Olmo]{capozziello2013hybrid}
Salvatore Capozziello, Tiberiu Harko, Francisco~SN Lobo, and Gonzalo~J Olmo.
\newblock Hybrid modified gravity unifying local tests, galactic dynamics and
  late-time cosmic acceleration.
\newblock \emph{International Journal of Modern Physics D}, 22\penalty0
  (12):\penalty0 1342006, 2013{\natexlab{d}}.

\bibitem[Brans and Dicke(1961)]{brans1961mach}
Carl Brans and Robert~H Dicke.
\newblock Mach's principle and a relativistic theory of gravitation.
\newblock \emph{Physical Review}, 124\penalty0 (3):\penalty0 925, 1961.

\bibitem[Kaluza(1921)]{kaluza1921unitatsproblem}
Theodor Kaluza.
\newblock Zum unit{\"a}tsproblem der physik.
\newblock \emph{Sitzungsber. Preuss. Akad. Wiss. Berlin.(Math. Phys.)},
  1921\penalty0 (966972):\penalty0 45, 1921.

\bibitem[Klein(1926{\natexlab{a}})]{klein1926quantentheorie}
Oskar Klein.
\newblock Quantentheorie und f{\"u}nfdimensionale relativit{\"a}tstheorie.
\newblock \emph{Zeitschrift f{\"u}r Physik}, 37\penalty0 (12):\penalty0
  895--906, 1926{\natexlab{a}}.

\bibitem[Klein(1926{\natexlab{b}})]{klein1926atomicity}
Oskar Klein.
\newblock The atomicity of electricity as a quantum theory law.
\newblock \emph{Nature}, 118:\penalty0 516, 1926{\natexlab{b}}.

\bibitem[Williams(2015)]{williams2015field}
LL~Williams.
\newblock Field equations and lagrangian for the kaluza metric evaluated with
  tensor algebra software.
\newblock \emph{Journal of Gravity}, 2015, 2015.

\bibitem[Faraoni(2004)]{faraoni2004cosmology}
Valerio Faraoni.
\newblock \emph{Cosmology in scalar-tensor gravity}, volume 139.
\newblock Springer Science \& Business Media, 2004.

\bibitem[Capozziello et~al.(1997)Capozziello, Marmo, Rubano, and
  Scudellaro]{capozziello1997nother}
Salvatore Capozziello, Giuseppe Marmo, Claudio Rubano, and Paolo Scudellaro.
\newblock N{\"o}ther symmetries in bianchi universes.
\newblock \emph{International journal of modern physics D}, 6\penalty0
  (04):\penalty0 491--503, 1997.

\bibitem[Barrow(1995)]{barrow1995universe}
John~D Barrow.
\newblock Why the universe is not anisotropic.
\newblock \emph{Physical Review D}, 51\penalty0 (6):\penalty0 3113, 1995.

\bibitem[Rothman and Ellis(1986)]{rothman1986can}
Tony Rothman and GFR Ellis.
\newblock Can inflation occur in anisotropic cosmologies?
\newblock \emph{Physics Letters B}, 180\penalty0 (1):\penalty0 19--24, 1986.

\bibitem[Demianski et~al.(1992)Demianski, De~Ritis, Rubano, and
  Scudellaro]{demianski1992scalar}
M~Demianski, R~De~Ritis, C~Rubano, and P~Scudellaro.
\newblock Scalar fields and anisotropy in cosmological models.
\newblock \emph{Physical Review D}, 46\penalty0 (4):\penalty0 1391, 1992.

\bibitem[Linde(1982)]{linde1982new}
Andrei~D Linde.
\newblock A new inflationary universe scenario: a possible solution of the
  horizon, flatness, homogeneity, isotropy and primordial monopole problems.
\newblock \emph{Physics Letters B}, 108\penalty0 (6):\penalty0 389--393, 1982.

\bibitem[Ford(1987)]{ford1987cosmological}
Larry~H Ford.
\newblock Cosmological-constant damping by unstable scalar fields.
\newblock \emph{Physical Review D}, 35\penalty0 (8):\penalty0 2339, 1987.

\bibitem[Ryan and Shepley(2015)]{ryan2015homogeneous}
Michael~P Ryan and Lawrence~C Shepley.
\newblock \emph{Homogeneous relativistic cosmologies}.
\newblock Princeton University Press, 2015.

\bibitem[Tsamparlis and
  Paliathanasis(2011{\natexlab{b}})]{tsamparlis2011geometric}
Michael Tsamparlis and Andronikos Paliathanasis.
\newblock The geometric nature of lie and noether symmetries.
\newblock \emph{General Relativity and Gravitation}, 43\penalty0 (6):\penalty0
  1861--1881, 2011{\natexlab{b}}.

\bibitem[Misner(1969)]{misner1969quantum}
Charles~W Misner.
\newblock Quantum cosmology. i.
\newblock \emph{Physical Review}, 186\penalty0 (5):\penalty0 1319, 1969.

\bibitem[Barut and Raczka(1986)]{barut1986theory}
Asim~Orhan Barut and Ryszard Raczka.
\newblock \emph{Theory of group representations and applications}, volume~2.
\newblock World Scientific, 1986.

\bibitem[Capozziello et~al.(2012{\natexlab{c}})Capozziello, De~Laurentis, and
  Odintsov]{CapHd}
Salvatore Capozziello, Mariafelicia De~Laurentis, and Sergei~D Odintsov.
\newblock Hamiltonian dynamics and noether symmetries in extended gravity
  cosmology.
\newblock \emph{The European Physical Journal C}, 72\penalty0 (7):\penalty0
  1--21, 2012{\natexlab{c}}.

\bibitem[Bohm(1952{\natexlab{a}})]{bohma}
David Bohm.
\newblock A suggested interpretation of the quantum theory in terms of" hidden"
  variables. i.
\newblock \emph{Physical Review}, 85\penalty0 (2):\penalty0 166,
  1952{\natexlab{a}}.

\bibitem[Bohm(1952{\natexlab{b}})]{bohmb}
David Bohm.
\newblock A suggested interpretation of the quantum theory in terms of" hidden"
  variables. ii.
\newblock \emph{Physical Review}, 85\penalty0 (2):\penalty0 180,
  1952{\natexlab{b}}.

\bibitem[Demorest et~al.(2010)Demorest, Pennucci, Ransom, Roberts, and
  Hessels]{demorest}
PB~Demorest, Tim Pennucci, SM~Ransom, MSE Roberts, and JWT Hessels.
\newblock A two-solar-mass neutron star measured using shapiro delay.
\newblock \emph{Nature}, 467\penalty0 (7319):\penalty0 1081--1083, 2010.

\bibitem[Rawls et~al.(2011)Rawls, Orosz, McClintock, Torres, Bailyn, and
  Buxton]{rawls}
Meredith~L Rawls, Jerome~A Orosz, Jeffrey~E McClintock, Manuel~AP Torres,
  Charles~D Bailyn, and Michelle~M Buxton.
\newblock Refined neutron star mass determinations for six eclipsing x-ray
  pulsar binariesthis paper includes data gathered with the 6.5 m magellan
  telescopes located at las campanas observatory, chile.
\newblock \emph{The Astrophysical Journal}, 730\penalty0 (1):\penalty0 25,
  2011.

\bibitem[Van~Kerkwijk et~al.(2011)Van~Kerkwijk, Breton, and Kulkarni]{van}
MH~Van~Kerkwijk, RP~Breton, and SR~Kulkarni.
\newblock Evidence for a massive neutron star from a radial-velocity study of
  the companion to the black-widow pulsar psr b1957+ 20.
\newblock \emph{The Astrophysical Journal}, 728\penalty0 (2):\penalty0 95,
  2011.

\bibitem[Astashenok et~al.(2015{\natexlab{a}})Astashenok, Capozziello, and
  Odintsov]{astash}
Artyom~V Astashenok, Salvatore Capozziello, and Sergei~D Odintsov.
\newblock Magnetic neutron stars in f(r) gravity.
\newblock \emph{Astrophysics and Space Science}, 355\penalty0 (2):\penalty0
  333--341, 2015{\natexlab{a}}.

\bibitem[Astashenok et~al.(2014)Astashenok, Capozziello, and Odintsov]{astash2}
Artyom~V Astashenok, Salvatore Capozziello, and Sergei~D Odintsov.
\newblock Maximal neutron star mass and the resolution of the hyperon puzzle in
  modified gravity.
\newblock \emph{Physical Review D}, 89\penalty0 (10):\penalty0 103509, 2014.

\bibitem[Astashenok et~al.(2015{\natexlab{b}})Astashenok, Capozziello, and
  Odintsov]{astash3}
Artyom~V Astashenok, Salvatore Capozziello, and Sergei~D Odintsov.
\newblock Extreme neutron stars from extended theories of gravity.
\newblock \emph{Journal of Cosmology and Astroparticle Physics}, 2015\penalty0
  (01):\penalty0 001, 2015{\natexlab{b}}.

\bibitem[Ek{\c{s}}i et~al.(2014)Ek{\c{s}}i, G{\"u}ng{\"o}r, and
  T{\"u}rko{\u{g}}lu]{eksi}
Kaz{\i}m~Yavuz Ek{\c{s}}i, Can G{\"u}ng{\"o}r, and Murat~Metehan
  T{\"u}rko{\u{g}}lu.
\newblock What does a measurement of mass and/or radius of a neutron star
  constrain: Equation of state or gravity?
\newblock \emph{Physical Review D}, 89\penalty0 (6):\penalty0 063003, 2014.

\bibitem[Berti et~al.(2015)Berti, Barausse, Cardoso, Gualtieri, Pani, Sperhake,
  Stein, Wex, Yagi, Baker, et~al.]{berti}
Emanuele Berti, Enrico Barausse, Vitor Cardoso, Leonardo Gualtieri, Paolo Pani,
  Ulrich Sperhake, Leo~C Stein, Norbert Wex, Kent Yagi, Tessa Baker, et~al.
\newblock Testing general relativity with present and future astrophysical
  observations.
\newblock \emph{Classical and Quantum Gravity}, 32\penalty0 (24):\penalty0
  243001, 2015.

\bibitem[Oliveira et~al.(2015)Oliveira, Velten, Fabris, and
  Casarini]{oliveira2015neutron}
AM~Oliveira, HES Velten, JC~Fabris, and L~Casarini.
\newblock Neutron stars in rastall gravity.
\newblock \emph{Physical Review D}, 92\penalty0 (4):\penalty0 044020, 2015.

\bibitem[Palenzuela and Liebling(2015)]{palenzuela2015constraining}
Carlos Palenzuela and Steve Liebling.
\newblock Constraining scalar-tensor theories of gravity from the most massive
  neutron stars.
\newblock \emph{arXiv preprint arXiv:1510.03471}, 2015.

\bibitem[Cisterna et~al.(2015)Cisterna, Delsate, and Rinaldi]{cisterna}
Adolfo Cisterna, T{\'e}rence Delsate, and Massimiliano Rinaldi.
\newblock Neutron stars in general second order scalar-tensor theory: The case
  of nonminimal derivative coupling.
\newblock \emph{Physical Review D}, 92\penalty0 (4):\penalty0 044050, 2015.

\bibitem[Cisterna et~al.(2016)Cisterna, Delsate, Ducobu, and
  Rinaldi]{cisterna2016slowly}
Adolfo Cisterna, T{\'e}rence Delsate, Ludovic Ducobu, and Massimiliano Rinaldi.
\newblock Slowly rotating neutron stars in the nonminimal derivative coupling
  sector of horndeski gravity.
\newblock \emph{Physical Review D}, 93\penalty0 (8):\penalty0 084046, 2016.

\bibitem[Wojnar and Velten(2016)]{wojnar2016equilibrium}
Aneta Wojnar and Hermano Velten.
\newblock Equilibrium and stability of relativistic stars in extended theories
  of gravity.
\newblock \emph{arXiv:1604.04257}, 2016.

\bibitem[Velten et~al.(2016)Velten, Oliveira, and Wojnar]{velten2016free}
Hermano Velten, Adriano~M Oliveira, and Aneta Wojnar.
\newblock A free parametrized tov: Modified gravity from newtonian to
  relativistic stars.
\newblock \emph{Proceedings of Science (MPCS2015) 025, arXiv:1601.03000}, 2016.

\bibitem[Weinberg(1972)]{weinberg}
Steven Weinberg.
\newblock \emph{Gravitation and cosmology: principles and applications of the
  general theory of relativity}, volume~1.
\newblock Wiley New York, 1972.

\bibitem[Glendenning(2012)]{glende}
Norman~K Glendenning.
\newblock \emph{Compact stars: Nuclear physics, particle physics and general
  relativity}.
\newblock Springer Science and Business Media, 2012.

\bibitem[Oppenheimer and Volkoff(1939)]{OV}
J~Robert Oppenheimer and George~M Volkoff.
\newblock On massive neutron cores.
\newblock \emph{Physical Review}, 55\penalty0 (4):\penalty0 374, 1939.

\bibitem[Tolman(1939)]{Tolman:1939dn}
Richard~C Tolman.
\newblock Static solutions of einstein's field equations for spheres of fluid.
\newblock \emph{Physical Review}, 55\penalty0 (4):\penalty0 364, 1939.

\bibitem[Tolman(1987)]{tolman1987relativity}
Richard~Chace Tolman.
\newblock \emph{Relativity, thermodynamics, and cosmology}.
\newblock Courier Corporation, 1987.

\bibitem[Nice et~al.(2005)Nice, Splaver, Stairs, Oliver, Jessner, Kramer,
  Cordes, et~al.]{nice20052}
David~J Nice, Eric~M Splaver, Ingrid~H Stairs, L~Oliver, Axel Jessner, Michael
  Kramer, James~M Cordes, et~al.
\newblock A 2.1m? pulsar measured by relativistic orbital decay.
\newblock \emph{The Astrophysical Journal}, 634\penalty0 (2):\penalty0 1242,
  2005.

\bibitem[Capozziello et~al.(2014{\natexlab{b}})Capozziello, Lobo, and
  Mimoso]{mim}
Salvatore Capozziello, Francisco~SN Lobo, and Jos{\'e}~P Mimoso.
\newblock Energy conditions in modified gravity.
\newblock \emph{Physics Letters B}, 730:\penalty0 280--283, 2014{\natexlab{b}}.

\bibitem[Capozziello et~al.(2015{\natexlab{d}})Capozziello, Lobo, and
  Mimoso]{mim2}
Salvatore Capozziello, Francisco~SN Lobo, and Jos{\'e}~P Mimoso.
\newblock Generalized energy conditions in extended theories of gravity.
\newblock \emph{Physical Review D}, 91\penalty0 (12):\penalty0 124019,
  2015{\natexlab{d}}.

\bibitem[Mimoso et~al.(2015)Mimoso, Lobo, and Capozziello]{mim3}
Jos{\'e}~P Mimoso, Francisco~SN Lobo, and Salvatore Capozziello.
\newblock Extended theories of gravity with generalized energy conditions.
\newblock In \emph{Journal of Physics: Conference Series}, volume 600, page
  012047. IOP Publishing, 2015.

\bibitem[Koivisto(2006{\natexlab{b}})]{koivisto}
Tomi Koivisto.
\newblock A note on covariant conservation of energy--momentum in modified
  gravities.
\newblock \emph{Classical and Quantum Gravity}, 23\penalty0 (12):\penalty0
  4289, 2006{\natexlab{b}}.

\bibitem[Bamba(2016)]{bamba}
Kazuharu Bamba.
\newblock Thermodynamic properties of modified gravity theories.
\newblock \emph{arXiv preprint arXiv:1604.02632}, 2016.

\bibitem[Capozziello et~al.(1996)Capozziello, De~Ritis, Rubano, and
  Scudellaro]{capozziello1996nother}
S~Capozziello, R~De~Ritis, C~Rubano, and P~Scudellaro.
\newblock N{\"o}ther symmetries in cosmology.
\newblock \emph{La Rivista del Nuovo Cimento (1978-1999)}, 19\penalty0
  (4):\penalty0 1--114, 1996.

\bibitem[Paliathanasis(2015)]{thesis}
Andronikos Paliathanasis.
\newblock Symmetries of differential equations and applications in relativistic
  physics.
\newblock \emph{PhD Thesis, University of Athens (2014) ,arXiv:1501.05129},
  2015.

\bibitem[Ingarden and Jamio{\l}kowski(1980)]{ingarden}
Roman~Stanis{\l}aw Ingarden and Andrzej Jamio{\l}kowski.
\newblock \emph{Mechanika klasyczna}.
\newblock Pa{\'n}stwowe Wydawnictwo Naukowe, 1980.

\bibitem[Landau and Lifshitz(1994)]{landau}
D.D. Landau and E.M. Lifshitz.
\newblock \emph{The Classical Theory of Fields}.
\newblock Butterworth Heiemann, 4th edition, 1994.

\bibitem[Yano(1957)]{yano}
Kentaro Yano.
\newblock The theory of lie derivatives and its applications.
\newblock 1957.

\bibitem[Paliathanasis and Tsamparlis(2012)]{pal_heat}
Andronikos Paliathanasis and Michael Tsamparlis.
\newblock Lie point symmetries of a general class of pdes: The heat equation.
\newblock \emph{Journal of Geometry and Physics}, 62\penalty0 (12):\penalty0
  2443--2456, 2012.

\bibitem[Govinder(2001)]{govinder}
KS~Govinder.
\newblock Lie subalgebras, reduction of order, and group-invariant solutions.
\newblock \emph{Journal of mathematical analysis and applications},
  258\penalty0 (2):\penalty0 720--732, 2001.

\bibitem[Paliathanasis et~al.(2015)Paliathanasis, Tsamparlis, and
  Mustafa]{and_bianchi}
A.~Paliathanasis, M.~Tsamparlis, and M.~T. Mustafa.
\newblock Symmetry analysis of the klein–gordon equation in bianchi i
  spacetimes.
\newblock \emph{International Journal of Geometric Methods in Modern Physics},
  12\penalty0 (03):\penalty0 1550033, 2015.

\bibitem[Wainwright and Ellis(2005)]{wainwright2005dynamical}
John Wainwright and George Francis~Rayner Ellis.
\newblock \emph{Dynamical systems in cosmology}.
\newblock Cambridge University Press, 2005.

\bibitem[Hrycyna and Szyd{\l}owski(2015)]{hrycyna2015cosmological}
Orest Hrycyna and Marek Szyd{\l}owski.
\newblock Cosmological dynamics with non - minimally coupled scalar field and a
  constant potential function.
\newblock \emph{Journal of Cosmology and Astroparticle Physics}, 2015\penalty0
  (11):\penalty0 013, 2015.

\bibitem[Szyd{\l}owski and Hrycyna(2009)]{szydlowski2009scalar}
Marek Szyd{\l}owski and Orest Hrycyna.
\newblock Scalar field cosmology in the energy phase-space—unified
  description of dynamics.
\newblock \emph{Journal of Cosmology and Astroparticle Physics}, 2009\penalty0
  (01):\penalty0 039, 2009.

\bibitem[Sami et~al.(2012)Sami, Shahalam, Skugoreva, and
  Toporensky]{sami2012cosmological}
M~Sami, M~Shahalam, M~Skugoreva, and A~Toporensky.
\newblock Cosmological dynamics of a nonminimally coupled scalar field system
  and its late time cosmic relevance.
\newblock \emph{Physical Review D}, 86\penalty0 (10):\penalty0 103532, 2012.

\bibitem[Coley(1999)]{coley1999dynamical}
Alan~A Coley.
\newblock Dynamical systems in cosmology.
\newblock \emph{arXiv preprint gr-qc/9910074}, 1999.

\bibitem[Carloni et~al.(2005)Carloni, Dunsby, Capozziello, and
  Troisi]{carloni2005cosmological}
Sante Carloni, Peter~KS Dunsby, Salvatore Capozziello, and Antonio Troisi.
\newblock Cosmological dynamics of rn gravity.
\newblock \emph{Classical and Quantum Gravity}, 22\penalty0 (22):\penalty0
  4839, 2005.

\bibitem[Perko(2013)]{perko}
Lawrence Perko.
\newblock \emph{Differential equations and dynamical systems}, volume~7.
\newblock Springer Science and Business Media, 2013.

\bibitem[Hirsch et~al.(2012)Hirsch, Smale, and Devaney]{hirsch}
Morris~W Hirsch, Stephen Smale, and Robert~L Devaney.
\newblock \emph{Differential equations, dynamical systems, and an introduction
  to chaos}.
\newblock Academic press, 2012.

\bibitem[Szyd{\l}owski(2012)]{szydlowski2012generic}
Marek Szyd{\l}owski.
\newblock Generic scenarios of the accelerating universe.
\newblock \emph{Astrophysics and Space Science}, 339\penalty0 (2):\penalty0
  389--399, 2012.

\end{thebibliography}




\end{document}